\documentclass[preprint]{aastex}
\usepackage{rotating}

\shorttitle{Inconspicuous star clusters in the LMC}
\shortauthors{Samyaday Choudhury et al.}

\begin{document}


\title{Deep Washington photometry of inconspicuous star cluster candidates in the Large Magellanic Cloud}


\author {Samyaday Choudhury$^{1,2}$, Annapurni Subramaniam$^1$, Andr\'{e}s E. Piatti$^{3,4}$}
%   \newauthor % starts a new line in the
%              % author environment

\affil{$^1$Indian Institute of Astrophysics, 2B Koramangala, Bangalore, India 560034\\
  $^2$Indian Institute of Science, Bangalore, India 560012\\
  $^3$Observatorio Astron\'omico, Universidad Nacional de C\'ordoba, Laprida 854, X5000BGR, C\'ordoba, Argentina\\
  $^4$Consejo Nacional de Investigaciones Cient\'{\i}ficas y T\'ecnicas, Av. Rivadavia 1917, C1033AAJ,
Buenos Aires, Argentina}
\email{samyaday@iiap.res.in}


%\altaffiltext{1}{Visiting Astronomer, Cerro Tololo Inter-American Observatory.
%CTIO is operated by AURA, Inc.\ under contract to the National Science
%\altaffiltext{2}{Society of Fellows, Harvard University.}
%\altaffiltext{3}{present address: Center for Astrophysics,
%    60 Garden Street, Cambridge, MA 02138}
%\altaffiltext{4}{Visiting Programmer, Space Telescope Science Institute}
%\altaffiltext{5}{Patron, Alonso's Bar and Grill}


\begin{abstract}
We present deep Washington photometry of 45 poorly populated star cluster candidates in the Large Magellanic Cloud (LMC). We have performed a systematic study to estimate the parameters of the cluster candidates by matching theoretical isochrones to the cleaned and de-reddened cluster color-magnitude diagrams (CMDs). We were able to estimate the basic parameters for 33 clusters, out of which, 23 are identified as single clusters and 10 are found to be members of double clusters. Other 12 cluster candidates have been classified as possible clusters/asterisms. About 50$\%$ of the true clusters are in the 100-300 Myr age range, while some are older or younger. We have discussed the distribution of age, location, reddening with respect to field as well as size of true clusters. The sizes and masses of the studied sample are found to be similar to that of open clusters in the Milky Way. Our study adds to the lower end of cluster mass distribution in the LMC, suggesting that the LMC apart from hosting rich clusters also has formed small, less massive open clusters in the 100-300 Myr age range.
\end{abstract}





%%%%%%%%%%%%%%%%%%%%%%%%%%%%%%%%%%%%%%%%%%%%%%%%%%%%%%%%%%%%%%%%%%%%%%%%%%%%%%%

\section{INTRODUCTION}

Star clusters in the Large Magellanic Cloud (LMC) have been the target of detailed study to understand several processes, such as star formation, chemical evolution of the galaxy etc. \citep{olsetal91, pu00, groetal06}. The LMC hosts a large number of star clusters, and the most recent and extensive catalog of known clusters in the Magellanic Clouds is by \cite{betal08} (B08). However the cluster sample as mentioned by the authors is still incomplete. Most of the previous studies of LMC star clusters have targeted rich clusters which stand out from the field due to their high stellar density. Two of the profound studies of LMC star clusters using color-magnitude diagrams (CMDs) are by \cite{pu00} (PU00) and \cite{gletal10} (G10). PU00 estimated the ages of about 600 clusters utilizing the Optical Gravitational Lensing  Experiment (OGLE) II data, whereas G10 identified 1193 star clusters and estimated their ages utilizing the Magellanic Cloud Photometric Survey (\cite{zetal04}, MCPS) data. Both PU00 and G10 had carried out their work primarily for young clusters aged less than 1 Gyr, aiming to understand the cluster formation history. Another method that is generally employed to estimate the masses and ages of clusters for a sufficiently large number of samples, is the use of integrated photometry e.g. \cite{huntetal03, pop12}. \cite{pop12} (P12) estimated the age and mass for 920 LMC clusters based on previously published broadband photometry and the stellar cluster analysis package, MASSCLEANage.

Apart from rich clusters, the LMC also hosts a large number of clusters which have relatively less number of stars, similar to the open clusters of our Galaxy. Despite the above mentioned extensive studies, these category of clusters are in general either unstudied or poorly studied due to lack of deep photometric data. As these sparse clusters are also part of the cluster system of the LMC, it is necessary to study them, in order to understand the cluster formation and survival processes. The recent works of \cite{p12}, \cite{paletal13} and \cite{p14} were directed towards increasing the sample of poorly studied/unstudied clusters in the LMC. They used the cluster CMDs to estimate ages for such cluster candidates in the LMC using deep Washington photometry.

In this study, we attempt to increase our understanding  of inconspicuous stars clusters in the LMC using deep Washington photometric data. The current study thus aims at increasing the number of studied clusters, and disentangling the possible asterisms from genuine clusters. We have carried out a homogeneous analysis of 45 LMC clusters using deep photometric data in the Washington system. As is well-known, the Washington photometric system has been widely applied to studies of intermediate-age and old clusters in the Galaxy and in the Magellanic Clouds (e.g., \citealt{getal97, gs99, petal03, p12}). Particularly, the depth reached by the present photometric data helps us to trace the poorly populated clusters as well as trace the fainter end of the main-sequence of sparse clusters.

The paper is organized as follows. Section 2 deals with the acquisition and reduction of the aforementioned Washington photometric data. Section 3 describes the methods adopted for estimating the cluster parameters (radius, reddening and age). In section 4, we present the results derived from our analysis and discuss the same in section 5. We summarize our findings in section 6.

%%%%%%%%%%%%%%%%%%%%%%%%%%%%%%%%%%%%%%%%%%%%%%%%%%%%%%%%%%%%%%%%%%%%%%%%%%%%%%%

%% In a manner similar to \objectname authors can provide links to dataset
%% hosted at participating data centers via the \dataset{} command.  The
%% second curly bracket argument is printed in the text while the first
%% parentheses argument serves as the valid data set identifier.  Large
%% lists of data set are best provided in a table (see Table 3 for an example).
%% Valid data set identifiers should be obtained from the data center that
%% is currently hosting the data.
%%
%% Note that AASTeX interprets everything between the curly braces in the 
%% macro as regular text, so any special characters, e.g. "#" or "_," must be 
%% preceded by a backslash. Otherwise, you will get a LaTeX error when you 
%% compile your manuscript.  Special characters do not 
%% need to be escaped in the optional, square-bracket argument.

%% In this section, we use  the \subsection command to set off
%% a subsection.  \footnote is used to insert a footnote to the text.


%% This section also includes several of the displayed math environments
%% mentioned in the Author Guide.

%%%%%%%%%%%%%%%%%%%%%%%%%%%%%%%%%%%%%%%%%%%%%%%%%%%%%%%%%%%%%%%%%%%%%%%%%%%%%%%

\section{THE DATA}

This paper is a continuation of our series of studies about LMC cluster candidates \citep{petal09, p11, petal11, p12, paletal13, p14} using the $CT_1$ Washington photometric system \citep{c76,g96}. In this study, we focus on 45 LMC star cluster candidates for which Washington $C$ and Kron-Cousins $R$ data were retrieved from the National Optical Astronomy Observatory (NOAO) Science Data Management (SDM) Archives\footnote{http://www.noao.edu/sdm/archives.php.}. The cluster sample was selected from the cataloged clusters identified by \cite{p11} in the 21 LMC fields observed at the Cerro-Tololo Inter-American Observatory (CTIO) 4-m Blanco telescope with the Mosaic II camera attached (36 $\times$ 36 arcmin$^2$ field on to a 8K $\times$ 8K CCD detector array) through program 2008B-0912 (PI: D. Geisler). The volume of images includes calibration frames (zeros, sky-flats, etc.), and standard and program fields observed through the Washington $C$, and Kron-Cousins $R,I$ filters. Note that the $R$ filter has a significantly higher throughput as compared to the standard Washington $T_1$ filter so that $R$ magnitudes can be accurately transformed to yield $T_1$ magnitudes \citep{g96}. 

The data were reduced following the procedures documented by the NOAO Deep Wide Field Survey team \citep{jetal03} by utilizing the {\sc mscred} package in IRAF\footnote{IRAF is distributed by the National Optical Astronomy Observatories, which is operated by the Association of Universities for Research in Astronomy, Inc., under contract with the National Science Foundation.}. The different tasks performed went through overscan, trimming and cross-talk corrections, bias subtraction, obtained an updated world coordinate system (WCS) database, flattened all data images, etc., once the calibration frames (zeros, sky- and dome- flats, etc.) were properly combined. Nearly 90 independent measures of standard stars were derived per filter for each of the three nights (2008, Dec. 18-20) during which the observations were carried out, in order to obtain the coefficients of the transformation equations :

\begin{equation}
c = a_1 + T_1 + (C-T_1) + a_2\times X_C + a_3\times (C-T_1),
\end{equation}

\begin{equation}
r = b_1 + T_1 + b_2\times X_R + b_3\times (C-T_1),
\end{equation}

\begin{equation}
i = c_1 + T_1 - (T_1-T_2) + c_2\times X_I + c_3\times (T_1-T_2),
\end{equation}

\noindent where $a_i$, $b_i$, and $c_i$  ($i$ = 1, 2, and 3) are the fitted coefficients, and $X$ represents the effective airmass. Capital and lowercase letters represent standard and instrumental magnitudes, respectively. These equations were solved with the {\sc fitparams} task in IRAF and mean color terms ($a_3$,$b_3$,$c_3$) resulted to be -0.090$\pm$0.003 in $C$, -0.020$\pm$0.001 in $T_1$, and 0.060 $\pm$ 0.004 in $T_2$ while typical airmass coefficients ($a_2$,$b_2$,$c_2$) resulted in 0.31, 0.09 and 0.06 for $C$, $T_1$ and $T_2$, respectively. The nightly rms errors from the transformation to the  standard system were 0.021, 0.023 and 0.017  mag for $C$, $T_1$ and $T_2$, respectively, indicating these nights were of excellent photometric quality. 

The star-finding and point-spread-function (PSF) fitting routines in the {\sc daophot/allstar} suite of programs \citep{setal90} were used with the aim of performing the stellar photometry. For each frame, we selected $\sim$ 960 stars to fit a quadratically varying PSF, once the neighbors were eliminated using a preliminary PSF derived from the brightest, least contaminated $\sim$ 240 stars. Both groups of PSF stars were interactively selected. We then used the {\sc allstar} program to apply the resulting PSF to the identified stellar objects and to create a subtracted image which was used to find and to measure magnitudes of additional fainter stars. This procedure was repeated three times
for each frame. Finally, we standardized the resulting instrumental magnitudes and combined all the independent measurements using the stand-alone {\sc daomatch} and {\sc daomaster} programs, kindly provided by Peter Stetson. The final information gathered for each cluster consists of a running number per star, of the X and Y coordinates, of the measured $T_1$ magnitudes and $C-T_1$ and $T_1-T_2$ colors, and of the observational errors $\sigma(T_1)$, $\sigma(C-T_1)$ and $\sigma(T_1-T_2)$.

% Note that $T_1-T_2$ colors were included in this paper for completeness purposes, since they are thought to be useful for breaking the age-metallicity degeneracy when studying composite field star populations of the LMC. 

%\shortcite{petal12} performed artificial star tests over the whole image data set used in this paper. Such experiments consumed much time, but the authors gained a detailed knowledge of the quality and the scope of the data in the analysis of each image. This is also the best way to assess real photometric errors. {\bf The resultant completeness fractions showed that the 50$\%$ completeness level is located at $C$ $\sim$ 23.5-24.5 and $T_1$ $\sim$ 23.0-24.0, depending on the crowding and exposure time. In the subsequent analysis, we only analyse data of cluster candidates to the magnitude where the completeness level begins to fall below 100$\%$.} 

%%%%%%%%%%%%%%%%%%%%%%%%%%%%%%%%%%%%%%%%%%%%%%%%%%%%%%%%%%%%%%%

\section{ESTIMATION OF CLUSTER PARAMETERS}

For each of the selected cluster candidate fields, we made use of the measured stars within a radius of approximately 130$\arcsec$ around the central coordinates provided by B08. The cluster candidates analyzed here are relatively faint objects and are mostly unstudied. Furthermore, most of them are in turn relatively small angular size objects projected towards densely populated star fields. Our first step in the analysis is estimating the radius of each cluster from their radial density profiles (RDPs). Then, the second step is to remove the field stars within the cluster radius that contaminate the cluster Color-Magnitude diagram (CMD). Finally, we dealt with estimating the age and reddening of the genuine clusters by visually fitting isochrones to the cleaned cluster CMDs. The set of isochrones used in this study come from \cite{metal08}, with a metallicity of Z = 0.008, as judged from the observed LMC metallicity range for the last 3 Gyr \citep{pg13}. 

\subsection{Estimation of cluster center and cluster radius}

We assume the presence of a star cluster when a stellar density enhancement is identified in the spatial distribution of field stars. For this purpose, we first created finding charts for all clusters using measured stars with sizes proportional to their $T_1$ magnitudes. The cluster centers were estimated through an iterative method, starting from an eye estimated center (X$_e$, Y$_e$), for stars brighter than $T_1$ = 22.0 mag. We computed the average of the coordinates (X, Y) for all the stars distributed within 200 pixels around (X$_e$, Y$_e$), to estimate the central coordinate of the cluster, (X$_c$, Y$_c$). Iterations were carried out till the difference in the estimated center of two consecutive iterations is less than 10 pixels (2.7 arcsec). The number of stars per unit area in rings of 10 pixel width around the cluster center are used to build the RDPs. The RDPs were visually fitted with King (1962) profile: 
\begin{equation}
\rho(r) = \frac{\rho_0}{1 + (r/r_c)^2} + \rho_b ,
\end{equation}
where $\rho_0$ is the central density, $\rho_b$ is the background density and r$_c$ is the core radius. We fix the value of $\rho_0$ to visually fit the peak, and the value of $\rho_b$ to visually fit the background field density of the RDP at large radial distance. The r$_c$ value is then adopted so as to obtain the best visual fit of King profile to the RDP. We adopted cluster radii (r) is the distance from the cluster center at which the cluster density becomes equal to the background field density, which is taken as three times r$_c$. This is found to hold for most of the clusters. The clusters studied here are sparse and hence we have used this radius to include most of the cluster members. The error in the estimation of cluster radius is expected to be about $\pm$ 10$\arcsec$, which is about 3 times the bin size used in estimating the RDP.

In clusters where there is incompleteness of bright stars in the central cluster region due to saturation, we were unable to obtain their centers from the method described above. For these clusters we adopted either the central coordinates given by B08 or the eye estimated centers from the densest visible cluster regions. Likewise, when we were unable to reliably determine a cluster radius by visually fitting a King profile to the RDP or were unable to estimate an RDP, we chose the radius which brings out the cluster features clearly. For some of the clusters it was difficult to define a circular area, instead we used a rectangular region where the cluster might be most probably located. It is to be noted that the convention followed while mentioning the rectangular dimension is, size along X coordinate times size along Y coordinate in arcseconds.

\subsection{Cleaning the cluster Color Magnitude Diagram}

In order to analyze the cluster CMDs using stars located within the adopted cluster radii, one has to remove the contamination due to the field stars by performing a statistical field star subtraction. For field star subtraction, we selected field stars within an annular region of area equal to that of the cluster, with the inner radius of the annulus to be around twice or more than the cluster radius. The field stars in the cluster area are then removed by taking each star in the field CMD and finding the nearest one in the cluster CMD, considering a grid of magnitude-color bins with different sizes, starting with [$\Delta$$T_1$, $\Delta$$(C-T_1)$] = [0.01, 0.005] up to a maximum of [0.4, 0.2], where the units are in magnitude. In order to minimize effects due to field star density fluctuations, we repeated the decontamination procedure using different annular regions for each cluster and then compared the different resultant cleaned cluster CMDs. The cleaned cluster CMDs thus primarily show the  cluster features with minimum inescapable field characteristics. 

In cases where we could not consider an annular field region, the field stars were removed by selecting field regions (not necessarily circular), of equal cluster area in different parts of the observed field, located away from the cluster, and performing the cleaning procedure  described above. The cluster features that stay irrespective of the field used are considered as genuine cluster features and are used for estimating parameters.

\subsection{Estimating ages and reddenings for the cluster sample}

We determine the ages of the clusters by a visual fit of theoretical isochrones from \cite{metal08} with LMC metallicity (Z=0.008) to the cleaned cluster CMDs. For visually fitting theoretical isochrones to the observed CMDs, the $(C-T_1)$ colors and $T_1$ magnitudes need to be corrected for reddening and distance  modulus, respectively. \cite{smsn10} have created a reddening map for the LMC field using OGLE III  data \citep{uetal08}. They provide $E(V-I)$ color excesses for small regions within the galaxy. For a given cluster, we find the closest region in the reddening map and assume $E(V-I)$ of the field as the cluster reddening. The average of the distance between the clusters and their closest adopted field regions in the reddening map is approximately 6 arcmin. Finally, the theoretical isochrones were shifted to the observation plane according to equations \ref{eq:ct1obs} and \ref{eq:t1obs}:

\begin{equation}
(C-T_1)_{observed} = (C-T_1)_o + E(C-T_1) ,
\label{eq:ct1obs}
\end{equation}
where $E(C-T_1)$ = 1.97$E(B-V)$ \citep{gs99} and $E(B-V)$ = $E(V-I)$/1.25 \citep{bb88}. The expected error in reddening is less than $\pm$ 0.05 mag, which includes the photometric error and the error in the estimation of field reddening.

\begin{equation}
T_{1_{observed}} = M_{T_1} + 2.62E(B-V) + (m-M)_o , 
\label{eq:t1obs}
\end{equation}
as given by \cite{gs99}.

We assume a true distance modulus of $(m-M)_o$ = 18.50 for all the cluster sample, recently obtained by \cite{saetal10}. The cleaned cluster CMDs were matched with different isochrones after incorporating the corrections due to reddening and distance modulus. The age of the isochrone that visually provides the best match to the observed cluster CMD was adopted as the cluster age. However, whenever a cluster exhibits a dispersion in its CMD features, particularly near the turn-off, we over plotted additional isochrones, in order to take into account the observed spread. \cite{p14} discusses the error in the estimation of age. In general, the observed dispersion seen in the cluster CMDs can be encompassed with a spread of $\Delta$log($t$) $\sim$ 0.10. We discuss cases with a large spread in age or a large error in the age estimation separately.

%%%%%%%%%%%%%%%%%%%%%%%%%%%%%%%%%%%%%%%%%%%%%%%%%%%%%%%%%%%%

\begin{figure} 
\begin{center} 
%\resizebox{\hsize}{!}{\includegraphics*[width=10cm, height=5.0cm, angle=-90]{fig3.eps}}
\includegraphics[height=5.0in,width=5.0in]{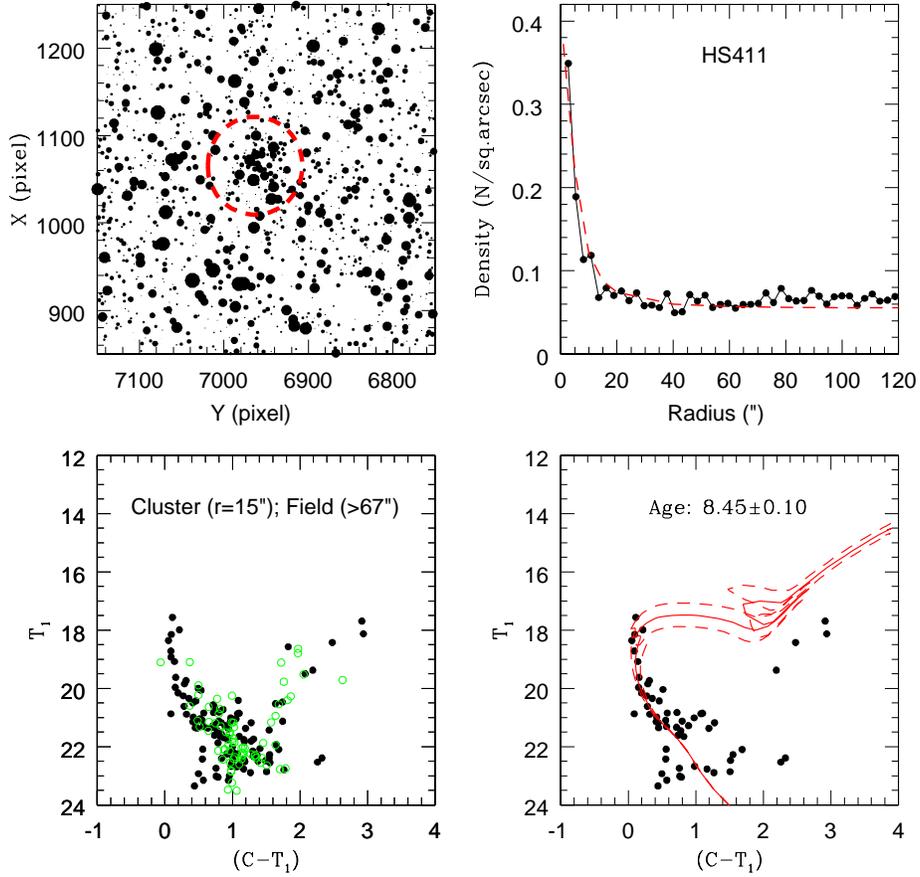}
\caption{\small 
HS411: (i) Top left - Spatial distribution of the stars in cluster field (North is up and East is left) along with the estimated cluster size (red dashed circle). (ii) Top right- King profile over plotted (red dashed line) to the RDP (black solid line). (iii) Bottom left- Uncleaned cluster CMD (black filled circles) within the cluster radius with the field CMD over plotted (green open circles). The cluster radius and the inner radius of the field are indicated. (iv) Bottom right-Isochrone over plotted (red solid and dashed lines) to cluster CMD after removal of field stars.\label{fig:true_cluster}}  
\end{center} 
\end{figure}  

\begin{sidewaystable} 
{\small
\begin{center}
%\begin{minipage}{200mm}
\caption{Estimated parameters for single clusters}
\label{table:tab1}
\begin{tabular}{|l|l|l|l|l|l|l|l|l|l|}
\hline \hline
Cluster  & $\alpha$ & $\delta$ & (X$_c$,Y$_c$)              & r                   & $E(C-T_1)$  & log(t)   & Lit.$^a$  &  Cross ID       \\
name     & (h m s)  & ($^\circ$ $'$ $''$) & (pixels)        & ($''$)              & (mag)    &          &           &           \\
\hline
BRHT45a  & 04 56 54 & -68 00 08 & (1833, 3604)              & 27                  & 0.15     & 8.10     & 8.00 (1)  & HS72, \\
         &          &           &                           &                     &          &          &           & KMHK326     \\
%         &          &           &      R=130pixel                                                                                            \\
BSDL77   & 04 50 29 & -67 19 33 & (5421, 5679)              & 24                  & 0.00     & 8.90     &           &         \\
BSDL268  & 04 55 52 & -69 42 21 & (5040, 3290) \bf{$^{*}$}  & (64.$''$8$\times$54$''$)    & 0.20     & 7.95     & 7.50 (1)  &         \\
BSDL631  & 05 06 34 & -68 25 38 & (509, 3280)               & 15                  & 0.00     & 8.35     & 7.50 (1)  & OGLE109        \\
%         &          &           &      R=130pixel                                                                                           \\
H88-33   & 04 55 41 & -67 47 00 & (4791, 2141)              & 18                  & 0.13     & 8.20-8.50      & 8.13 (3)    &  KMHK286        \\ 
H88-131  & 05 06 41 & -67 50 32 & (2841, 1234)              & 24                  & 0.06     & 9.00     &           & KMHK544       \\
%         &          &           &      R=130pixel                                                                                            \\
H88-265  & 05 18 05 & -69 10 18 & (1521, 3570) \bf{$^{*}$}  & 20 \bf{$^{**}$}     & 0.10     & 8.30     & 7.90 (2)  & OGLE323        \\
%         &          &           &      R=130pixel                                                                                           \\
H88-269  & 05 18 41 & -69 04 46 & (2760, 4314)              & 20 \bf{$^{**}$}     & 0.10     & 8.90     & 8.80 (2)  & OGLE337        \\ 
H88-320  & 05 41 58 & -69 02 51 & (5595, 2411)              & 27                  & 0.33     & 8.20     & 8.00 (1)  & KMHK1248        \\
%         &          &           &      R=100pixel                                                                                           \\ 
H88-331  & 05 44 11 & -69 20 00 & (1737, 5028)              & 24                  & 0.23     & 8.70     &           & KMHK1313         \\
%         &          &           &      R=130pixel                                                                                            \\
HS116    & 05 06 12 & -68 03 47 & (5389, 2885)              & 18                  & 0.08     & 8.55     &           & KMHK536         \\
%         &          &           &      R=100pixel                                                                                            \\
HS131    & 05 09 12 & -68 26 39 & (192, 6565)               & (27$''$$\times$27$''$)     & 0.16     & 9.10     &           &          \\
HS247    & 05 21 45 & -68 55 02 & (4874, 8051)              & 20 \bf{$^{**}$}     & 0.25     & 8.55     & 8.04 (3)         &          \\
%         &          &           &      R=130pixel                                                                                            \\
HS390    & 05 41 30 & -69 11 06 & (3764, 1829)              & 20                  & 0.45     & 8.25     & 7.92 (3)         & KMHK1239         \\
HS411    & 05 45 50 & -69 22 49 & (1065, 6963)              & 15                  & 0.34     & 8.45     &           & KMHK1345         \\
HS412    & 05 45 56 & -69 16 19 & (2546, 7088) \bf{$^{*}$}  & 25 \bf{$^{**}$}     & 0.34     & 8.10     & 8.10 (1)  & KMHK1347         \\
KMHK95   & 04 47 26 & -67 39 35 & (809, 1705)               & 21                  & 0.08     & 8.55     &           &          \\
%         &          &           &      R=150pixel                                                                                            \\
KMHK907  & 05 26 12 & -70 58 53 & (486, 3558)               & 21                  & 0.18     & 8.40     &           &          \\
%         &          &           &      R=130pixel                                                                                            \\
KMHK975  & 05 29 59 & -67 52 44 & (7467, 582)               & 21                  & 0.10     & 8.30     & 7.40 (1)  &          \\
%         &          &           &      R=130pixel                                                                                            \\
LW54     & 04 46 04 & -66 54 41 & (470, 7304)               & 18                  & 0.00     & 8.60     & 8.60 (1)  & KMHK72         \\
NGC1793  & 04 59 38 & -69 33 27 & (7068, 7815)              & 25 \bf{$^{**}$}     & 0.21     & 8.05     & 8.00 (1)  & SL163, \\
         &          &           &                           &                     &          &          &           & ESO56SC43,\\ 
         &          &           &                           &                     &          &          &           & KMHK405         \\
%         &          &           &      R=170pixel                                                                                            \\
SL397    & 05 20 12 & -68 54 15 & (5076, 6175) \bf{$^{*}$}  & 25 \bf{$^{**}$}     & 0.16     & 8.20     & 7.80 (1)  &          \\
SL579    & 05 34 13 & -67 51 23 & (7723, 5989)              & 18                  & 0.13     & 8.15     & 7.80 (1)  & KMHK1085         \\

\hline     
\end{tabular}
\tablecomments{For Table 1, 2 and 3: $^{*}$ implies cases where we adopted either the central coordinates given by B08 or the eye estimated centers from the densest visible cluster regions as the cluster center. $^{**}$ are cases where either we could not over plot King profile to the RDP or the RDP could not be estimated. For these cases, the estimated radius is the one at which the cluster profile becomes prominent. Cases where one could not define a circular area, the possible rectangular area of cluster is mentioned (dimension along X coordinate times that along Y coordinate)}.\\
\tablerefs{\hspace{2.0ex} Lit.$^a$: (1) \cite{gletal10}; (2) \cite{pu00}; (3) \cite{pop12}}

%\end{minipage}
\end{center}
}
\end{sidewaystable} 

%-------------------------------
%-------------------------------

\begin{sidewaystable}
{\small
\begin{center}
%\begin{minipage}{200mm}
\caption{Double Clusters}
\label{table:tab2}
\begin{tabular}{|l|l|l|l|l|l|l|l|l|l|}
\hline \hline
Cluster  & $\alpha$ & $\delta$ & (X$_c$,Y$_c$)              & r                   & $E(C-T_1)$  & log(t)   & Lit.$^a$  &  Cross ID  \\
name     & (h m s)  & ($^\circ$ $'$ $''$) & (pixels)        & ($''$)              & (mag)    &          &           &            \\  
\hline

BSDL341  & 04 58 15 & -68 02 57 & (1219, 5349)              & 24                  & 0.17    & 8.45      & 7.64 (3)  &            \\
%        &          &           &      R=100pixel                                                                                            \\
H88-52   & 04 58 10 & -68 03 37 & (1031, 5229)              & 27                  & 0.08    & 9.05      & 8.73 (3)  &  KMHK365    \\
%        &          &           &      R=100pixel                                                                                            \\
\hline
HS154    & 05 10 56 & -67 37 36 & (5716, 6660)              & 24                  & 0.13    & 8.65     & 8.80 (2)  & H88-189, \\
         &          &           &                           &                     &         &          &           & KMHK625, \\
         &          &           &                           &                     &         &          &           & OGLE194 \\
HS156    & 05 11 11 & -67 37 37 & (5727, 6961)              & 21                  & 0.13    & 9.05     &           & H88-190, \\
         &          &           &                           &                     &         &          &           & KMHK632, \\
         &          &           &                           &                     &         &          &           & OGLE199 \\
%        &          &           &      R=130pixel                                                                                            \\
\hline
KMHK979  & 05 29 39 & -70 59 02 & (387, 7390) \bf{$^{*}$}   & 20 \bf{$^{**}$}     & 0.17    & 7.90      & 7.30 (1)  & GKK-O101  \\
HS329    & 05 29 46 & -71 00 02 & (150, 7475) \bf{$^{*}$}   & (37.$''$8$\times$40.$''$5)    & 0.00      & 8.90-9.00     &    & KMHK984 \\
\hline
SL230    & 05 06 34 & -68 21 47 & (1380, 3281)\bf{$^{*}$}   & 25 \bf{$^{**}$}     & 0.16    & 7.90      & 7.40 (1)  & BRHT29b, \\
         &          &           &                           &                     &         &           &           & OGLE107 \\
SL229    & 05 06 25 & -68 22 30 & (1230, 3088)\bf{$^{*}$}   & 21                  & 0.12    & 8.50      & 8.35 (2)  & BRHT29a, \\
         &          &           &                           &                     &         &           &           & OGLE105 \\
\hline
SL551    & 05 31 51 & -67 59 28 & (5956, 2989)              & 20 \bf{$^{**}$}     & 0.18    & 8.15      & 7.90 (1)  & BRHT38a, \\
         &          &           &                           &                     &         &           &           & KMHK1027,\\
         &          &           &                           &                     &         &           &           & GKK-O202 \\
BRHT38b  & 05 31 58 & -67 58 18 & (6200, 3140)\bf{$^{*}$}   & (27$''$$\times$32.$''$4)   & 0.16    & 8.25      & 8.00 (3)   & KMHK1032 \\
\hline     
\end{tabular}
\end{center}
%\end{minipage}
}
\end{sidewaystable}

%-------------------------------
%-------------------------------

\begin{table}
{\small
\begin{center}
%\begin{minipage}{200mm}
\caption{Possible clusters and Asterisms}
\label{table:tab3}
\begin{tabular}{|l|l|l|l|l|l|l|l|l|l}
\hline \hline
Cluster  & $\alpha$ & $\delta$ & (X$_c$,Y$_c$)              & r                   & $E(C-T_1)$  & log(t)   & Lit.$^a$  &  Cross ID  \\
name     & (h m s)  & ($^\circ$ $'$ $''$) & (pixels)        & ($''$)              & (mag)    &          &           &            \\  
\hline
BSDL677  & 05 07 54 & -67 55 44 & (7177, 5070)\bf{$^{*}$}   & 21                  & 0.08     & 8.25     &           &           \\
H88-235  & 05 15 47 & -69 11 31 & (1243, 780)               & 15 \bf{$^{**}$}     & 0.12     & 8.75     & 8.55 (3)  & OGLE277   \\
%        &          &           &      R=150pixel                                                                                            \\
H88-244  & 05 16 17 & -69 09 15 & (1749, 1410)\bf{$^{*}$}   & 25 \bf{$^{**}$}     & 0.25     & 8.30     & 8.10 (2)  & OGLE285        \\
H88-279  & 05 20 02 & -69 15 40 & (265, 5890) \bf{$^{*}$}   & 20 \bf{$^{**}$}     & 0.16     & 8.10     & 8.00 (2)  & OGLE361        \\
H88-288  & 05 21 15 & -69 01 43 & (3403,7486)               & 18                  & 0.25     & 8.40     &  8.04 (3)  &         \\
%         &          &           &      R=150pixel                                                                                           \\ 
H88-289  & 05 21 20 & -69 00 30 & (3674, 7521)\bf{$^{*}$}   & 20 \bf{$^{**}$}     & 0.25     & 8.45     & 7.80 (3)  &           \\
H88-307  & 05 40 26 & -69 14 55 & (2905, 559)\bf{$^{*}$}    & (54$''$$\times$54$''$)      & 0.30     & 8.25     &           &           \\
H88-316  & 05 41 39 & -69 13 46 & (3159, 1950)\bf{$^{*}$}   & (54$''$$\times$54$''$)      & 0.30     & 8.25     & 8.00 (1)  &           \\
%HS130    & 05 09 15 & -67 41 59 & (4792, 4474)\bf{$^{*}$}   & 13 \bf{$^{**}$}     & 0.09     & 8.30     &           & KMHK588   \\
KMHK378  & 04 58 22 & -69 48 11 & (3608, 6451)              & 15                  & 0.14     & 8.45     & 7.40 (1)  &           \\
%        &          &           &      R=150pixel                                                                                            \\
KMHK505  & 05 04 33 & -67 58 32 & (6609, 791)               & 18                  & 0.11     & 8.75     &           &          \\
OGLE298  & 05 16 53 & -69 09 00 & (1800, 2135)\bf{$^{*}$}   & 15                  & 0.25     & 8.30     & 7.30 (2)  &           \\
SL269    & 05 09 35 & -67 48 38 & (3291, 4878)              & 25 \bf{$^{**}$}     & 0.11     & 8.25     &           & KMHK598,\\
         &          &           &                           &                     &          &          &           & GKK-O216   \\
\hline     
\end{tabular}
%\end{minipage}
\end{center}
}
\end{table} 

%--------------------------
%--------------------------

\begin{figure} 
\begin{center} 
\includegraphics[height=3.2in,width=3.2in]{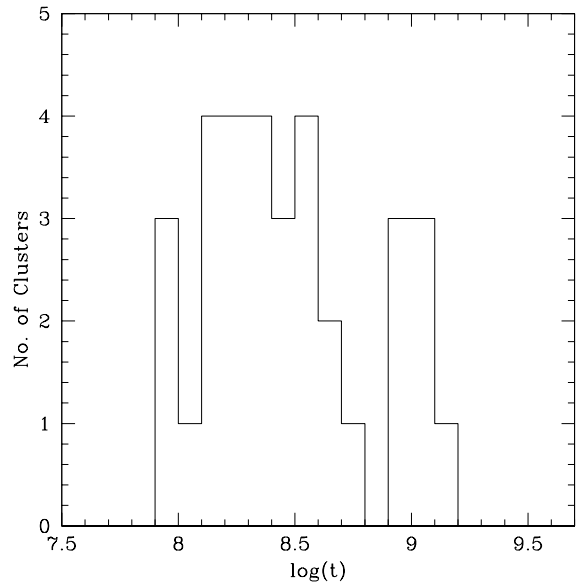}
\caption{\small Age distribution of 33 true clusters.\label{fig:hist_age}}  
\end{center} 
\end{figure}

\begin{figure} 
\begin{center} 
\includegraphics[height=3.5in,width=5.0in]{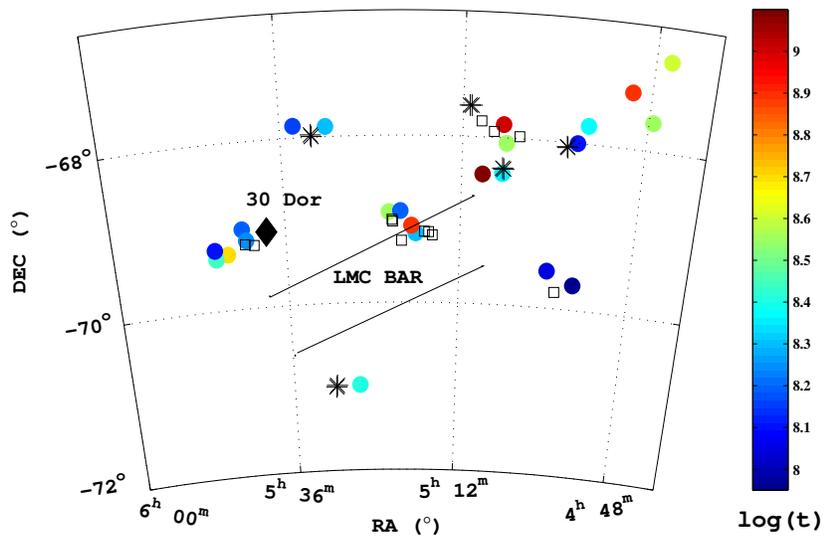}
\caption{\small 
Spatial distribution of 45 studied clusters in the LMC: Color coded circles denote the 23 true single clusters. Black asterisks denote the location of 5 pairs of double cluster and the black open boxes denote the location of 12 possible clusters/asterisms. 30 Doradus (black diamond) and the LMC bar (parallel lines) are also shown. \label{fig:spatial_age}}
\end{center} 
\end{figure}

\section[]{RESULTS}
%{\bf
We classified the studied clusters into two groups namely true clusters and possible clusters/asterisms. True clusters are the ones which have prominent cluster features (upper MS and/or MSTOs) and where one could satisfactorily estimate the cluster parameters like radius, age and reddening. There are 33 such clusters out of which 23 fields contain single clusters (Table \ref{table:tab1}) and 5 fields contain a pair of clusters (Table \ref{table:tab2}). There are 12 cases where the cluster features are either poor (only a few stars in their upper MS and MSTO) or the cluster features are suspicious/missing. Due to the difficulty in confirming the presence of an actual cluster in those fields and getting a satisfactory estimation of their cluster parameters, they are categorized as possible clusters/asterisms (Table \ref{table:tab3}). We have presented the finding chart, cluster CMD (before and after field star correction) and the estimated RDP (wherever possible), for all clusters, in the Appendix (available in the online version). The true single and double clusters are presented in Appendix A and B respectively, whereas the possible cluster/asterisms are presented in Appendix C. In Tables 1, 2 and 3, we have presented the coordinates, RA and Dec, the center of the cluster (to correlate with the finding chart), estimated cluster radius in arcseconds, reddening $E(C-T_1)$,
 age, earlier age estimate and cross IDs. In Table \ref{table:tab2}, the 5 double clusters are listed with members of each pair grouped together.

As an example, the derived parameters and the corresponding diagram for a true single cluster, HS411 is presented in Figure \ref{fig:true_cluster}. In the multi panel plot, the top left panel shows the schematic chart with the red dashed circle denoting the estimated extent of the cluster. The RDP of HS411 is shown with a King profile in the top right panel. The bottom left panel shows the CMD of stars within the estimated radius, and field stars located in adopted field region. The inner radius of the field region chosen is also indicated
in the figure. The bottom right panel shows isochrone  visually fitted to the cleaned CMD of HS411 cluster. Similar multi panel plots have been created for all the analyzed clusters unless stated otherwise (see Appendix A and B).

The true clusters, listed in Table \ref{table:tab1} and \ref{table:tab2}, stand out either in terms of number density or features in the CMD, or both. There are clusters, for which either the RDP did not show a strong peak (BSDL631, H88-265, H88-269 and HS247) or where we could not obtain an RDP (BSDL268, NGC1793 and SL397), suggesting a poor density enhancement in the cluster region. This is likely to be due to saturation effects caused by the presence of bright stars near the cluster center, resulting in missing stars and incompleteness in the central region for these clusters. As our data could not confirm the density enhancement, we tried to identify the density enhancement using other optical data. The OGLE III is one of the complete and relatively deep surveys of the inner region of the LMC (4$^\circ$-5$^\circ$) with good spatial resolution. We extracted OGLE III fields corresponding to each of these clusters and created similar finding charts using V magnitude. Though the OGLE III data helps for young clusters with bright stars, the OGLE III data is not of much help for relatively older clusters. We have discussed these cases in Appendix A.

The clusters in Table \ref{table:tab3},  possible clusters/asterisms, fall in to this group mainly due to inability of the present study to identify/detect any of the cluster characteristics for stars in the cluster region. This group has two types of clusters, one type could be clusters, but we could not estimate their parameters reliably, while the others could not even be detected/recognized reliably as clusters. We define possible cluster candidates as those with the following properties: 1) they have marginal spatial density enhancement with respect to the field, 2) cluster features in the CMD are very poorly defined with only a few stars in their upper MS and MSTO, and 3) there is identifiable difference in the evolutionary features of the CMD for the cluster and field regions. The first two properties make it difficult to clearly estimate the cluster parameters such as radius, age and reddening, whereas the third one suggests that there may be a cluster present in the location. Asterisms are those objects with marginal or no spatial apparent density enhancement with respect to the field, the cluster features in the CMD were either suspicious or completely missing, and the CMD of the cluster and the field region appear almost similar. As there is a very thin dividing line between the two types, we have put them as one group. In our study we report 12 cases which belong to this category. Detailed discussion of individual objects and their corresponding plots (similar to that of Figure \ref{fig:true_cluster}) are presented in Appendix C. We provide a limit on the possible spatial extent and age of these objects, if at all these objects are true clusters. Given the very sparse nature of the objects under this category, the age uncertainty could be larger ($\Delta$log($t$) $\sim$ 0.20). To avoid clutter in the CMD we have only shown isochrones of age $\pm$ 0.10 with respect to the best visually fitted isochrone. We also took help of OGLE III field for verification, but these clusters are too poor to confirm them as clusters, using this data. As our data is much deeper when compared to OGLE III and other survey data, the present data should have detected the presence of such poor and faint clusters. 8 out of 12 have been previously studied by either PU00, G10 or by P12. We need deeper data, like LSST to identify the true nature of these clusters and also to reliably estimate the parameters of the possible clusters.

The age distribution of the 33 true clusters is shown as a histogram in Figure \ref{fig:hist_age}. The age is considered on a logarithmic scale and the bin size is chosen to be equal to the typical age error in this study (i.e. $\Delta$log(t)= $\pm$ 0.10) so as to avoid any bias in the distribution due to error in age. For H88-33 and HS329 where we could only find out a range in age, their mean age is considered in the histogram. It is clearly seen that approximately 50$\%$ of the clusters are in the age range log(t)=8.0 $\textendash$ 8.5 (i.e. $\sim$ 100-300 Myr). Rest of them are either younger or older. BSDL268, KMHK979 and SL230 lie at the youngest end ($<$100 Myr) of the age distribution. These clusters could be younger than their estimated ages as some upper MS or MSTO stars are missing from the center of the cluster region due to saturation effect. The clusters H88-52, HS156, H88-131 and HS131 are aged around 1 Gyr and lie at the oldest age end of the histogram.

23 out of the 33 true clusters have previous age estimates. We have compared our age estimation with these previous studies of LMC clusters by PU00 and G10, who used CMDs to estimate ages using the OGLE II and the Magellanic Cloud Photometric Survey (MCPS) data respectively. The limiting magnitude of their data do not allow them to detect older clusters and constrain them to only younger clusters. We also compared our results with P12, who used integrated photometry to estimate ages of LMC clusters mentioned in \cite{huntetal03}. With our deep photometric data we were able to go faint enough to detect older MSTOs as well as identify clusters with poor cluster features. We have also included the study of LMC star cluster The estimated ages are compared with these previous estimates in the respective tables. The age comparison shows some agreement with our results as well as deviation. This is discussed in the next section.
%}

%%%%%%%%%%%%%%%%%%%%%%%%%%%%%%%%%%%%%%%%%%%%%%%%%%%%%%%%%%%%%%%%%%%%%%%%%%%%%%%

\begin{figure}[ht]
\centering
\begin{minipage}[b]{0.45\linewidth}
\includegraphics[height=3.0in,width=3.0in]{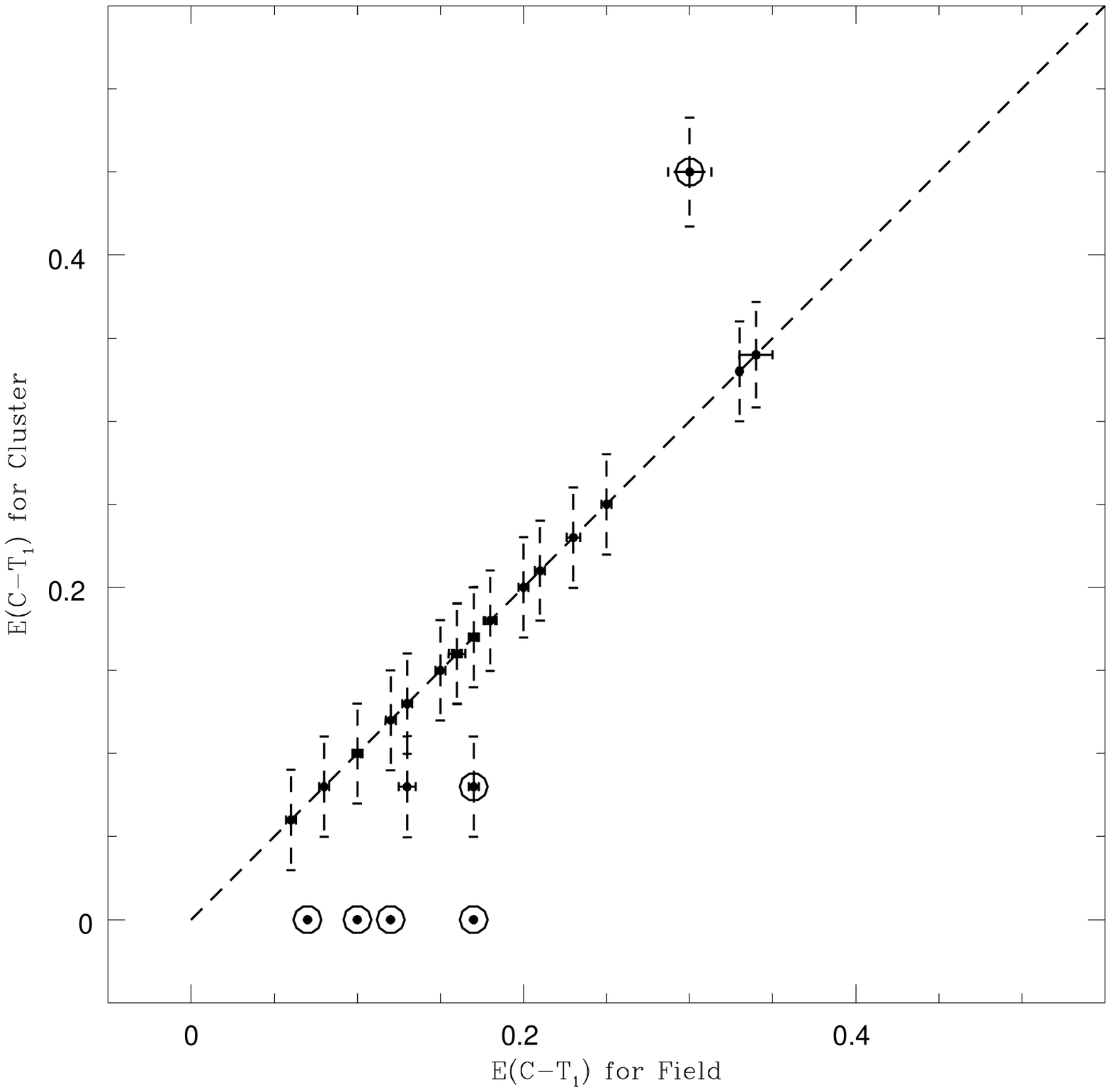}
\caption{\small Correlation between estimated cluster and field reddening values for 33 true clusters. The dashed line corresponds to the one-to-one relation. The deviations are marked with open circles.}
\label{fig:red_compare}  
\end{minipage}
\quad
\begin{minipage}[b]{0.45\linewidth}
\includegraphics[height=3.0in,width=3.0in]{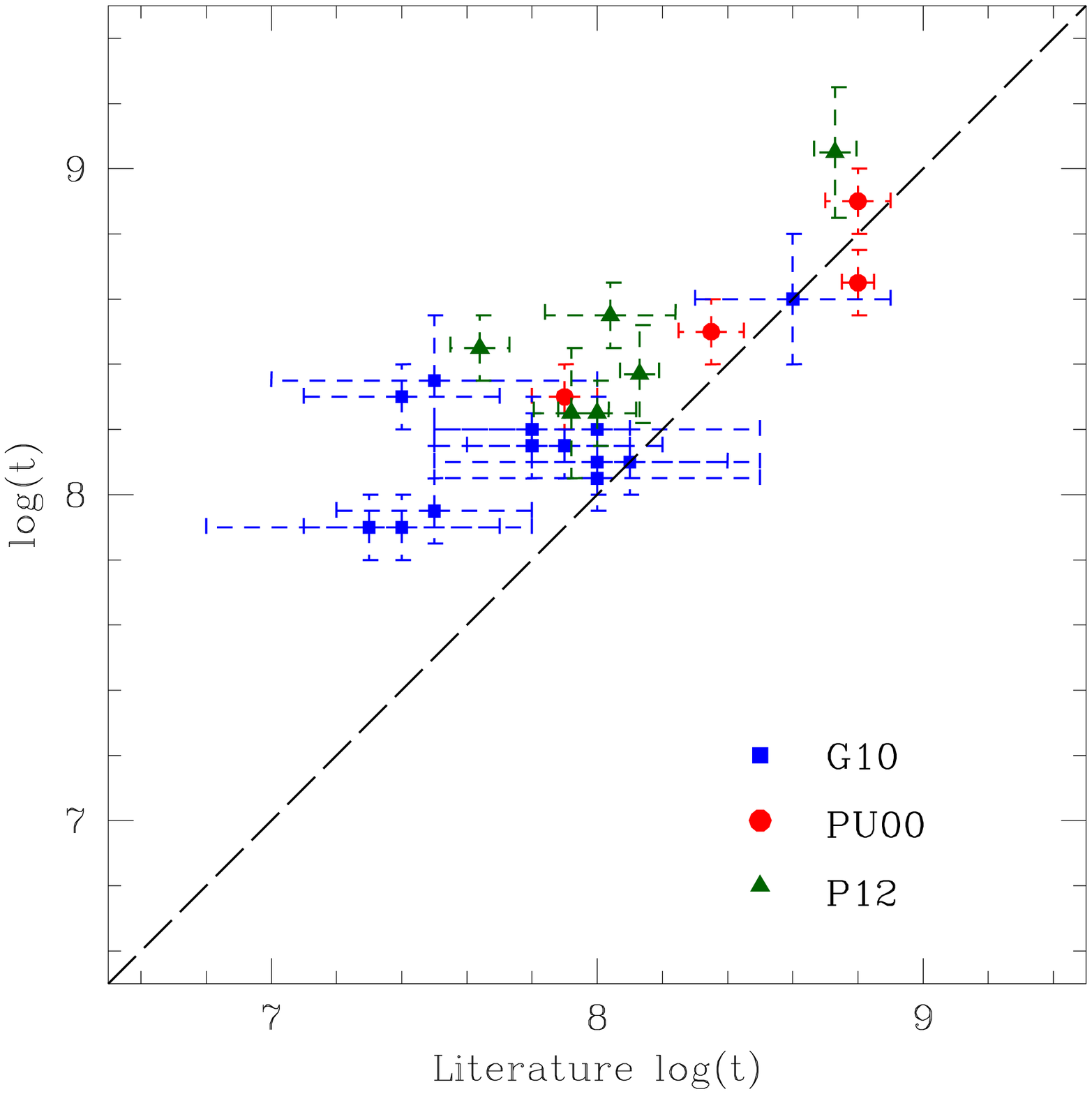}
\caption{\small Correlation between derived and published ages by G10, PU00 and P12 for 23 out of 33 true clusters. For values in G10 we have used the upper limit of the errors. For values of P12 we have used the mean of upper and lower limit of the errors. The dashed line corresponds to the one-to-one relation.}
\label{fig:age_compare}   
\end{minipage}
\end{figure}

\section[]{DISCUSSION}

%{\bf
We have presented a study of 45 inconspicuous poorly studied clusters in the LMC, based on deep Washington photometry. The data for all the 45 clusters in the Washington photometric system is presented/analyzed for the first time and the data have enough photometric depth to identify the turn-off of faint, poorly populated clusters. The data also cover substantial field region around the cluster to effectively remove the field contamination even in regions with varying density and reddening. The coverage as well as the depth of the data has also helped in the identification of possible asterisms and cluster candidates from the sample. We were able to estimate the basic parameters for 33 clusters,
out of which, 23 are identified as single clusters and 10 are found to be members of double cluster. We suggest that the rest of the 12 clusters studied here are possible cluster candidates or asterisms. 10 out of the 33 true clusters are previously unstudied, thus we report their sizes, reddenings and ages for the first time. The rest of the 23 clusters have been previously studied (by either PU00, G10 or P12) but our data is deep enough to derive accurate ages.

The spatial distribution of all the 45 clusters studied is shown in Figure \ref{fig:spatial_age}. The 23 true single clusters are represented by color coded filled circles according to their ages, whereas the 5 double clusters and the 12 cases of possible clusters/asterisms are depicted by black asterisks and black open squares respectively. The studied clusters are seen to be located mostly in the inner LMC with a few of them located towards the north west side. The figure also shows the location of the bar and 30 Dor. For clusters lying in and near such crowded regions there could be issues due to differential reddening as well as varying field density. While performing the field star removal from the CMD, we have taken care to choose the field regions carefully so as to minimize the effects of variation in density and reddening. This has helped in extracting the cluster features in the cluster CMD and the derivation of cluster parameters efficiently. 

The RDP method is used to estimate the radius of most of the clusters. In the case of a few clusters, either the RDP did not show a strong peak or we were unable to derive the RDP due to incompleteness of bright stars near the cluster center, and the cluster radii was chosen as the one at which the cluster features prominently shows up. We have compared the present estimation of radii for 33 true clusters with their previous estimations cataloged by B08, and we find the estimation to be comparable. B08 gives the dimension of the major and minor axis for these objects, using which we calculated their mean radius. It is seen that the clusters analyzed are typically small angular sized objects with radii in the range of 10$''$ - 40$''$ ($\sim$ 2 - 10 pc). We also find similarity in size for objects under the category of single clusters and double clusters. All the clusters studied here are small in size and are similar to the sizes of open clusters in our Galaxy. Thus this study helps to derive the parameters of open cluster like objects in the LMC.

As mentioned earlier, the reddening value for the clusters were adopted from the reddening map of field region \citep{smsn10}. We find that the average separation of clusters studied here from the nearest field region is about 6 arcmin. Though this adopted reddening was found satisfactory for most of the clusters, for a few cases the isochrones had to be further reddened/de-reddened with respect to the field reddening values in order to get proper visual fit of isochrone. Figure \ref{fig:red_compare} shows a comparison between the reddening values of 33 true clusters and their corresponding fields. The errors in the estimation of the field reddening, taken from \citep{smsn10} and the error in the reddening of the cluster (photometric error and field reddening error) are shown in the X and Y axes respectively. The range of reddening values for the clusters studied is about 0.05-0.45 mag. The cluster reddening is found to be very similar to the field reddening, except for 6 clusters (marked by open circles). Out of these 6 cases, 4 clusters have zero reddening (BSDL77, BSDL631, LW54 and HS329) and we have not indicated any error bar for them in the figure. H88-52 has less reddening compared to the corresponding field. The cluster HS390 has largest value of reddening in our sample and is found to be located near the 30 Dor region. The observed difference in the reddening for these 6 clusters may be due to the spatial variation in reddening and/or due to projection effect. The shift in the reddening values of these clusters with respect to the field may cause an additional error in the age estimation. We expect that the error in the age estimation for these clusters can be up to $\Delta$log($t$) $\sim$ 0.20.  

We studied the parameters of 5 double clusters listed in Table \ref{table:tab2}. A large number of double clusters in the LMC are identified by Dieball, M\"{u}ller \& Grebel (2002). These 5 pairs are also found to be mentioned in their catalog. The sizes of double clusters are found to be similar to the single clusters. We expect the reddening and age to be similar among the cluster members, for the double clusters to be candidate for binary pair. In 2 of the double clusters (BSDL341 $\&$ H88-52; KMHK979 $\&$ HS329) we find that the estimated reddening differs significantly. These clusters also have fairly large difference in age. The above two differences suggest that these clusters may be pairs because of projection. Two cluster pairs (HS154 $\&$ HS156; SL230 $\&$ SL229) have similar reddening, but the ages are not similar, suggesting that they might not be physical pairs. The cluster pair SL551 $\&$ BRHT38b have comparable reddening and age, within errors. Thus, these two clusters may be a candidate for a binary pair.

The ages of 23 true clusters are compared with the results of PU00, G10 and P12 in Figure \ref{fig:age_compare}. Out of 33 true clusters, 13 clusters are in common with G10 while 6 clusters are in common with PU00. A couple of cluster in our sample (BSDL631 and SL230) have been studied by both G10 and PU00. For BSDL631, PU00 estimated an age of log(t)$<$6.70 (with no error) and G10 as log(t)=7.50 (0.30$\leq$ error $<$0.50). In the case of SL230, the age estimation by PU00 (log(t)=7.30$\pm0.05$) and G10 (log(t)=7.40, with error$<$0.30) agree very well within their errors. For both these clusters we considered the results by G10 for comparison as it is more recent and the errors are appropriate for such poor clusters. The figure shows that the G10 clusters are primarily younger than log(t) $\sim$ 8.50. P12 used integrated photometry to obtain the ages of LMC star clusters, and 16 out of 33 true clusters are common with their study. Out of these 16 clusters, G10 has mentioned ages for 8 (BRHT45a, BSDL268, H88-320, KMHK975, NGC1793, SL397, SL579 and SL551), whereas PU00 has mentioned ages for 2 (H88-265 and H88-269) clusters. Integrated light provides information about the combined stellar population along the line of sight. The clusters studied here are small angular sized objects embedded within relatively denser field. Use of integrated photometry to estimate ages for such cases can produce poor results due to field star contamination, stochastic effects and relatively shallower photometric depth. Thus for comparison we have adopted the age estimates given by G10 and PU00 wherever available as their results are more reliable, and we considered only 6 clusters whose ages are given by integrated photometry and finds mention only in P12. The comparison suggests that, the present study estimates relatively older ages for clusters younger than log(t) $\sim$ 7.50. In the case of young clusters, there is a possibility that our data has missed out brighter stars and this might cause the above anomaly. The ages of older clusters are comparable. When we compared our age estimates with those of P12 for 16 clusters, we find that most of the common clusters are younger than log(t)=8.00. P12 estimated relatively younger ages compared to our estimates for these clusters. We have only a few older clusters and we find that P12 estimated significantly younger ages for these clusters.

The clusters studied here are of the age up to 1 Gyr and most of them are poor and inconspicuous clusters. We have also suggested that some of the clusters could be asterisms, and not true clusters. The estimates of the radial extent (2-10 pc) of these clusters suggest that they are similar to the open clusters in our Galaxy.  We simulated  CMDs of a few rich and young clusters using Marigo et al. (2008) isochrones.  Assuming the Salpter's mass function and incorporating observational error, we simulated CMDs for the mass range 10 - 0.5 M$_\odot$. The total mass simulated is adjusted to create the same number of stars within 3 magnitudes below the turn-off, as in the observed CMD, after incorporating Poisson error. This is expected to reduce the effect of incompleteness of fainter stars in the CMD. The estimated masses were found to be up to 1000 M$_\odot$ for rich clusters. This is also found to be comparable to the mass estimates of P12 for the same clusters. Adopting the masses of all common clusters from P12, the relatively rich clusters in our sample are up to 1000 M$_\odot$, whereas the poor clusters are only a few 100 M$_\odot$. Thus, the masses also suggest that these clusters are similar to the open clusters of our Galaxy. Also, we find that the mass limit at which the object is unable to be identified as a cluster is about a few 100 M$_\odot$. At this mass limit, the number of stars formed are unable to create either a notable density enhancement, or an identifiable cluster sequence in the CMD or both.

\cite{baetal13} have studied the star cluster formation history of the LMC, using some recent catalogs that include PU00, G10 and P12. Their Figure 3 shows a plot between the mass and age of all the clusters. The mass of the clusters range from a few hundred to a few thousand of  M$_\odot$ and are within the age range 10 Myr to 1 Gyr. The figure shows that the number of clusters at the higher end of mass distribution is relatively more compared to the lower end, where the contribution is primarily from G10. However the age limit for G10 clusters is constrained only to $\leq$ 300 Myrs. The estimated mass range of our studied sample contribute to this lower end of the cluster mass distribution, and also contain clusters beyond the age limit of G10. As mentioned earlier, a significant fraction of our clusters lie within the age range of (100-300) Myr which corresponds to the recent star formation in the last 200 Myrs. This suggests that the LMC has produced very low mass clusters, along with the massive and rich clusters in the recent past . Thus, in the context of understanding the cluster mass function in the LMC, our study had added many clusters to the lower mass limit of the distribution. The poor clusters are also of interest to understand the survival time of these clusters in the LMC. Table \ref{table:tab3} suggests that the possible clusters/asterisms are in the age range of log(t)= 8.10-8.80, probably suggesting their survival time. This time scale is also similar to that in our Galaxy (a few hundred Myr) \citep{boetal10}. 

\section[]{SUMMARY}

\begin{itemize}
\item The study is aimed to enlarge the number of objects confirmed as genuine star clusters and to estimate their fundamental parameters. We present Washington photometry of 45 star clusters distributed in the inner LMC, some of which are projected towards relatively crowded fields. 
\item Out of 45 clusters, 33 are found to be true (genuine) cluster candidates whereas the remaining 12 clusters could only be categorized as possible cluster/asterism. We successfully estimated the parameters of the true clusters and at the same time list the parameters of the other category, if at all they are clusters. The age distribution of the true clusters shows that about 50$\%$ fall within the age range (100-300) Myr while some are older or younger.
\item The physical sizes and masses of the studied clusters are found to be similar to that of open clusters in the Milky Way. Our study adds to the lower end of cluster mass distribution in the LMC. Thus the LMC apart from hosting rich clusters also contains such small, less massive open clusters particularly in the (100-300) Myr range.
\item The 12 poor cases in the category of possible clusters/asterism also draws attention in the sense that they can throw light on the survival time of such objects in the LMC. 
\end{itemize}

%}

%%%%%%%%%%%%%%%%%%%%%%%%%%%%%%%%%%%%%%%%%%%%%%%%%%%%%%%%%%%%%%%%%%%%%%%%%%%%%%%%%

%======================================================================================================================

\appendix

\section[]{SINGLE CLUSTERS}
%{\bf
\begin{figure*}
\begin{center}
 \includegraphics[height = .35\textheight, keepaspectratio]{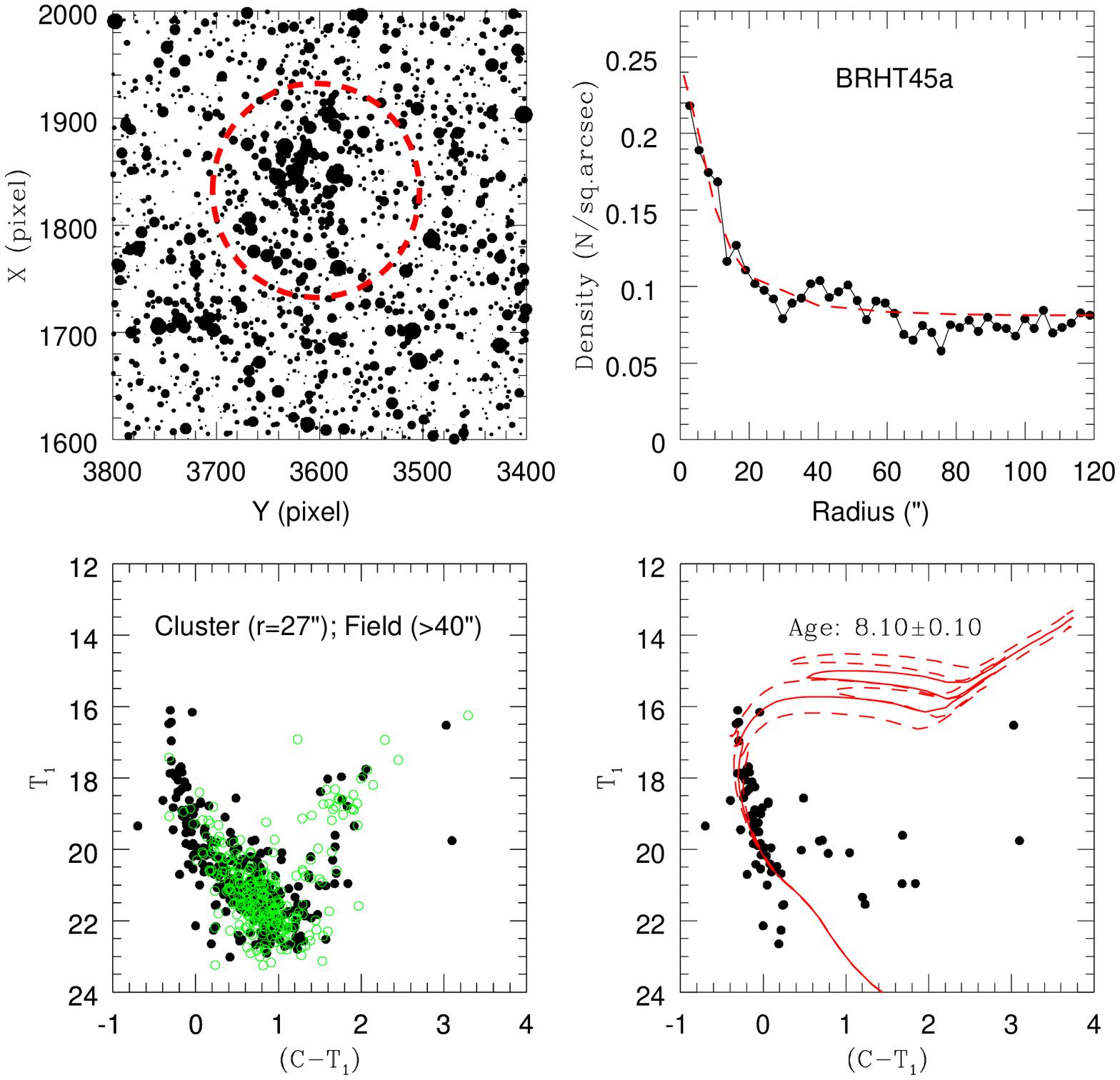}\hspace{1.0cm}
 \includegraphics[height = .35\textheight, keepaspectratio]{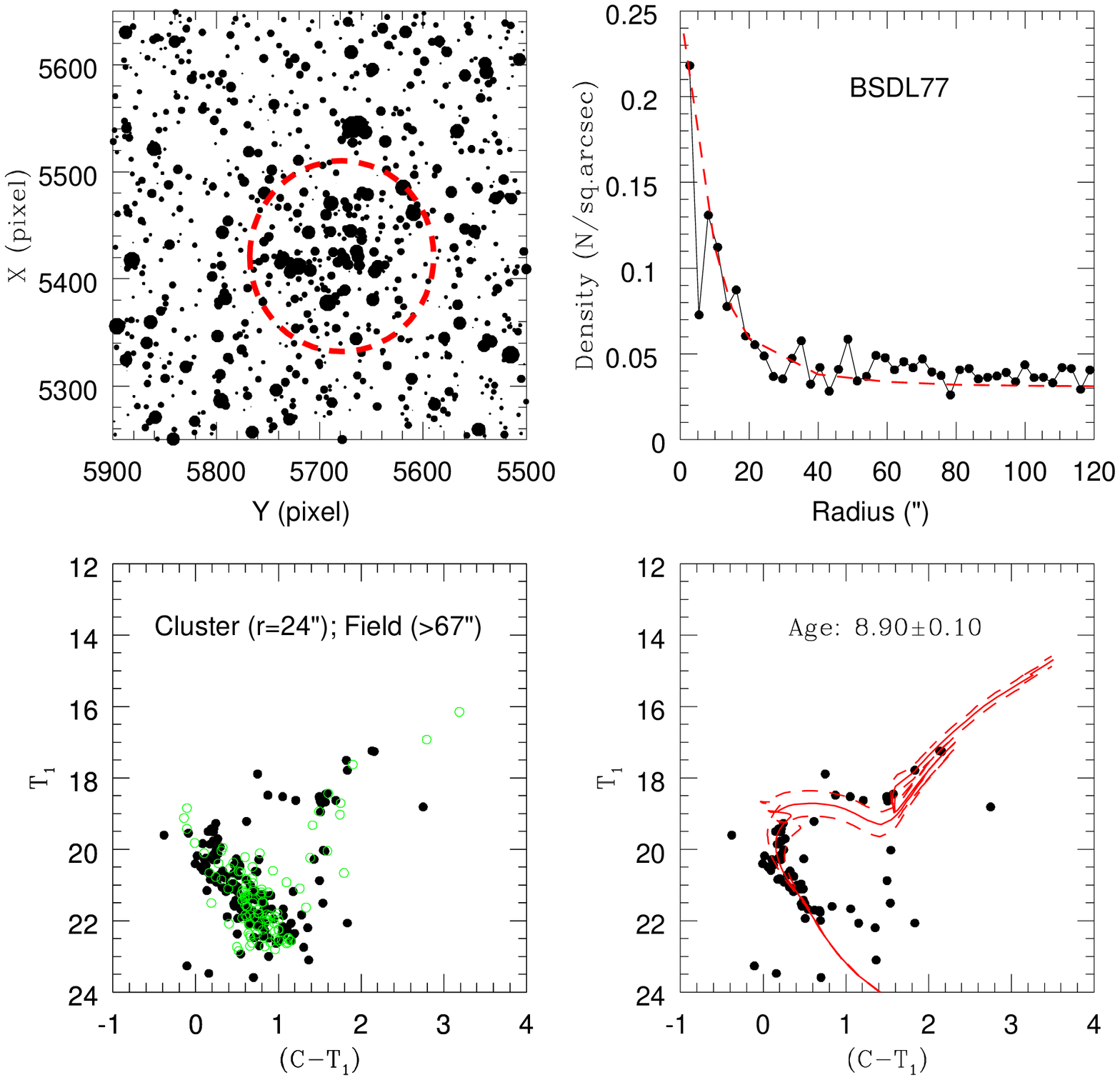} \\
 \includegraphics[height = .35\textheight, keepaspectratio]{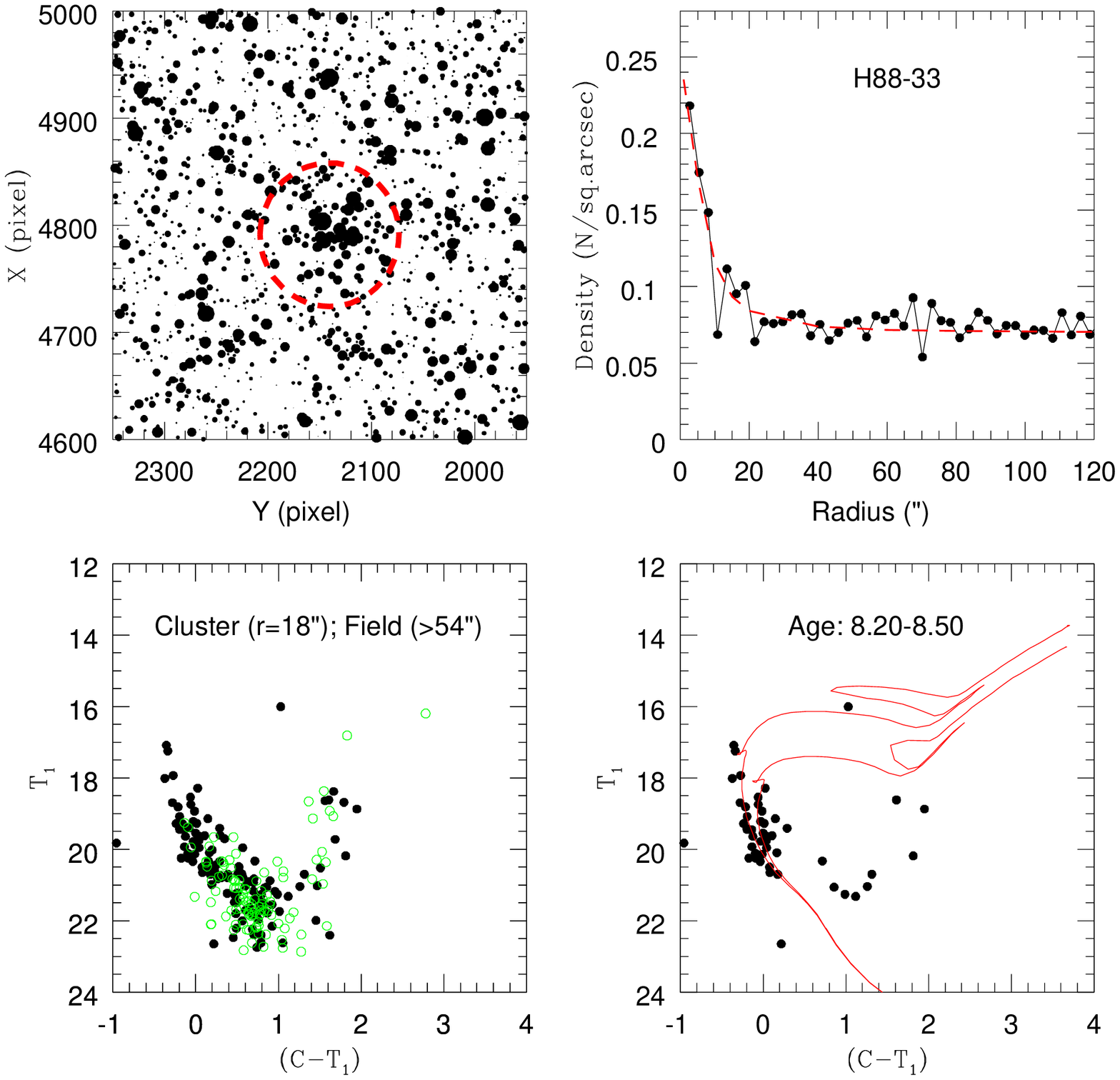}\hspace{1.0cm} 
 \includegraphics[height = .35\textheight, keepaspectratio]{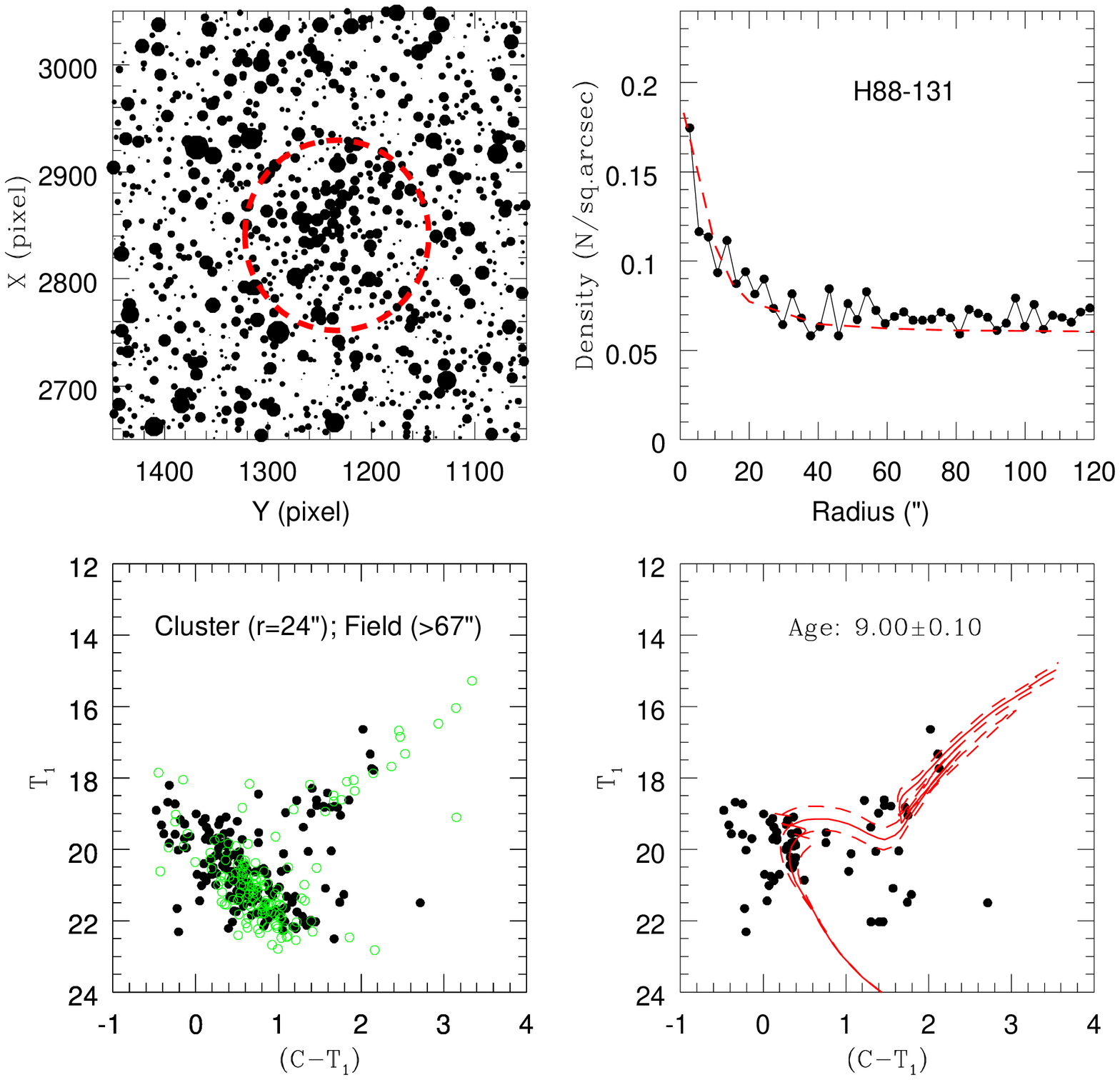} \\
 \includegraphics[height = .35\textheight, keepaspectratio]{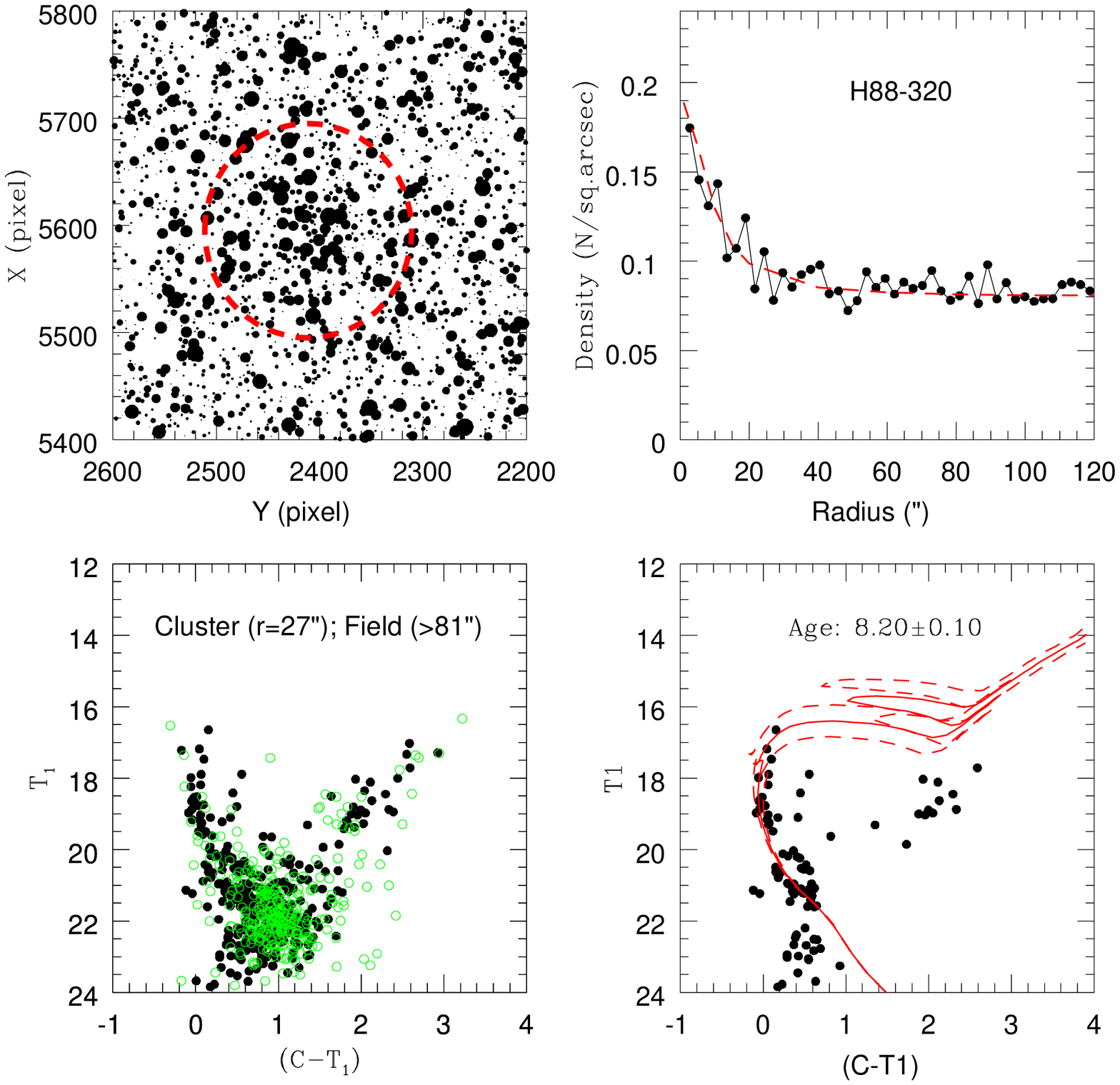}\hspace{1.0cm} 
 \includegraphics[height = .35\textheight, keepaspectratio]{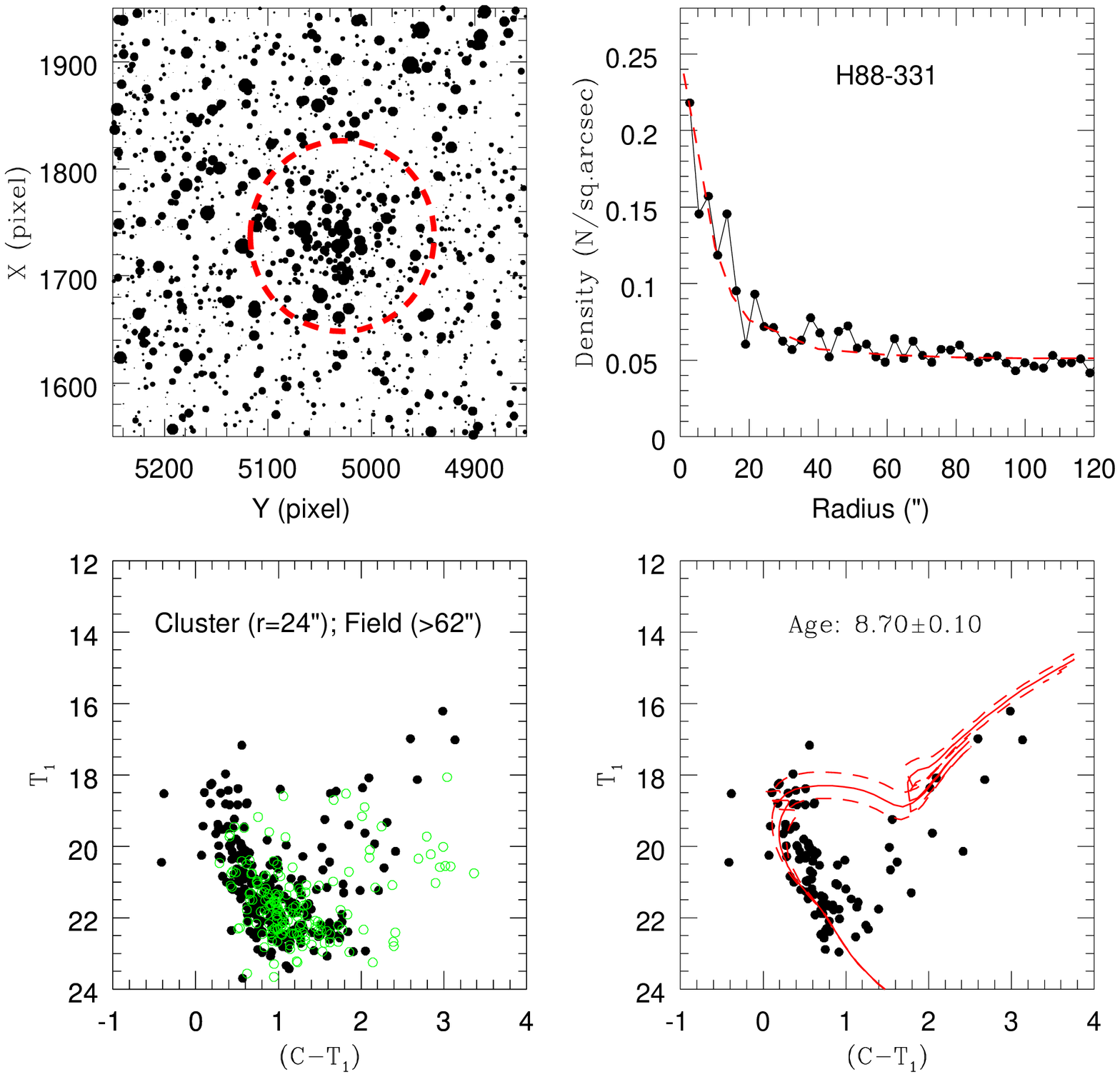} \\
\caption{\small {Single cluster candidates: BRHT45a, BSDL77, H88-33, H88-131, H88-320 and H88-331. Panel description for each cluster is same as Figure 1.}}
\label{single1}
\end{center}
\end{figure*}

\begin{figure*}
\begin{center}
 \includegraphics[height = .35\textheight, keepaspectratio]{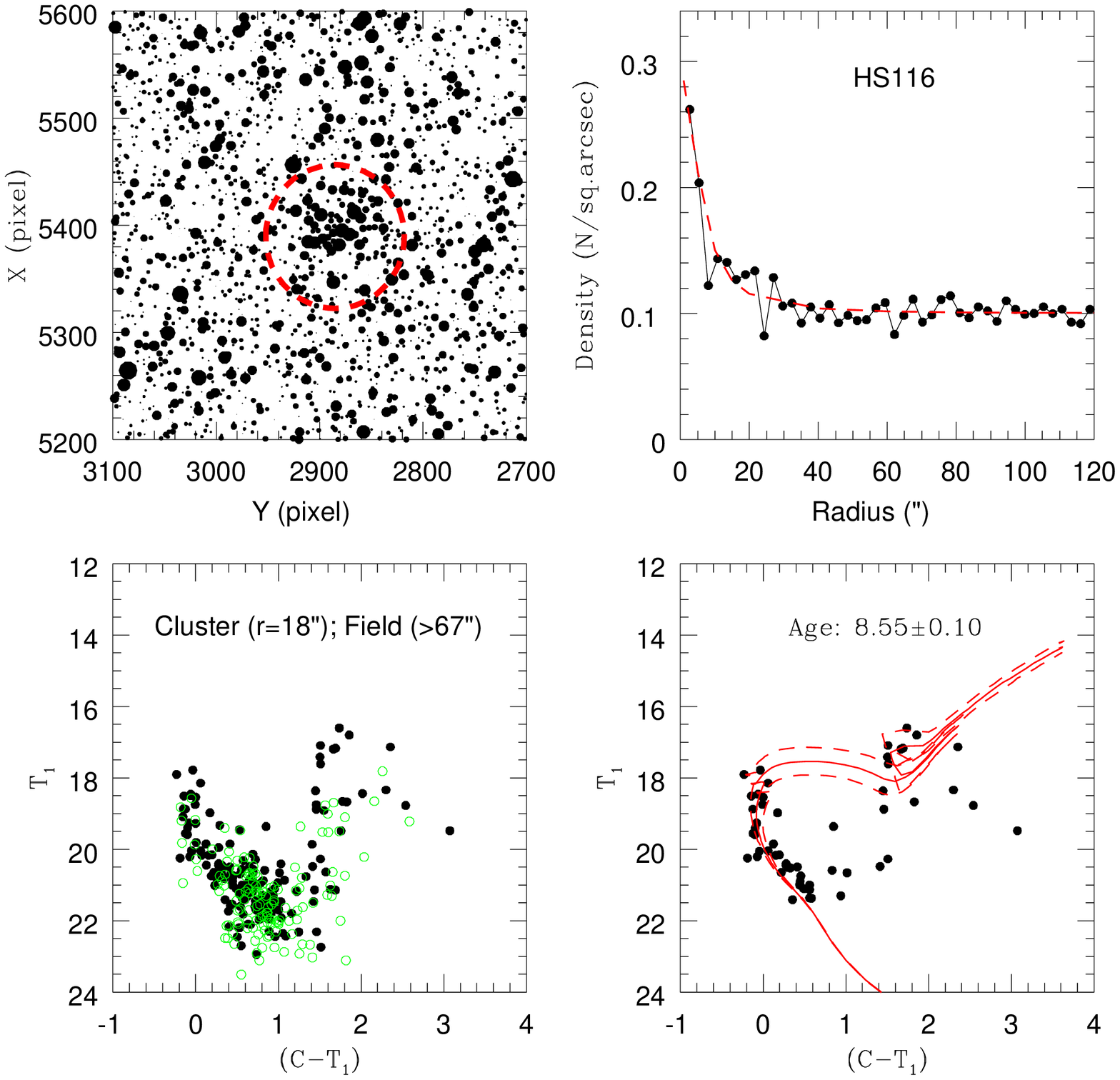}\hspace{1.0cm} 
 \includegraphics[height = .35\textheight, keepaspectratio]{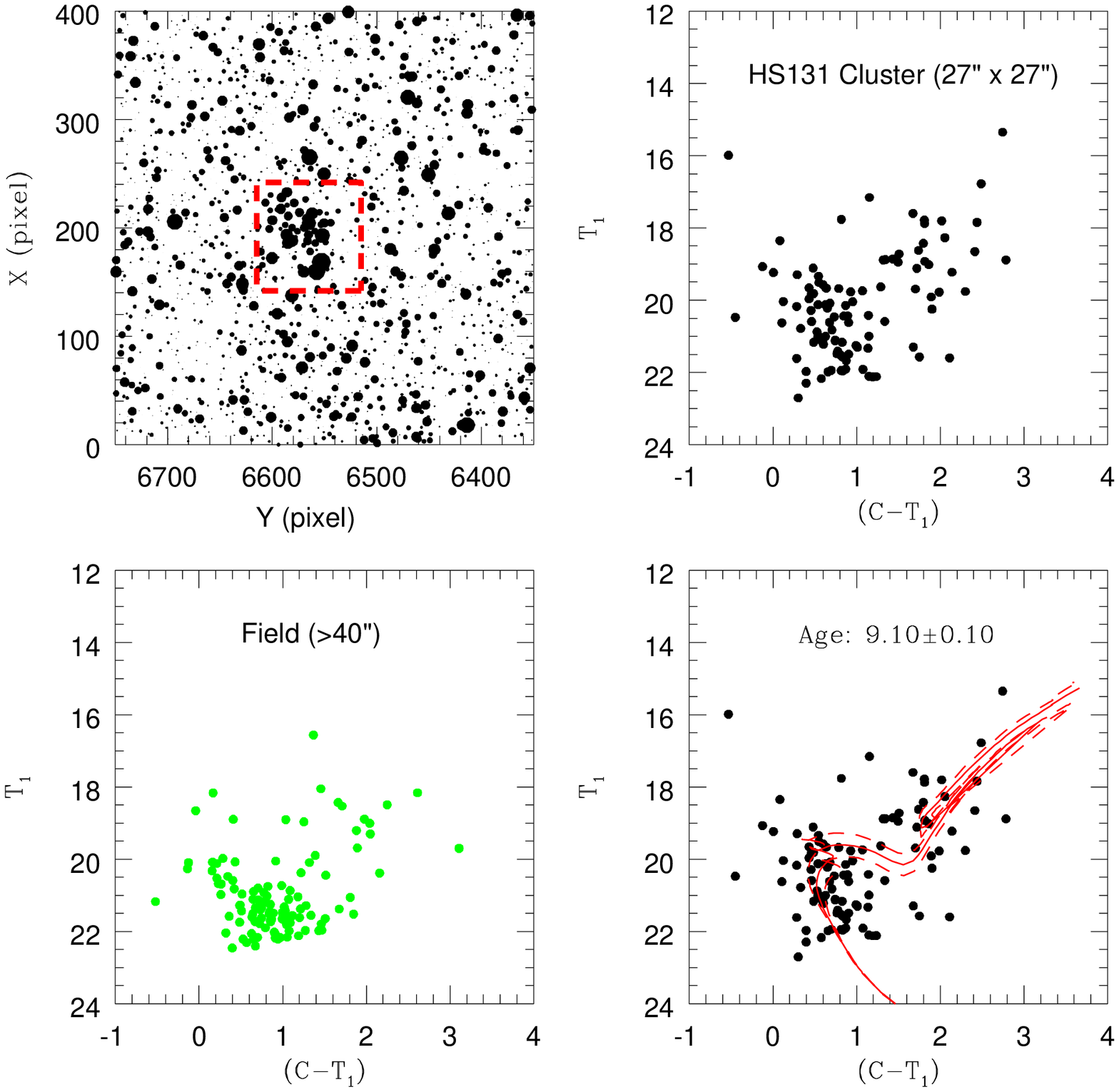} \\
 \includegraphics[height = .35\textheight, keepaspectratio]{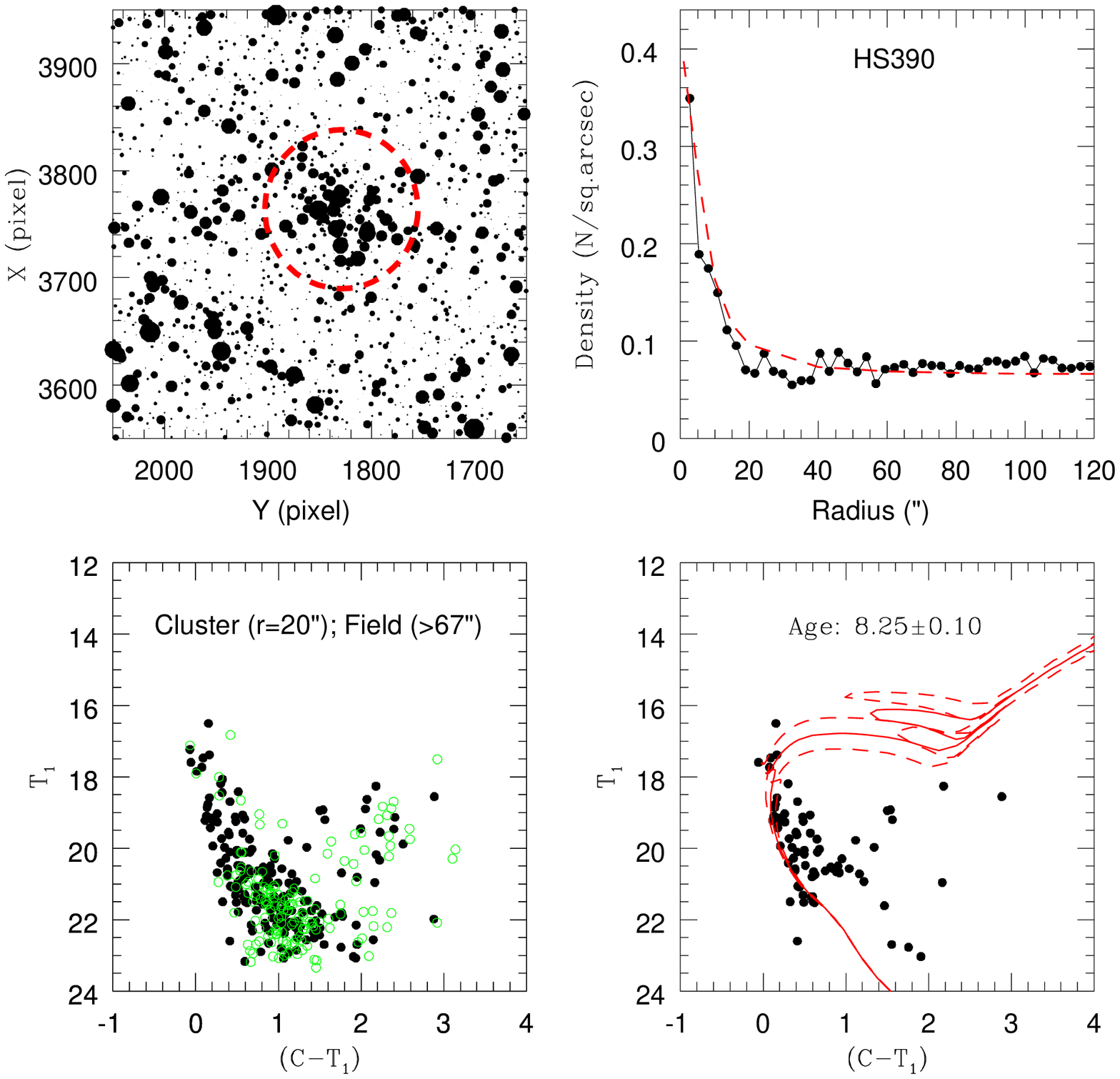}\hspace{1.0cm} 
 \includegraphics[height = .35\textheight, keepaspectratio]{hs411.eps} \\
 \includegraphics[height = .35\textheight, keepaspectratio]{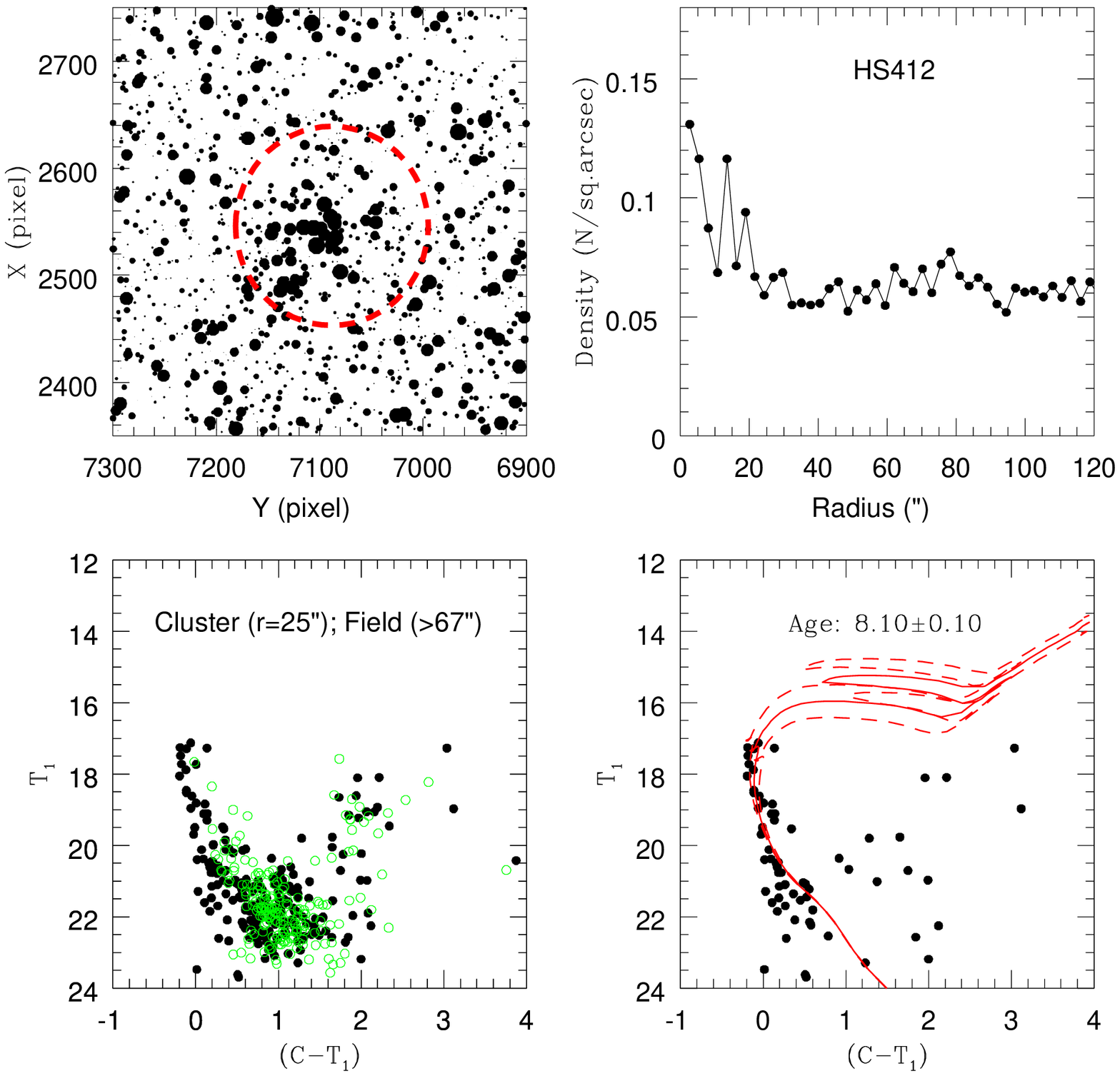}\hspace{1.0cm} 
 \includegraphics[height = .35\textheight, keepaspectratio]{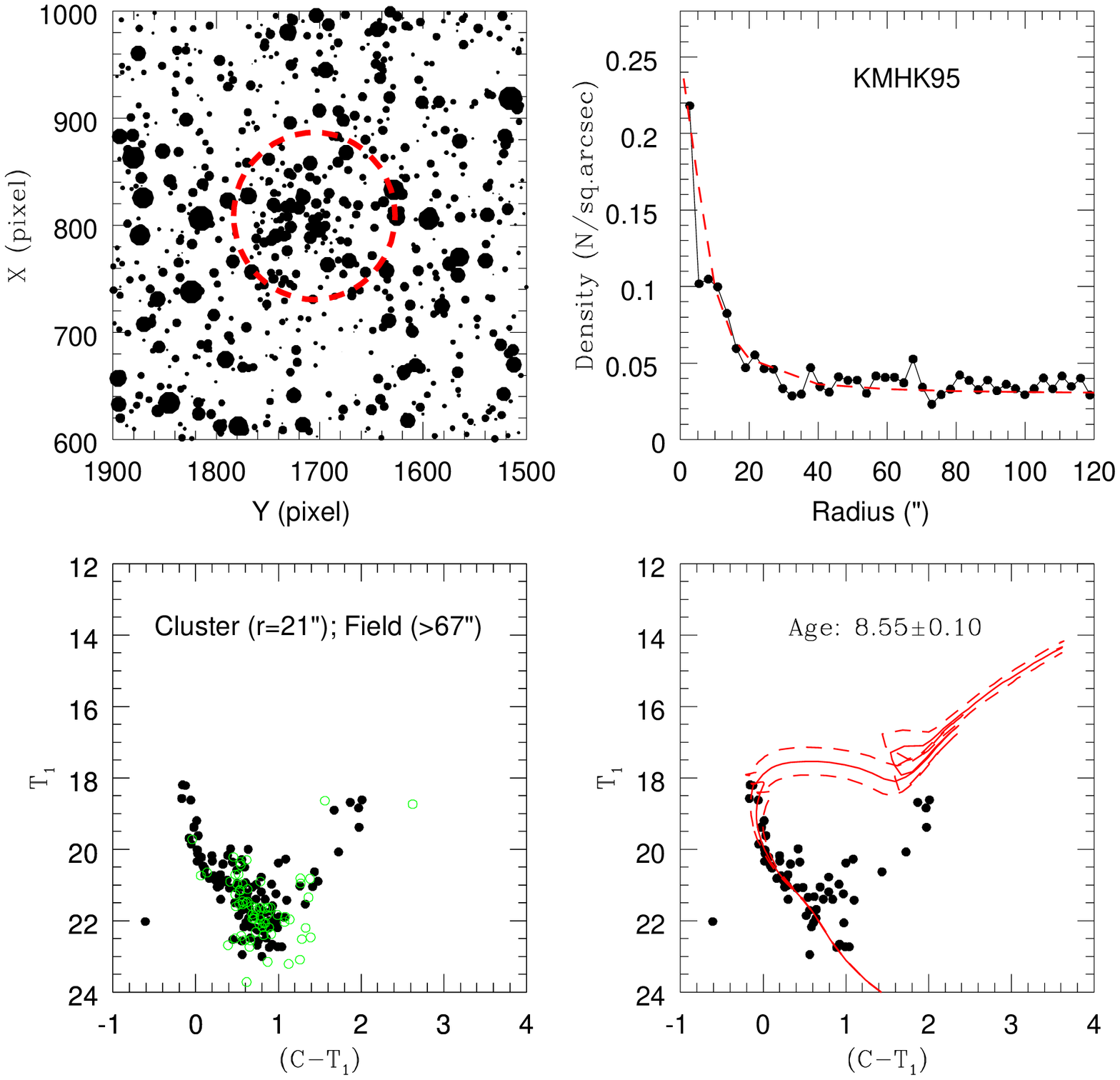} \\
\caption{\small {Single cluster candidates: For HS116, HS390, HS411, HS412 and KMHK95 the panel description for each cluster is same as Figure 1, except, that in case of HS412 no King profile over plot to RDP is shown. For HS131, the top-right panel shows the CMD of stars within the estimated cluster size (black filled circles). The bottom-left panel shows the CMD of the annular field (green filled circles) whereas the bottom-right panel shows isochrones over plotted to the unclean cluster CMD. The top-left panel description for HS131 is same as Figure 1.}}
\label{single2}
\end{center}
\end{figure*}

\begin{figure*}
\begin{center}
 \includegraphics[height = .35\textheight, keepaspectratio]{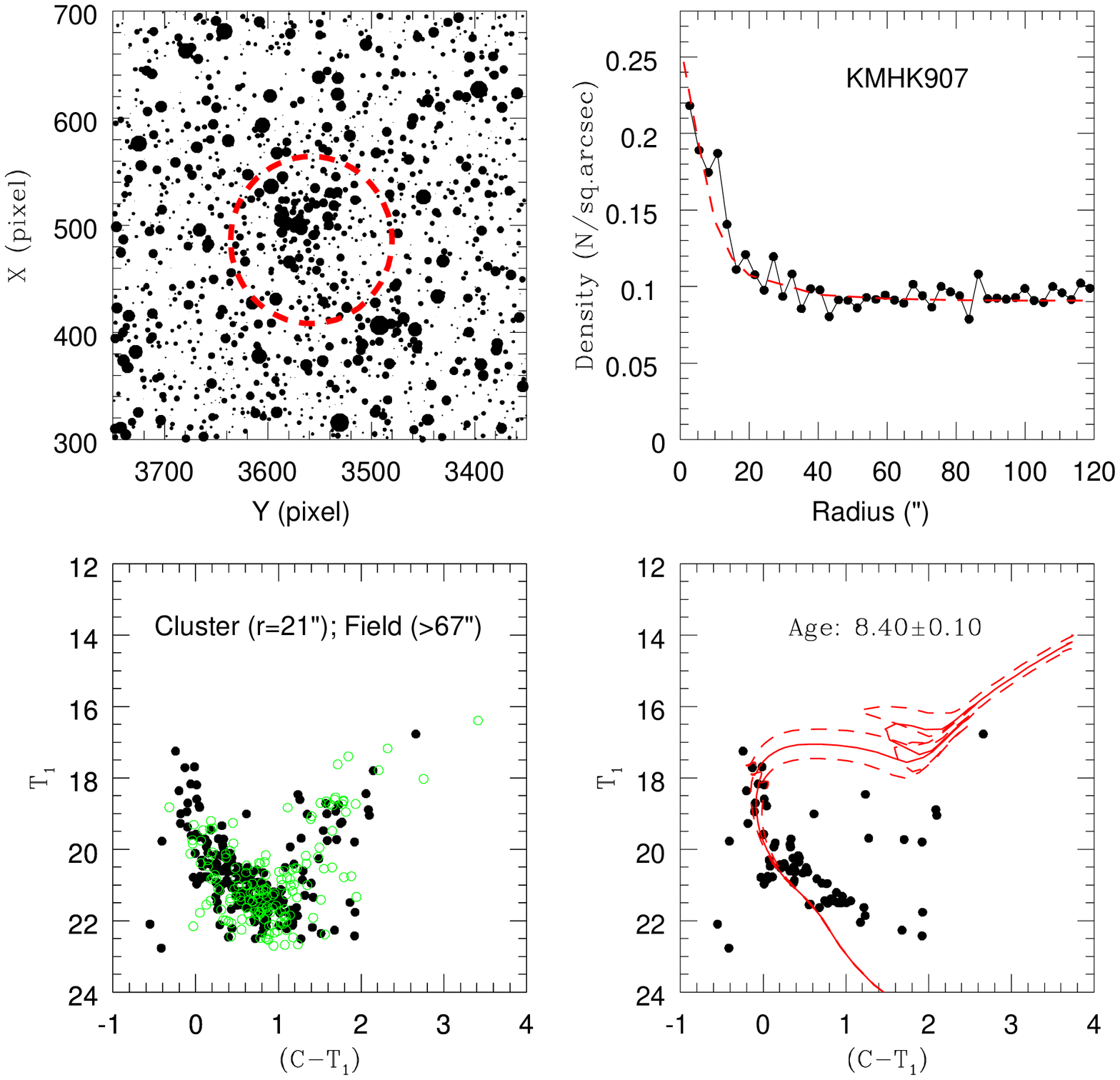}\hspace{1.0cm} 
 \includegraphics[height = .35\textheight, keepaspectratio]{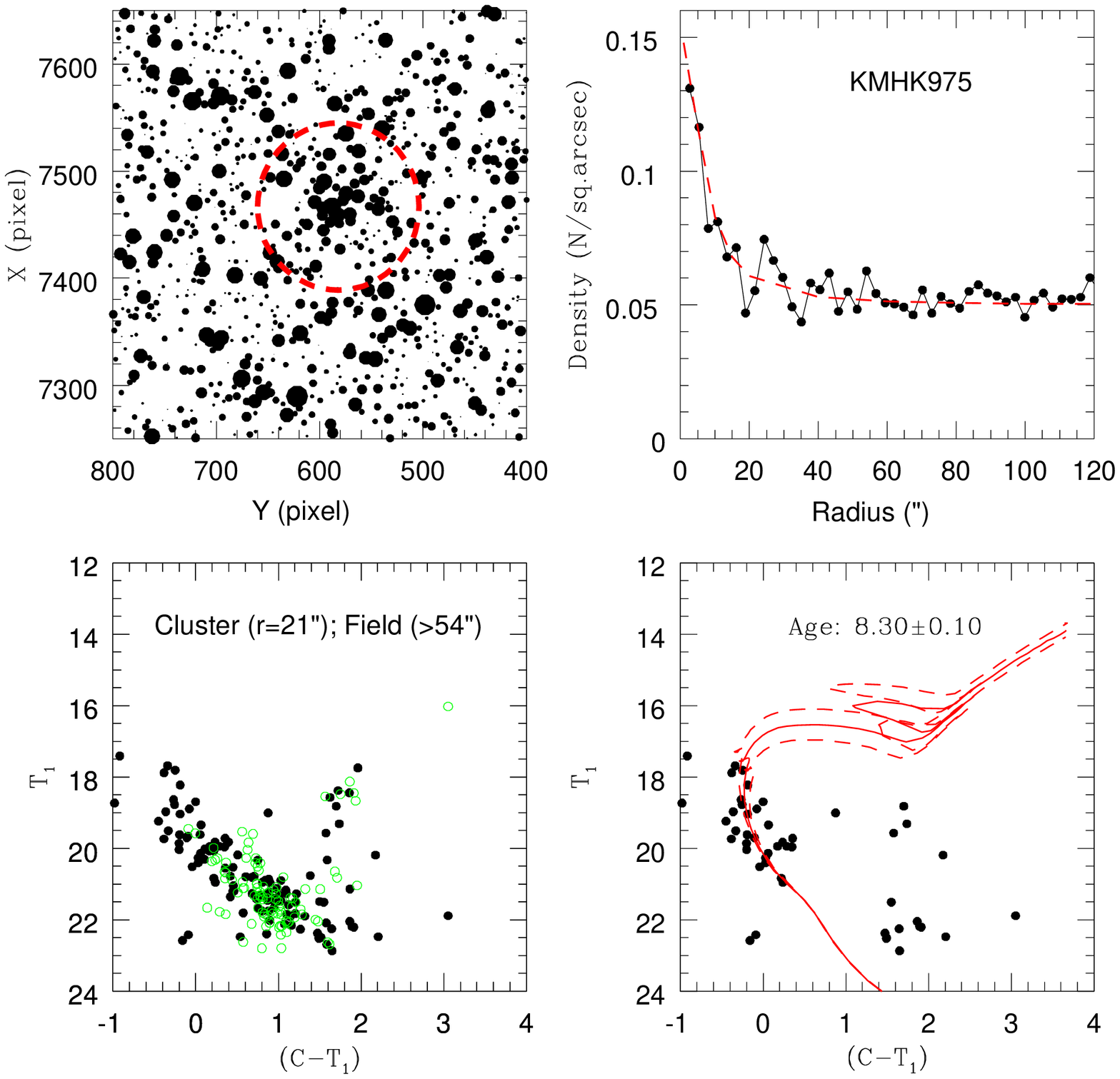} \\
 \includegraphics[height = .35\textheight, keepaspectratio]{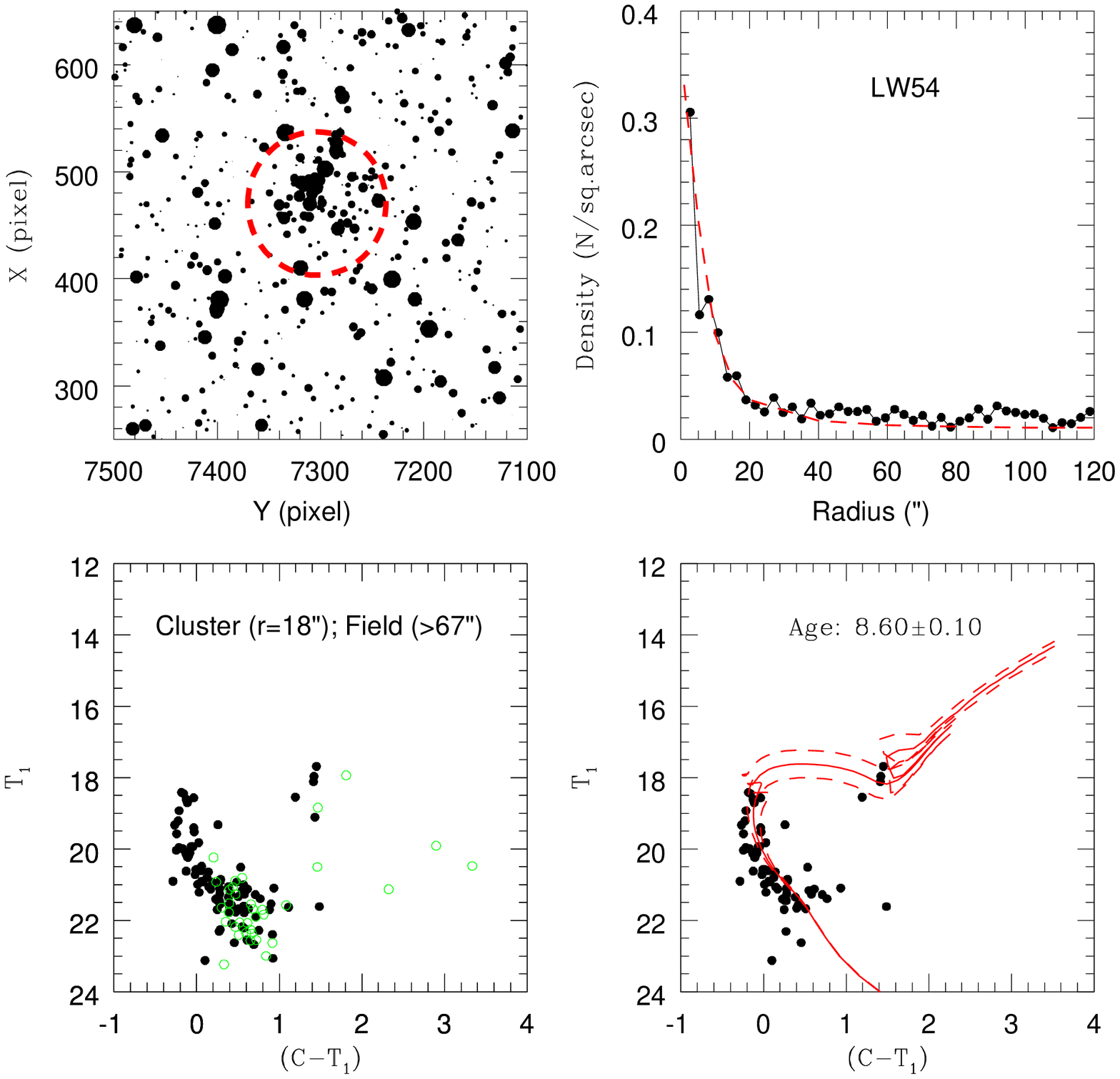}\hspace{1.0cm} 
 \includegraphics[height = .35\textheight, keepaspectratio]{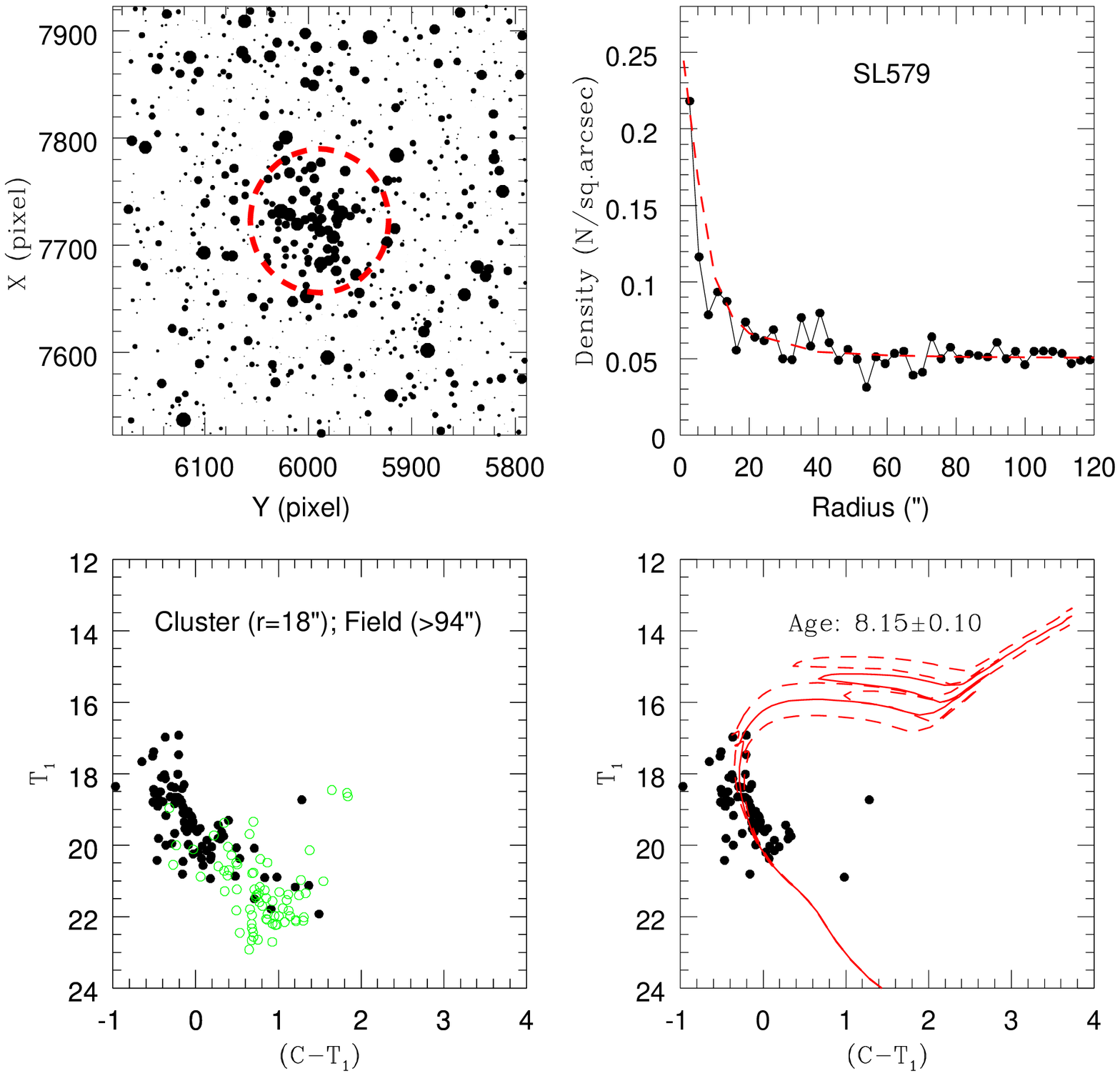} \\
\caption{\small {Single cluster candidates: KMHK907, KMHK975, LW54 and SL579. Panel description for each cluster is same as Figure 1.}}
\label{single3}
\end{center}
\end{figure*}

%----------------------------------------------
Notes on single clusters are presented here. Multi panel plots corresponding to each cluster are shown in Figure \ref{single1} (BRHT45a, BSDL77, H88-33, H88-131, H88-320 and H88-331), Figure \ref{single2} (HS116, HS131, HS390, HS411, HS412 and KMHK95) and Figure \ref{single3} (KMHK907, KMHK975, LW54 and SL579).

\begin{itemize}

\item BRHT45a is a bright, young ($\sim$125 Myr) cluster with prominent upper MS and MSTO. \cite{d02} reports a second cluster BRHT45b at coordinates (4$^h$ 56$^m$ 52$^s$, -68$^\circ$ 00$'$ 20$''$), which lies within the cluster radius (27$''$) of BRHT45a. G10 mentions that BRHT45b as a young cluster aged $\sim$40 Myr (log(t)=7.60, with 0.30$\leq$ error $<$0.50), which is similar to the age estimated by P12 using integrated photometry (log(t)=7.62$^{+0.18}_{-0.32}$). We in fact identify three clumps of stars within the cluster region and one of them could possibly be BRHT45b. \cite{p14} considered only the central clump as BRHT45b and estimated the age as $\sim$ 80 Myr (log(t)=7.90$\pm$0.10). However, given such a small spatial separation, it is difficult for us to identify BRHT45a and BRHT45b separately and estimate independent parameters for them. The age we have determined is possibly the age of the youngest or the dominant clump.

\item BSDL 77 is a compact cluster (aged $\sim$ 800 Myr) with a prominent MS, MSTO and a red giant branch. We also notice clumpy distribution of stars in the cluster region, which reflects in the radial density profile. This is one of the older clusters studied here.

\item H88-33 is a small compact cluster. The MSTO shows two possible turn-offs. As the cluster MS in very well populated, the scatter near the MSTO may be due to statistical effects. We have shown isochrones of log (t) = 8.20-8.50, suggesting that the age of the cluster is likely to be in this range.

\item H88-131 is a moderately large cluster as shown by the RDP. The field subtracted CMD shows a MS with a number of stars bluer than the MSTO. A few red giants can also be identified in the CMD. We have estimated the age of the cluster to be log(t) $\sim$ 1 Gyr. As the reddening to the cluster is very small, the stars seen bluer than the MS demand attention.

\item H88-320 is a fairly large cluster located in a relatively dense field, as shown by the RDP. The cluster MS is clearly identified in the field subtracted CMD and the age is estimated to be $\sim$160 Myr.

\item H88-331 is a dense and rich cluster. The MS has relatively large width and the MSTO also shows scatter. This may be due to the presence of differential reddening in the field. The age estimated for this cluster is $\sim$500 Myr.

\item HS116 is a small cluster in a relatively dense field. The field subtracted CMD shows the cluster features well which can be visually fitted with isochrones aged $\sim$350 Myr.

\item HS131: It is a dense rich cluster easily identified in the field. The RDP does not allow us to define the cluster radius and the features in the CMD also shows large scatter. A spatial plot of the evolved stars showed a density enhancement near the cluster center and we estimated the extent of the cluster to be  about (27$''$$\times$27$''$) around the cluster center. We thus considered all stars within this region and age of the cluster was estimated using visual fit of isochrones, especially to the RC and the RGB stars as $\sim$ 1.25 Gyr.

\item HS390 is a dense and slightly elongated cluster with a well populated MS and no giants. The estimated age is $\sim$180 Myr.

\item HS411 is one of the small clusters where we could identify a narrow and well populated MS. The cluster is aged $\sim$280 Myr.

\item HS412 shows clumpy distribution of stars in the cluster region, giving rise to RDP with multiple peaks. The cluster MS is prominent and we estimate the age to be $\sim$125 Myr.

\item KMHK95 is a moderately rich cluster with a well defined cluster MS. The cluster MS is clearly identified in the field subtracted CMD to estimate the age ($\sim$350 Myr).
 
\item KMHK907 is a bright, young cluster whose upper MS is prominently visible and its age is $\sim$250 Myr. The lower portion of the MS is broadened and the width increases with decrease in magnitude. This feature stays even after cleaning with field regions at different annular radii. This possibly is an effect of differential reddening or the presence of equal mass binaries in the lower MS which can be visually fitted by brightening the isochrones by 0.75 mag. \cite{d02} mentions the presence of another cluster in this field, BSDL1716 with coordinates (5$^h$ 26$^m$ 07$^s$, -70$^\circ$ 59$'$ 19$''$) and of similar size as KMHK907. B08 lists BSDL1716 as an association. We find no information about the age of BSDL1716 from PU00 or G10. Given its coordinates, it is possible that the bright clump seen in the spatial plot towards the south west direction (at a distance of $\sim$ 33$''$) of KMHK907 is BSDL1716. However, we could not find any prominent cluster feature in that specific location and hence we are unable to derive any cluster parameters for the same. 

\item KMHK975 is a small cluster where the CMD of the cluster region before field star subtraction shows a broad MS and a few RGs. The field subtracted CMD has only the upper part of the MS as the lower part has got subtracted away, even though the limiting magnitude of this region is about $T_1 \sim 23$ mag. The age estimated using the upper MS is $\sim$200 Myr.

\item LW54 is a compact and dense cluster which shows two concentrations of stars in the cluster region. The field region is very sparse as shown by the RDP. The cluster features are clearly seen in the CMD and are used to estimate the age ($\sim$400 Myr).

\item SL579 is a rich and dense cluster. The CMD of the region is relatively shallow with a limiting magnitude of $T_1 \sim 21$ mag. The cleaned CMD has a well populated MS which is used to derive the age ($\sim$140 Myr).

\end{itemize}

\begin{figure*}
\begin{center}
 \includegraphics[height = .35\textheight, keepaspectratio]{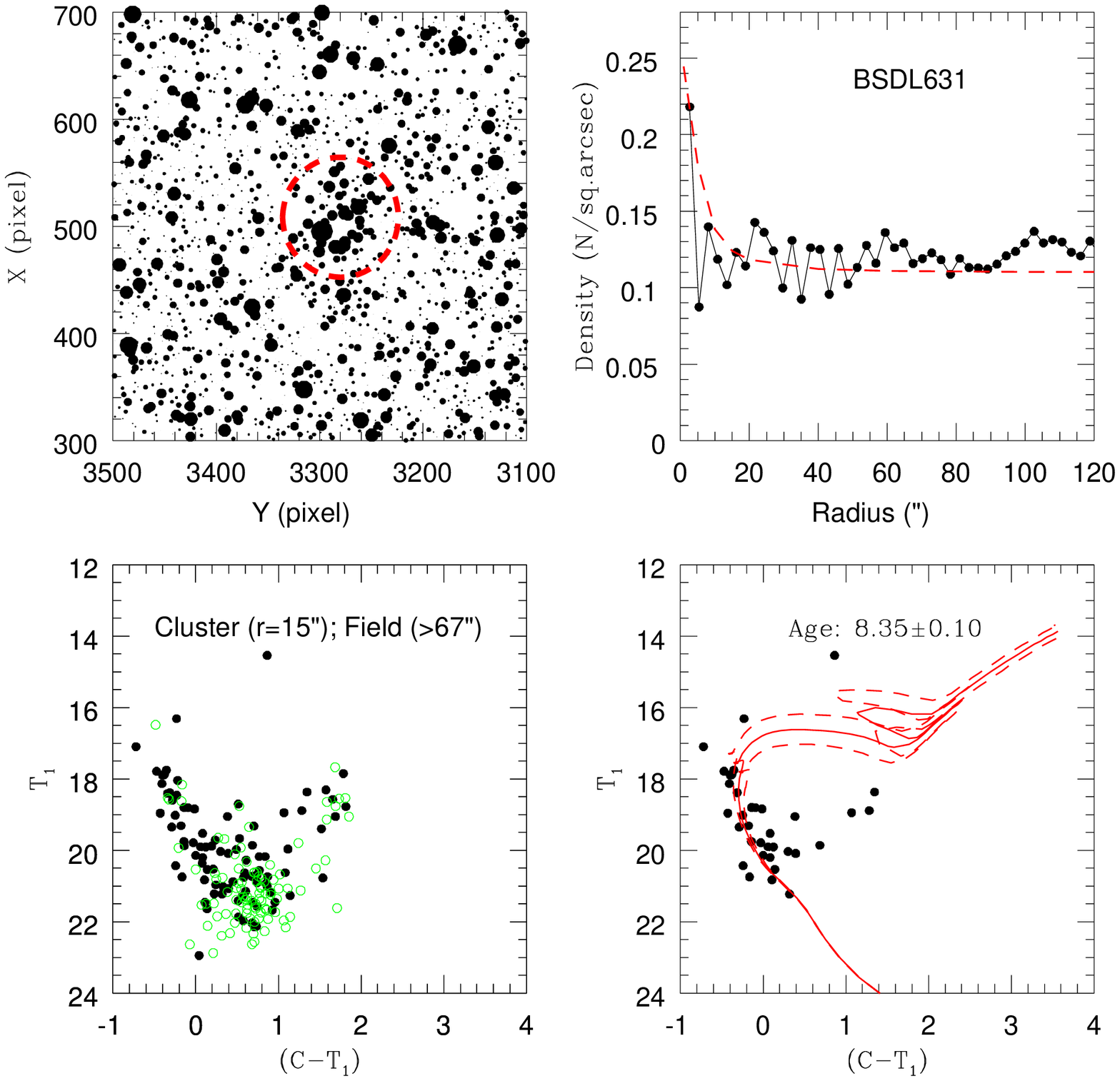}\hspace{1.0cm} 
 \includegraphics[height = .35\textheight, keepaspectratio]{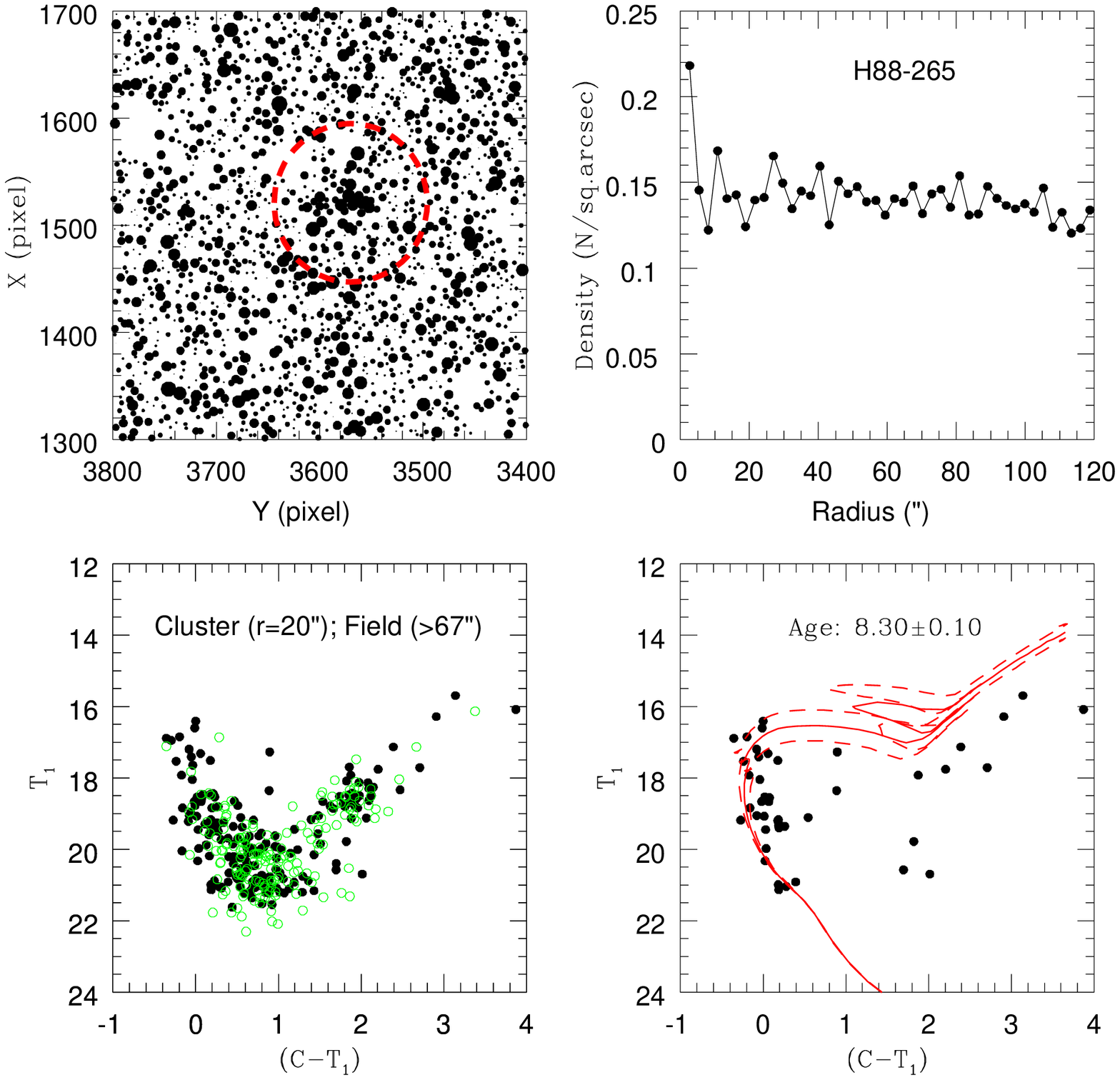} \\
 \includegraphics[height = .35\textheight, keepaspectratio]{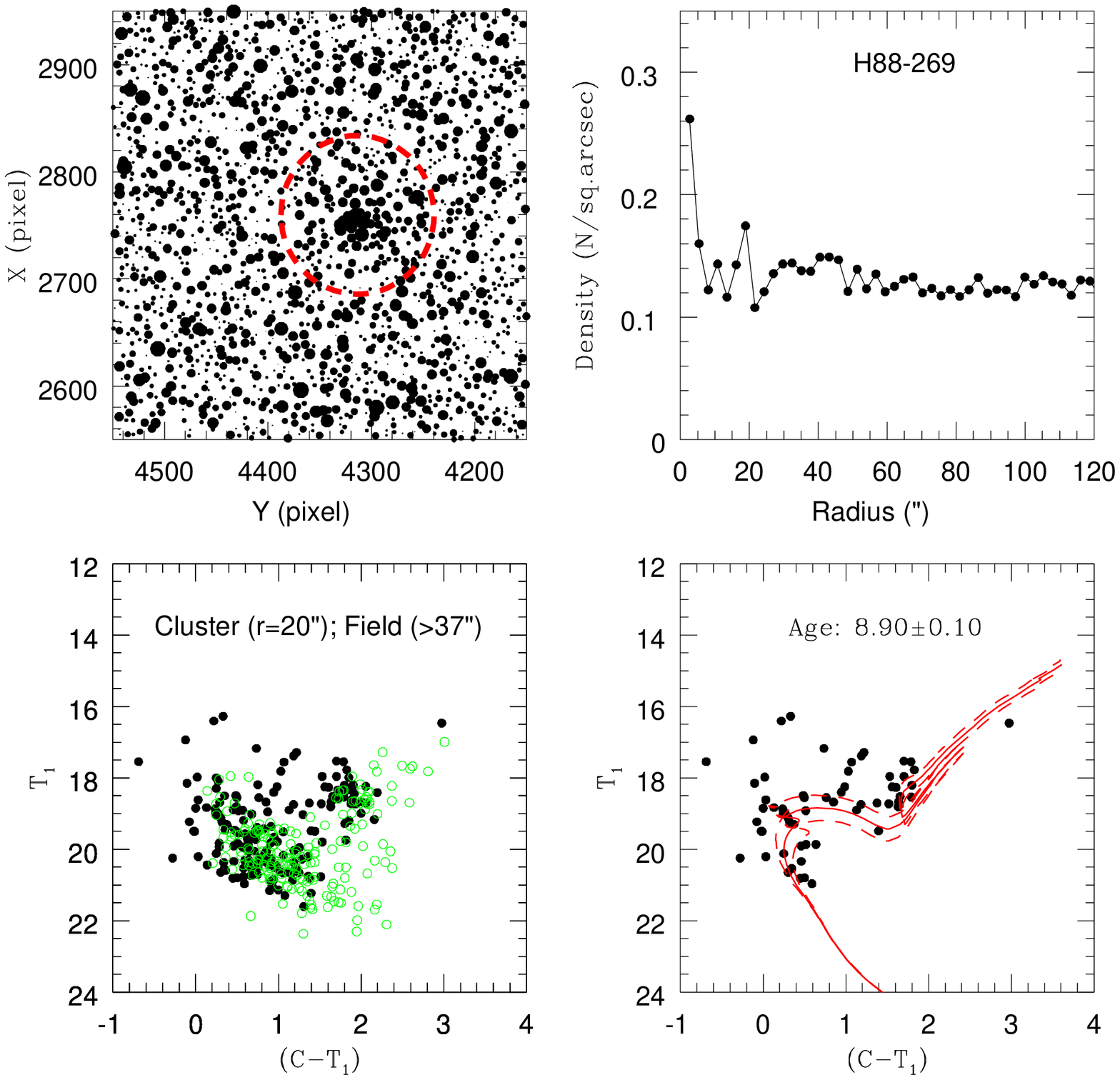}\hspace{1.0cm} 
 \includegraphics[height = .35\textheight, keepaspectratio]{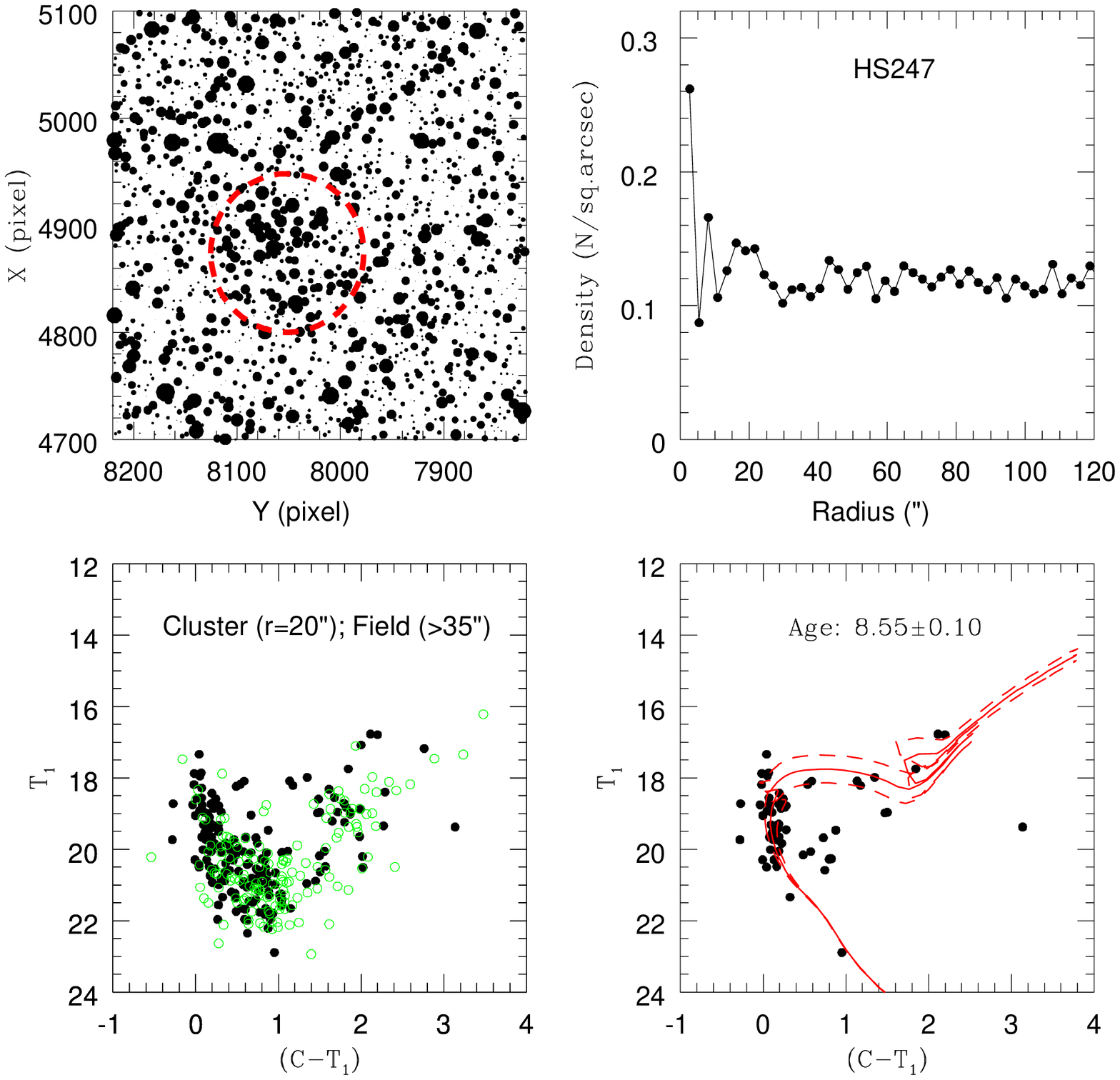} \\
\caption{\small {Single clusters with weak RDP: BSDL631, H88-265, H88-269 and HS247. Panel description for each cluster is same as Figure 1, except, that no King profile over plot to RDP is shown for H88-265, H88-269 and HS247.}}
\label{single4}
\end{center}
\end{figure*}

\begin{figure*}
\begin{center}
 \includegraphics[height = .35\textheight, keepaspectratio]{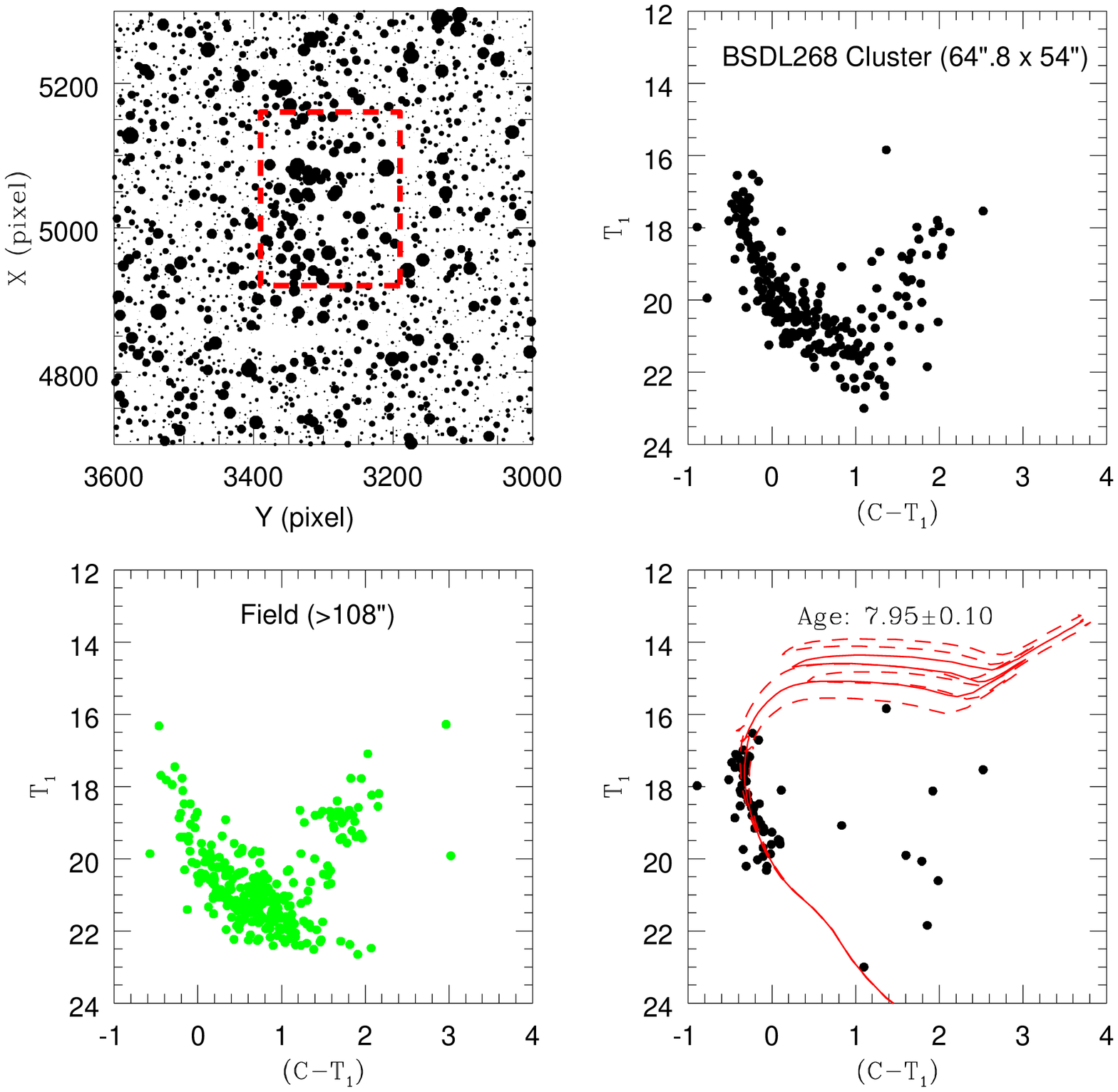}\hspace{1.0cm} 
 \includegraphics[height = .35\textheight, keepaspectratio]{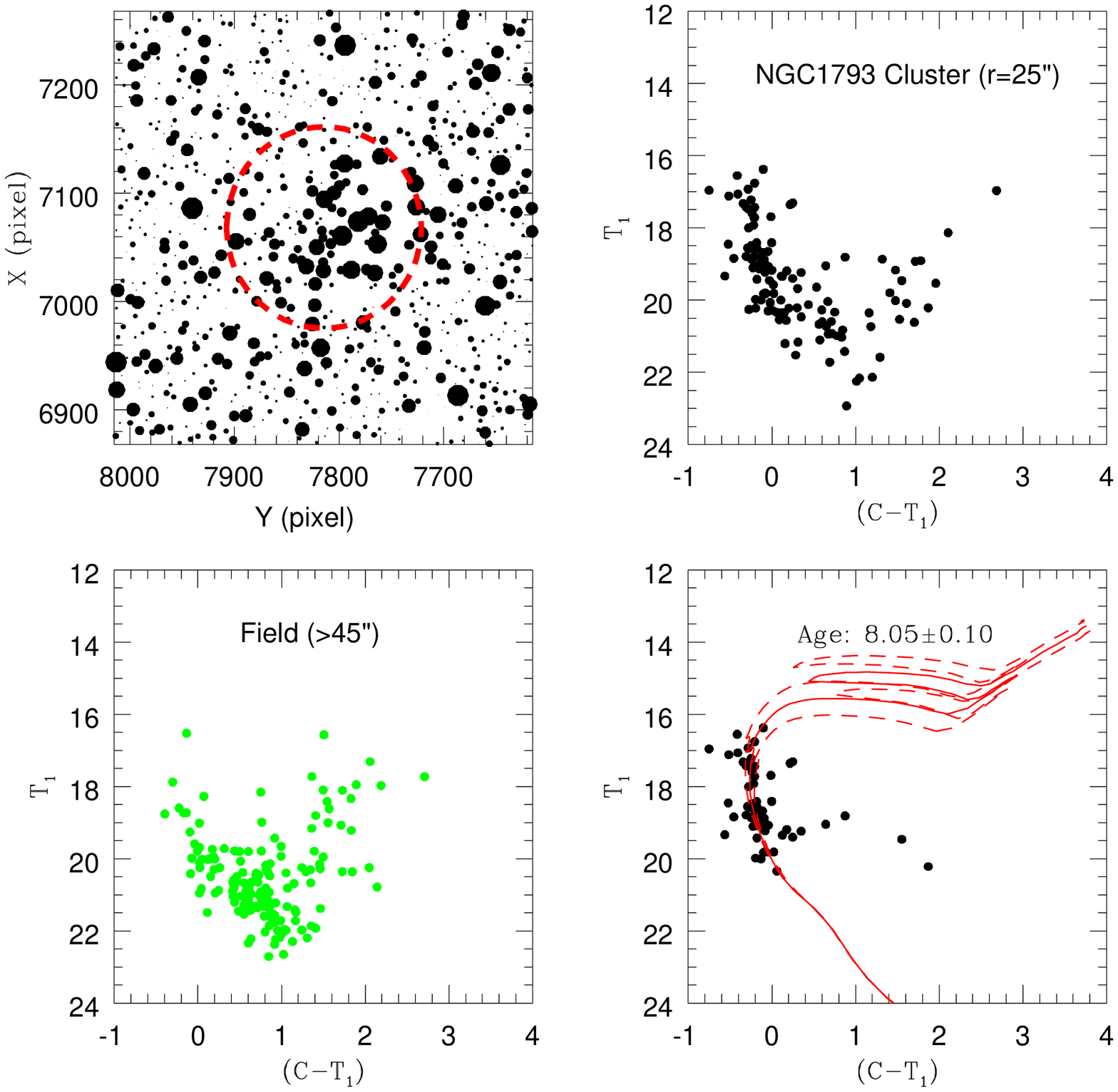} \\
 \includegraphics[height = .35\textheight, keepaspectratio]{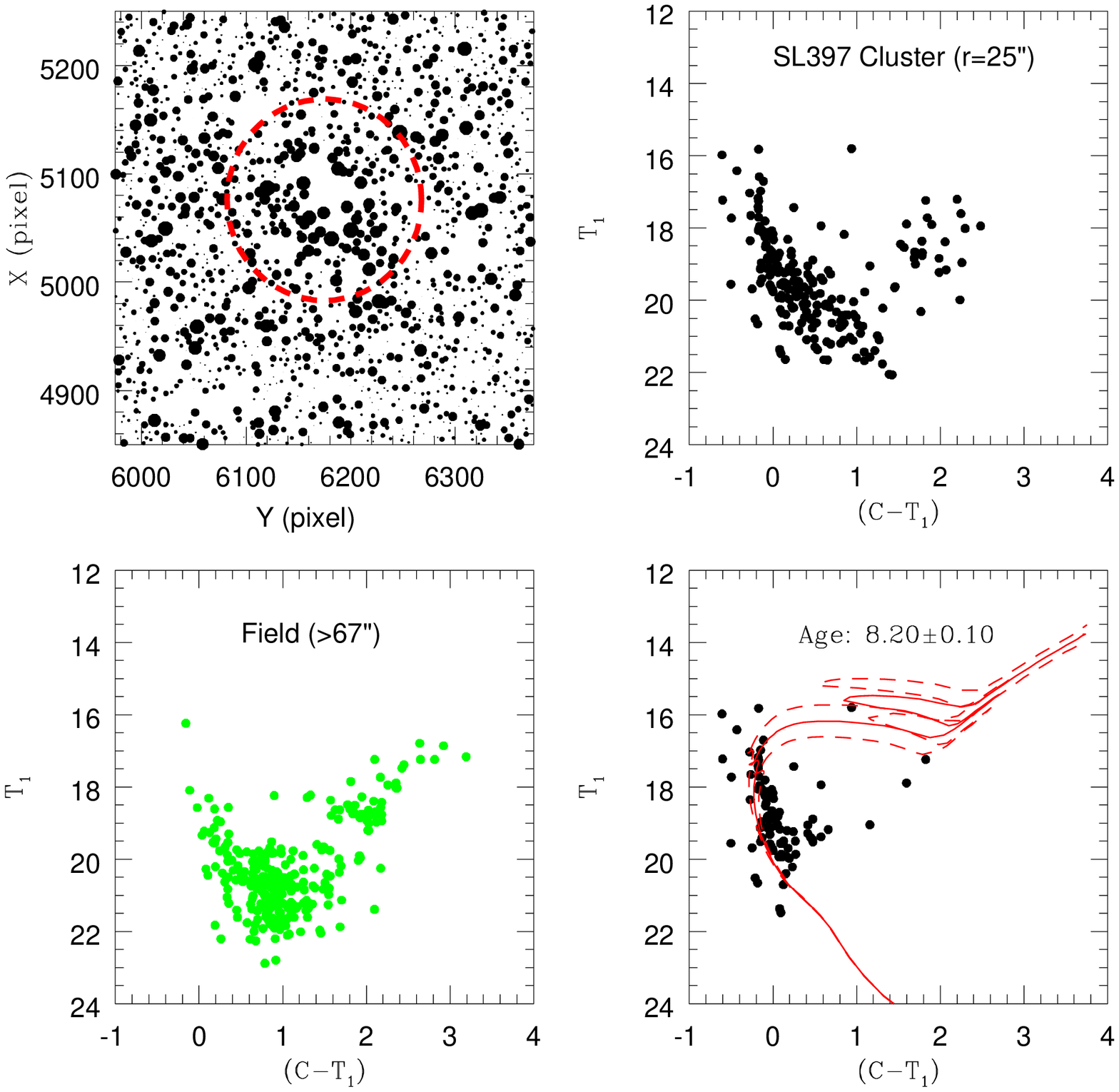} \\
 \caption{\small {Single cluster candidates with no RDP: BSDL268, NGC1793 and SL397. For each of these cases, the top-right panel shows the CMD of stars within the estimated cluster radius (black filled circles). Whereas, their bottom-left panel shows the CMD of the annular field (green filled circles). The top-left and bottom-right panel description for these clusters are same as Figure 1.}}
\label{single5}
\end{center}
\end{figure*}

\begin{figure}[ht]
\centering
\begin{minipage}[b]{0.45\linewidth}
\includegraphics[height=3.0in,width=3.0in]{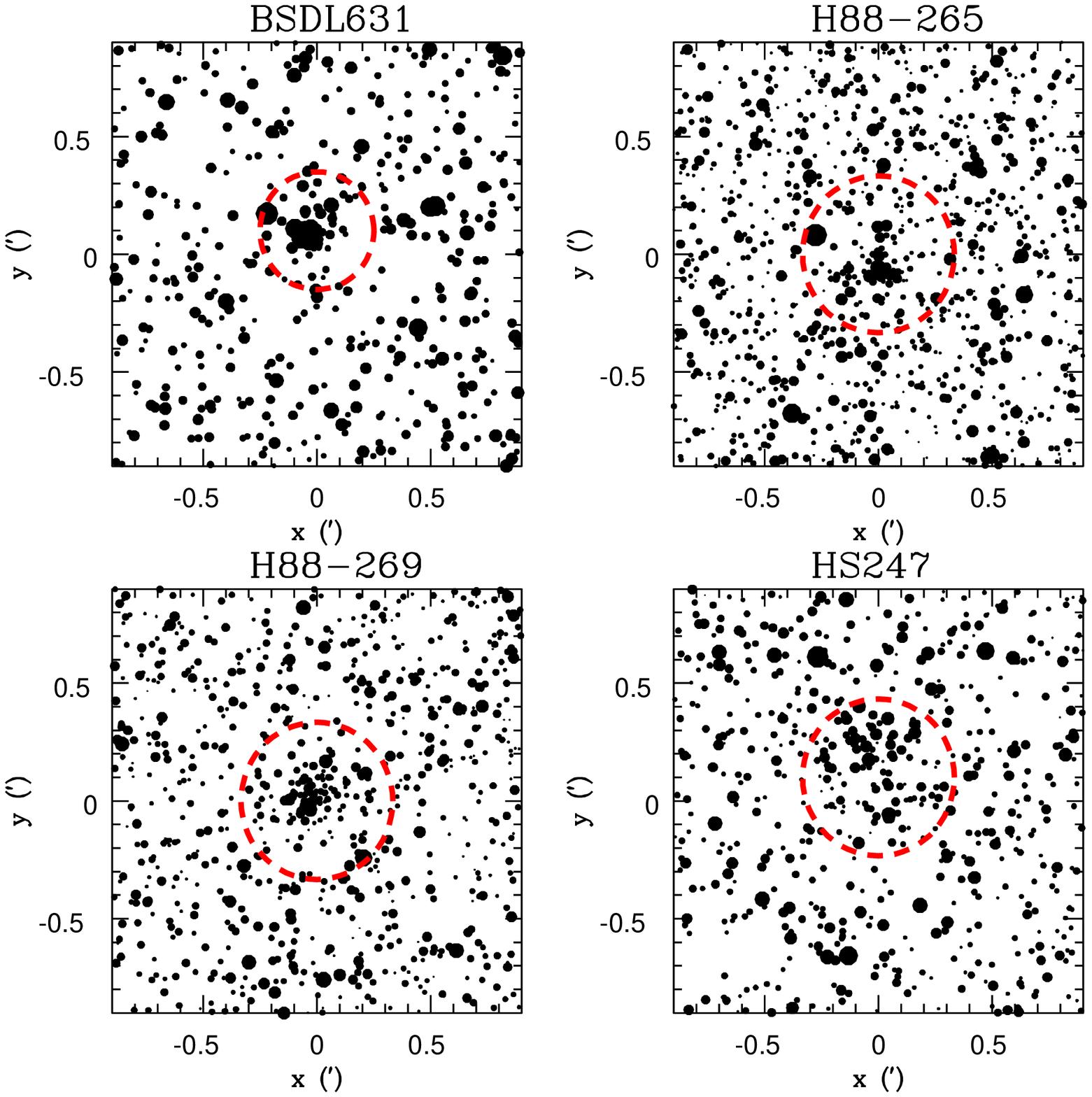}
\caption{\small {OGLE III fields of clusters with weak RDP: BSDL631, H88-265, H88-269 and HS247.The red dashed circle shows the derived size for these clusters using our Washington data.
\vskip 0.5cm
}}
\label{ogle1}
\end{minipage}
\quad
\begin{minipage}[b]{0.45\linewidth}
\includegraphics[height=3.0in,width=3.0in]{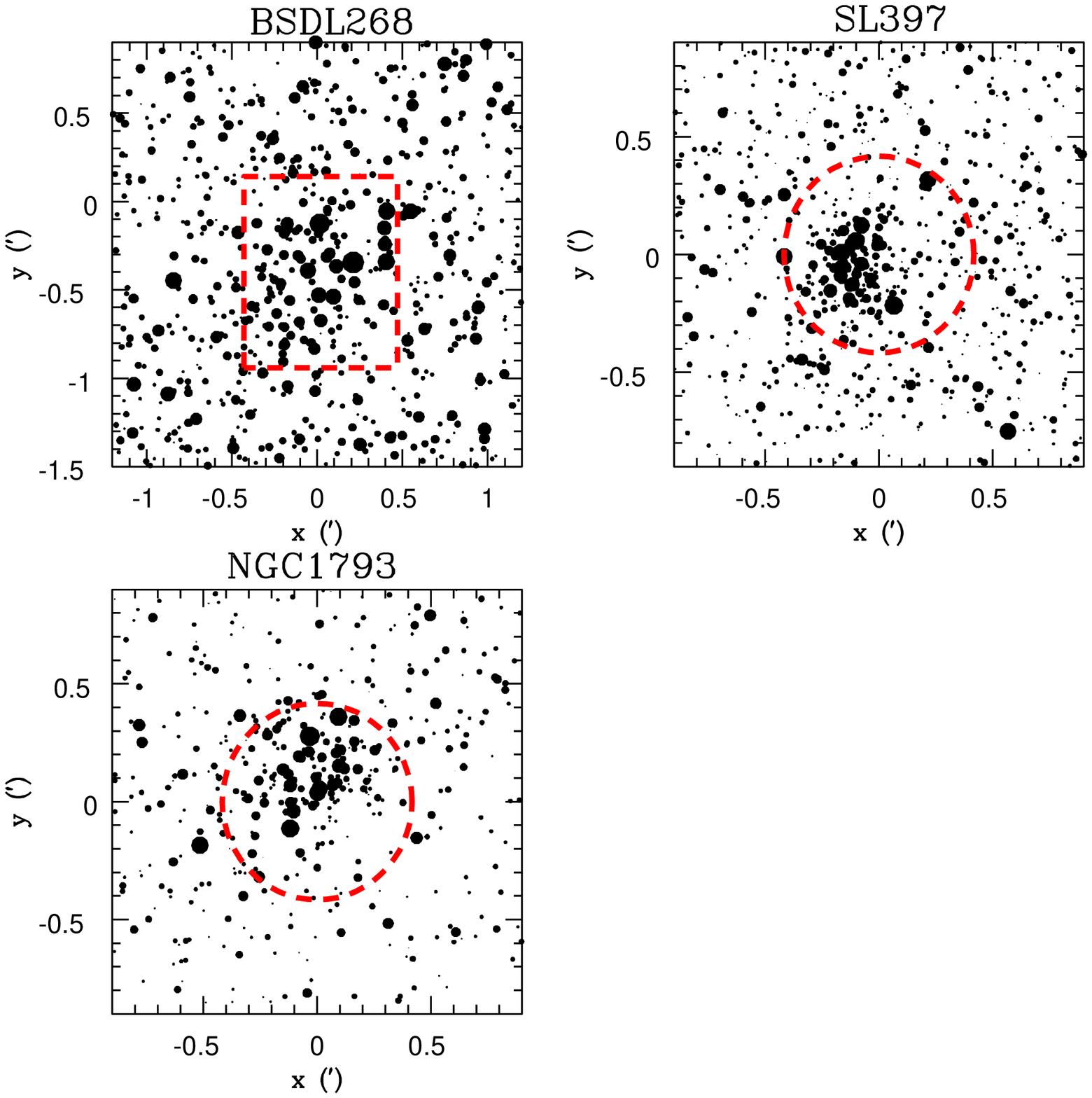} \\
\caption{\small {OGLE III fields of clusters with no RDP: BSDL268, NGC1793 and SL397. The red dashed circle/rectangle shows the derived size for these clusters using our Washington data.}}
\label{ogle2}
\end{minipage}
\end{figure}

%-------------------------------------------
We discuss below the cases of some clusters for which either the RDP did not show a strong peak (BSDL631, H88-265, H88-269 and HS247 in Figure \ref{single4}) or where we could not obtain an RDP (BSDL268, NGC1793 and HS329 in Figure \ref{single5}). Saturation effects caused by the presence bright stars near the cluster center have resulted in missing stars and incompleteness in the central region for these clusters. As our data could not confirm whether these clusters are true clusters based on spatial density enhancement, we tried to verify the existence of these clusters by identifying density enhancement using other optical data. The OGLE III is one of the complete and relatively deep surveys of the inner region of the LMC (4$^\circ$-5$^\circ$) with good spatial resolution \citep{uetal08}. We extracted the OGLE III fields corresponding to each of these clusters and created similar finding charts (Figure \ref{ogle1} for BSDL631, H88-265, H88-269 and HS247; Figure \ref{ogle2} for BSDL268, NGC1793 and HS329). It is to be noted that the OGLE III spatial plots presented in this section and hereafter (i.e. for a few cases in Appendix B and C), x and y denote the Cartesian coordinate system (units in arcminutes), with north on top and east to the left. The individual cases are discussed below.
%-------------------------------------------

%\begin{figure*}
%\begin{center}
% \includegraphics[height = .35\textheight, keepaspectratio]{weak_rdp.eps}
% \includegraphics[height = .35\textheight, keepaspectratio]{no_rdp.eps} \\
%\caption{\small {OGLE III fields of clusters with weak (BSDL631, H88-265, H88-269 and HS247) or no RDP (BSDL268, NGC1793 and SL397). The red dashed circle/rectangle shows the derived size for these clusters using our Washington data.}}
%\label{ogle1}
%\end{center}
%\end{figure*}

%-------------------------------------------

\begin{itemize} 

\item BSDL631: A small, compact cluster with symmetric distribution of bright stars is observed around the cluster center. The estimated RDP shows only a weak density enhancement within a radius $\leq$ 15$''$ mainly due to missing stars near the center. The OGLE III field compliments the Washington field showing a small but compact clump of bright stars present within the cluster region and is devoid of saturation effects. Based on the density enhancement seen in the OGLE III field (Figure \ref{ogle1}) and the MS feature identified in the CMD, we conclude that BSDL631 is very small, compact and young cluster aged $\sim$220 Myr. 

\item H88-265: A small but prominent clump of stars is observed around the cluster center in the spatial plot. A comparison of the CMD of the cluster region with the field region shows prominent MSTO and upper MS feature (brighter than 19.0 mag and bluer than 0.2 mag), which could belong to the cluster. A spatial distribution of the bright stars show significant clumping around the cluster center within a radius, r$\leq$ 20$''$ and is hence adopted as the cluster radius. The OGLE III spatial plot (Figure \ref{ogle1}) corresponding to H88-265 shows a similar small but prominent clump of bright stars around the cluster center, thus validating the presence this small and young cluster aged $\sim$200 Myr.

\item H88-269: The finding chart shows a small, dense clump of stars near the central region. The density enhancement observed is caused by stars near the MSTO as well as evolved stars, as seen in the cleaned cluster profile. The cluster CMD can be visually fitted with an isochrone of age log(t)=8.90$\pm$0.10. Although the RDP estimated for this cluster does not show a strong peak, we find the cluster profile to prominently show up within a radius of 20$''$. In fact, the field  seems to be have almost similar SFH as that of the cluster and suffers from differential reddening, thus making field star subtraction inefficient. To verify the existence of this cluster, we extracted the OGLE III field (Figure \ref{ogle1}) for this cluster and we were able to find a very significant density distribution of stars within the central region. We conclude that H88-269 is a small and compact cluster aged $\sim$800 Myr, immersed in a dense field of almost similar age.

\item HS247: We observe feeble density enhancement near the expected center of the cluster. However a strong upper MS (brighter than 20.0 mag and bluer than 0.4 mag) is observed when the CMDs of the cluster and the field region are compared. This prominent cluster feature is retained even after cleaning with alternate field regions. A spatial distribution of these bright stars show significant density distribution around the cluster center. The cluster radius (20$''$) is selected as the radius at which the cluster profile looks well populated. The OGLE III spatial plot (Figure \ref{ogle1}) corresponding to HS247 shows a density enhancement within the cluster region, supporting the existence of this cluster. We suggest that HS247 is a small, young cluster aged $\sim$ 350 Myr.

\item BSDL268: It is one of the youngest cluster in our sample aged $\sim$ 90 Myr. In the spatial plot we see a few bright stars clumped near the expected cluster center, with evidence of some missing stars. A comparison of the CMDs of the cluster and field region shows a bright upper MS brighter than 19.0 mag and bluer than 0.2 mag. Also, the cleaned CMD has a well populated MS. A spatial plot of these bright MS stars show a compact distribution within an area of about (64.$''$8$\times$54$''$), and the cluster center is chosen at the center of this distribution. The corresponding OGLE III spatial plot (Figure \ref{ogle2}) for the cluster shows strong density enhancement due to bright stars in the central region validating this young cluster candidate. 

\item NGC1793:  We considered an area of (54$''$$\times$54$''$) around the central region where the presence of a cluster is suspected. The CMD shows a strong upper MS brighter than 19.0 mag and bluer than 0.2 mag, which could belong to the cluster and SFH appears quite different from that of the field CMD. A radius of 25$''$ is selected for the cluster and the cluster feature is found to appear clearly in the cleaned CMD. The corresponding OGLE III spatial plot (Figure \ref{ogle2}) for the cluster is presented which shows a strong density enhancement due to bright stars in the central region validating this to be a true cluster aged $\sim$ 110 Myr. 

\item SL397: The spatial plot shows a clumpy distribution of bright stars located symmetrically around the expected cluster center,and possibly some missing stars. The bright stars form a prominent upper MS (brighter than 19.0 mag and bluer than 0.2 mag) for the cluster, even after cleaning the field stars. We were unable to estimate a RDP. The cluster radius (25$''$) is selected as the distance from the center at which the cluster features seem to be well populated. The corresponding OGLE III spatial plot (Figure \ref{ogle2}) for the cluster shows a strong density enhancement in the central region. This further confirms the object as a true young ($\sim$ 160 Myr) cluster.

\end{itemize}

%}
%-----------------------------------------------------

\begin{figure*}
\begin{center}
 \includegraphics[height = .35\textheight, keepaspectratio]{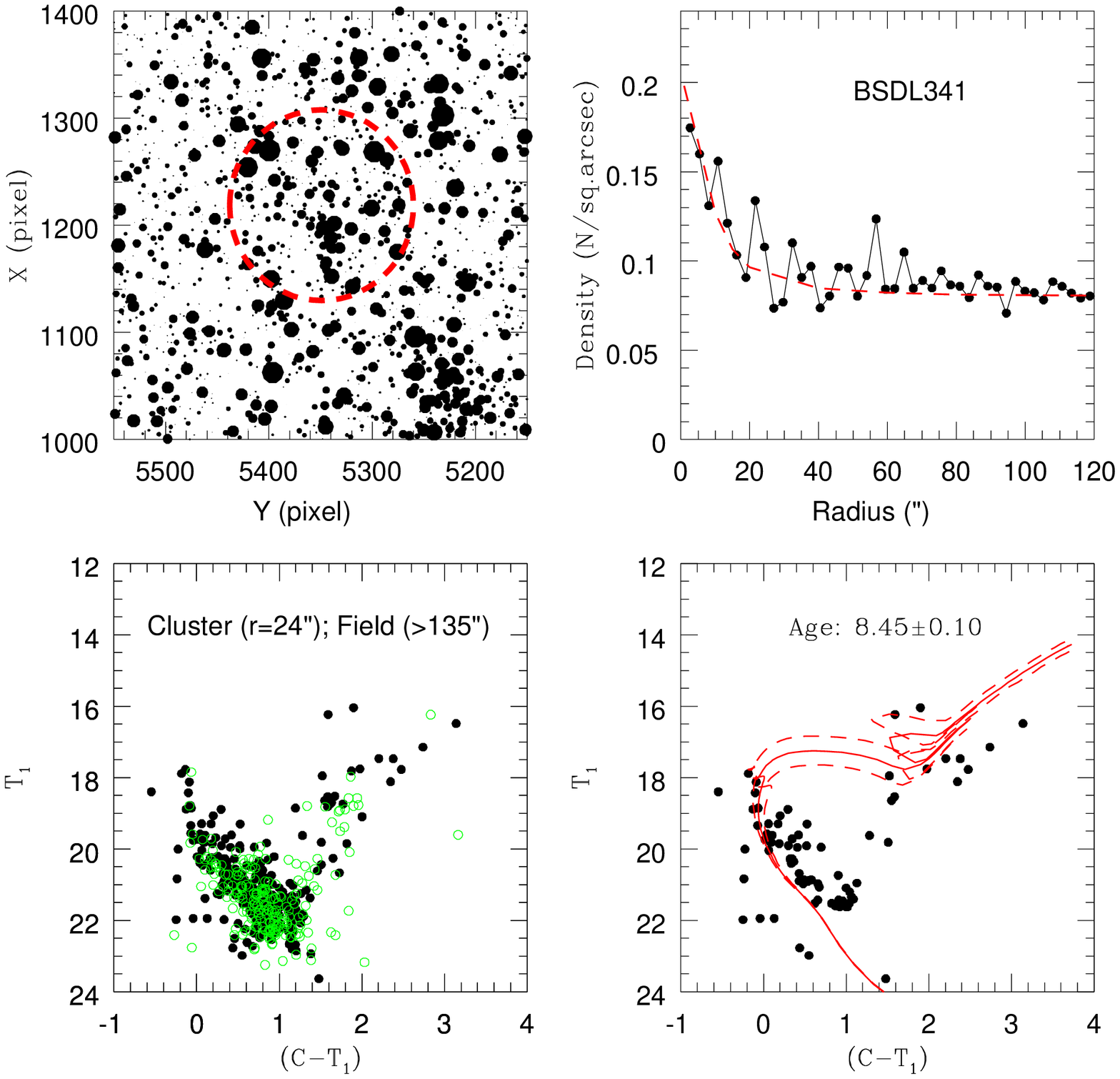}\hspace{1.0cm} 
 \includegraphics[height = .35\textheight, keepaspectratio]{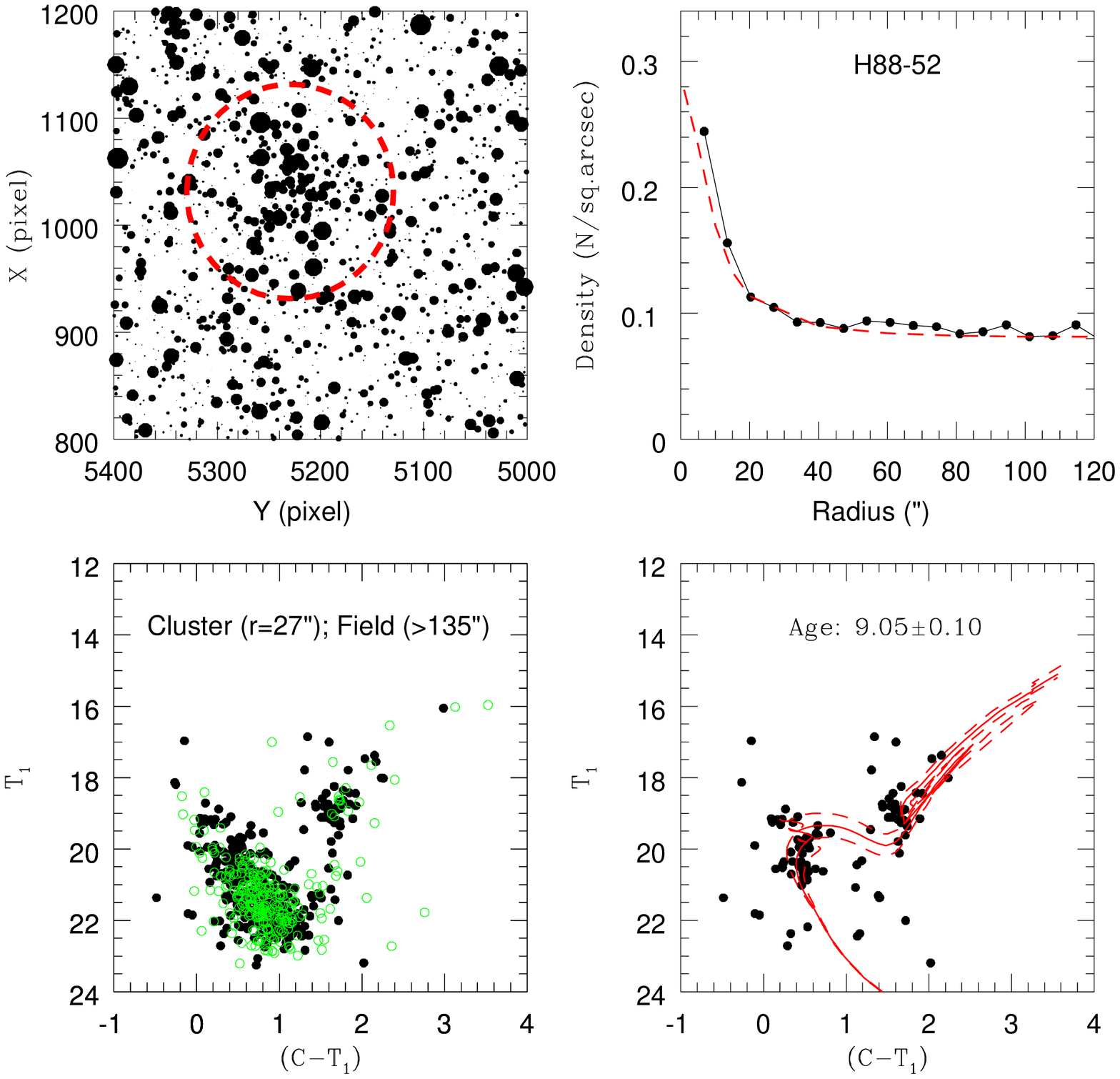} \\
 \includegraphics[height = .35\textheight, keepaspectratio]{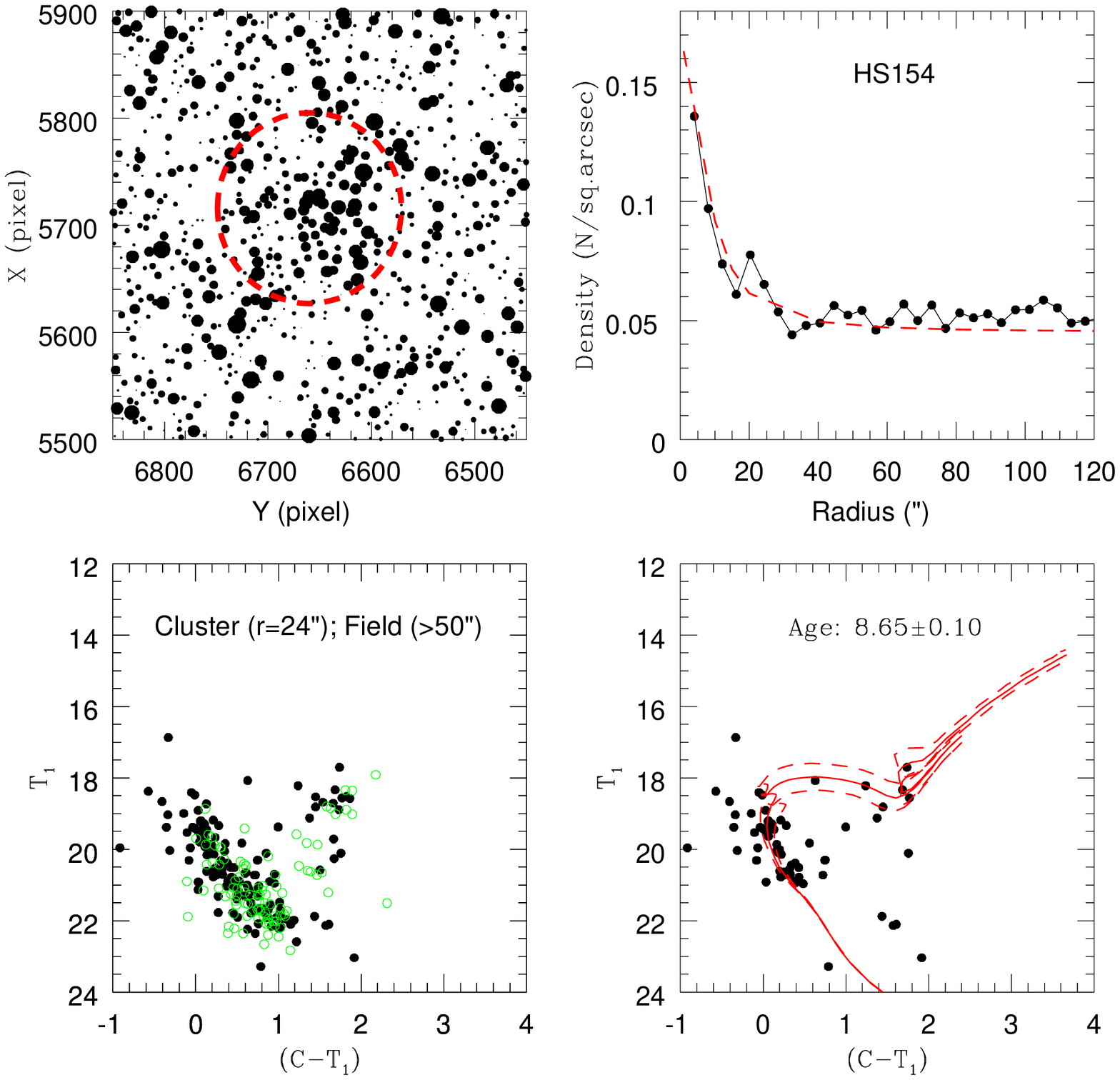}\hspace{1.0cm} 
 \includegraphics[height = .35\textheight, keepaspectratio]{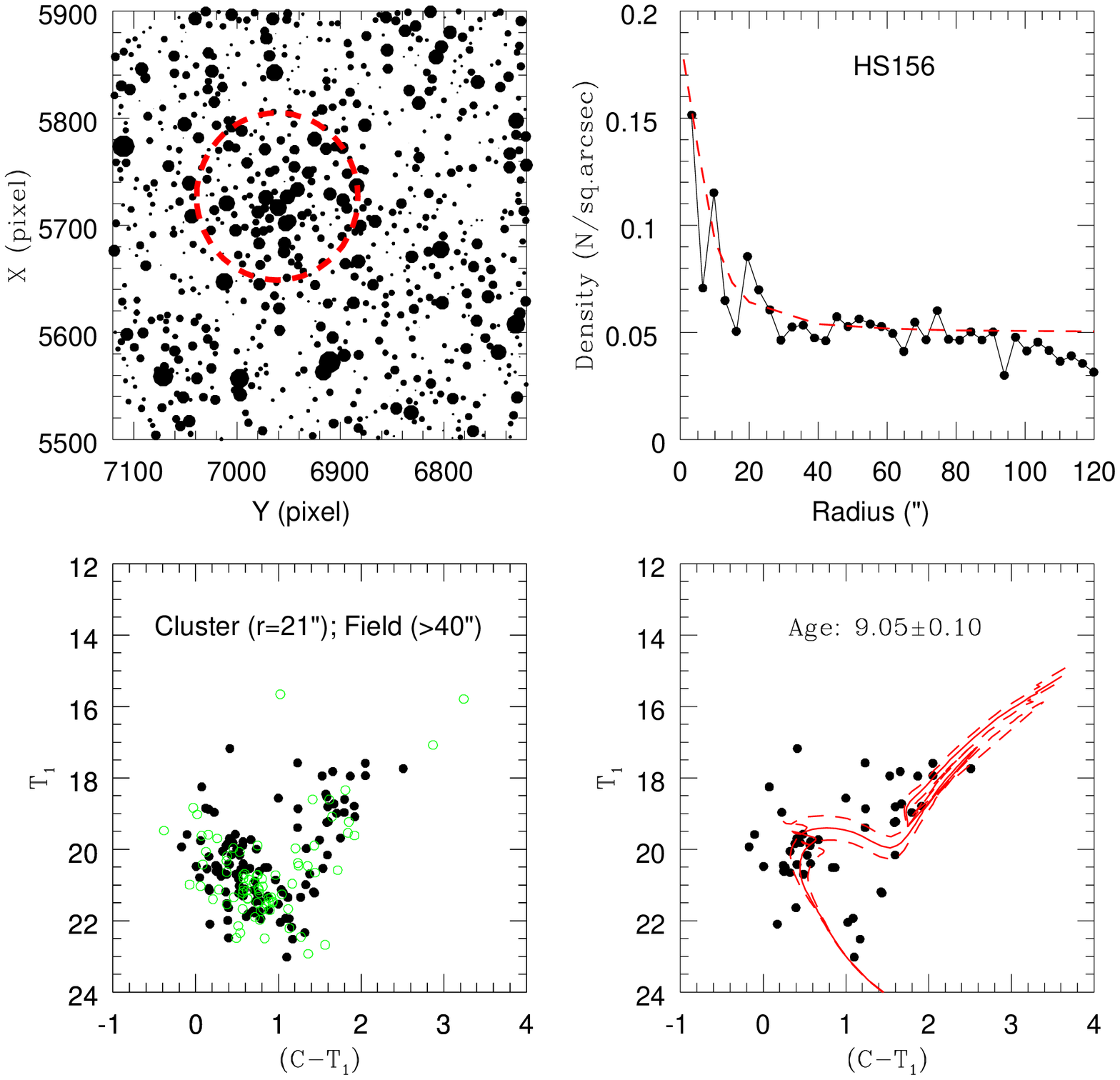} \\
 \includegraphics[height = .35\textheight, keepaspectratio]{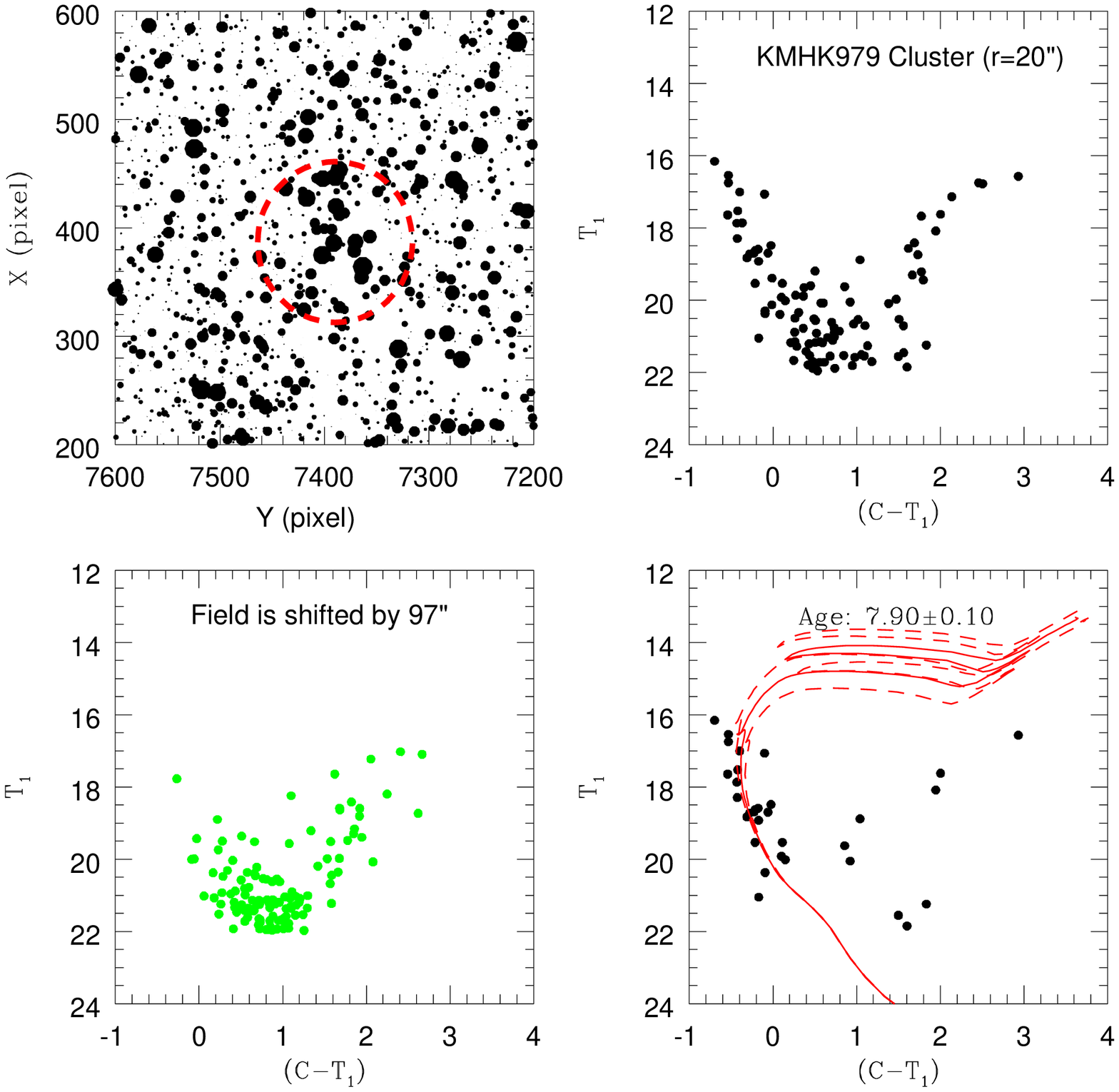}\hspace{1.0cm} 
 \includegraphics[height = .35\textheight, keepaspectratio]{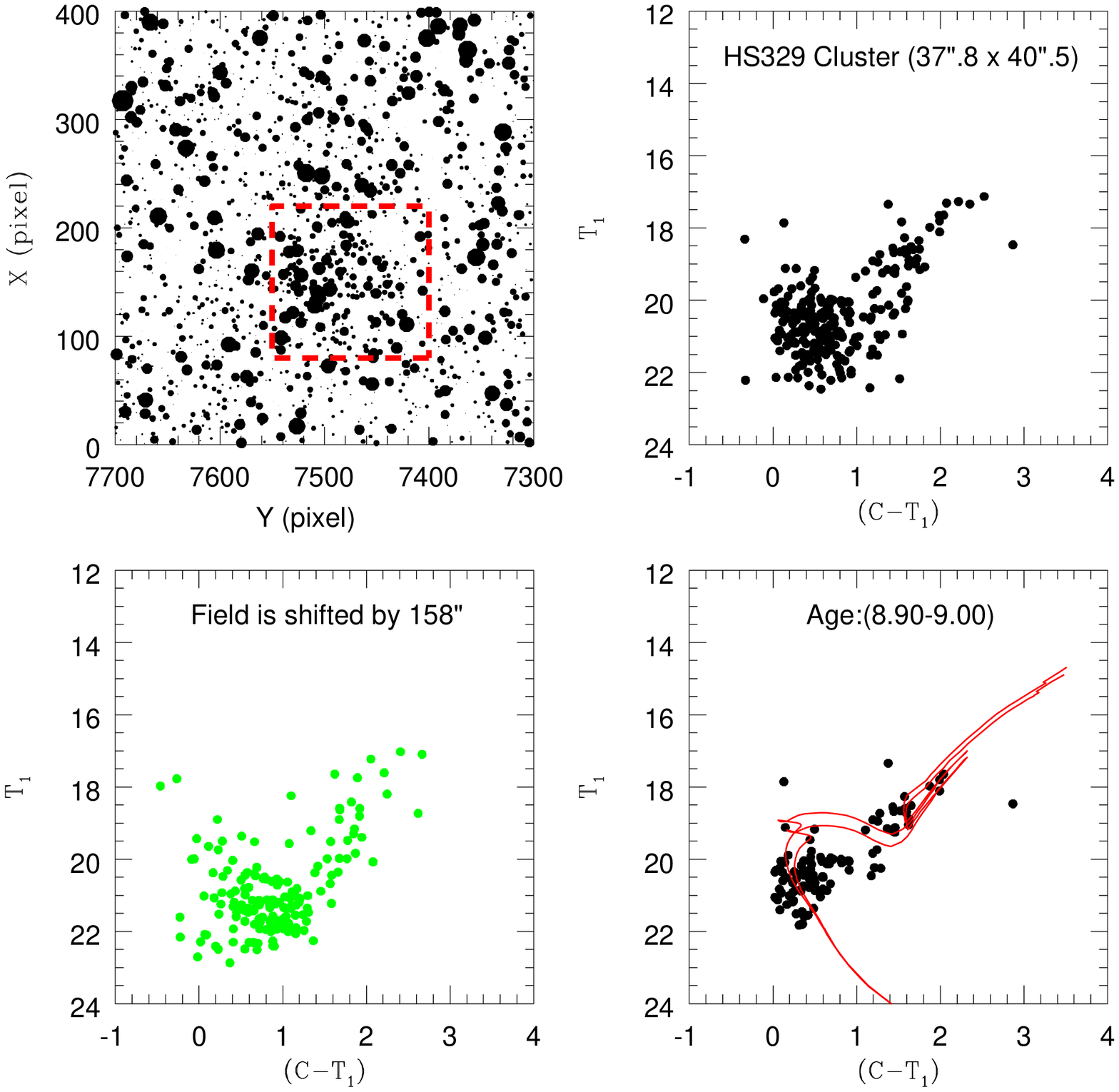} \\
\caption{\small {Double clusters: BSDL341 \& H88-52; HS154 \& HS156; KMHK979 \& HS329. The panel description for first two pairs is same as that of Figure 1. For KMHK979 \& HS329, the top-right panel shows the CMD of stars within the estimated cluster radius (black filled circles). Whereas, their bottom left panel shows the CMD of a non-annular, similar sized field region (green filled circles) located away from the clusters. The top-left and bottom-right panel description for them are same as Figure 1. }}
\label{double1}
\end{center} 
\end{figure*}

\begin{figure*}
\begin{center}
 \includegraphics[height = .35\textheight, keepaspectratio]{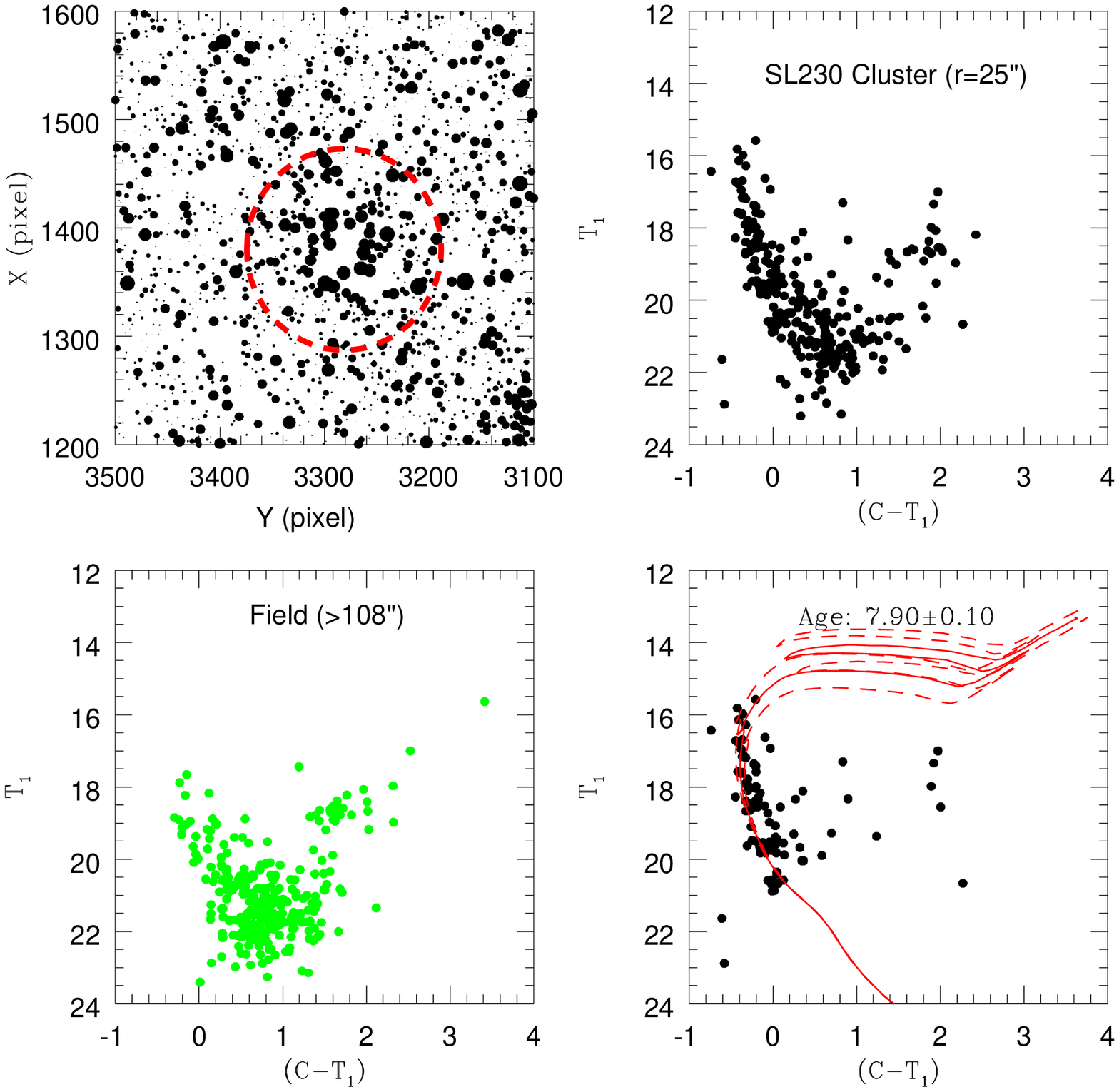}\hspace{1.0cm} 
 \includegraphics[height = .35\textheight, keepaspectratio]{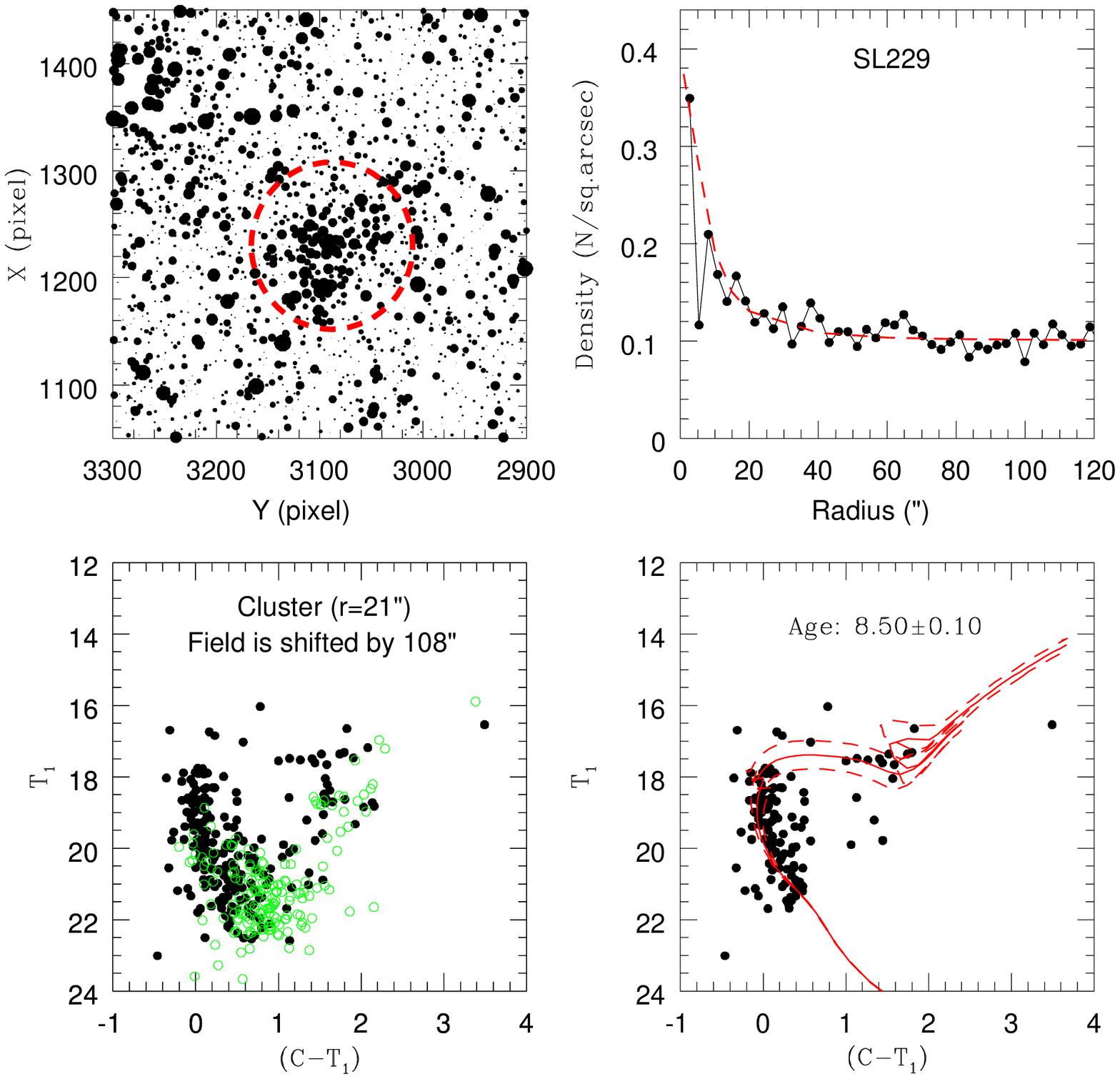} \\
 \includegraphics[height = .35\textheight, keepaspectratio]{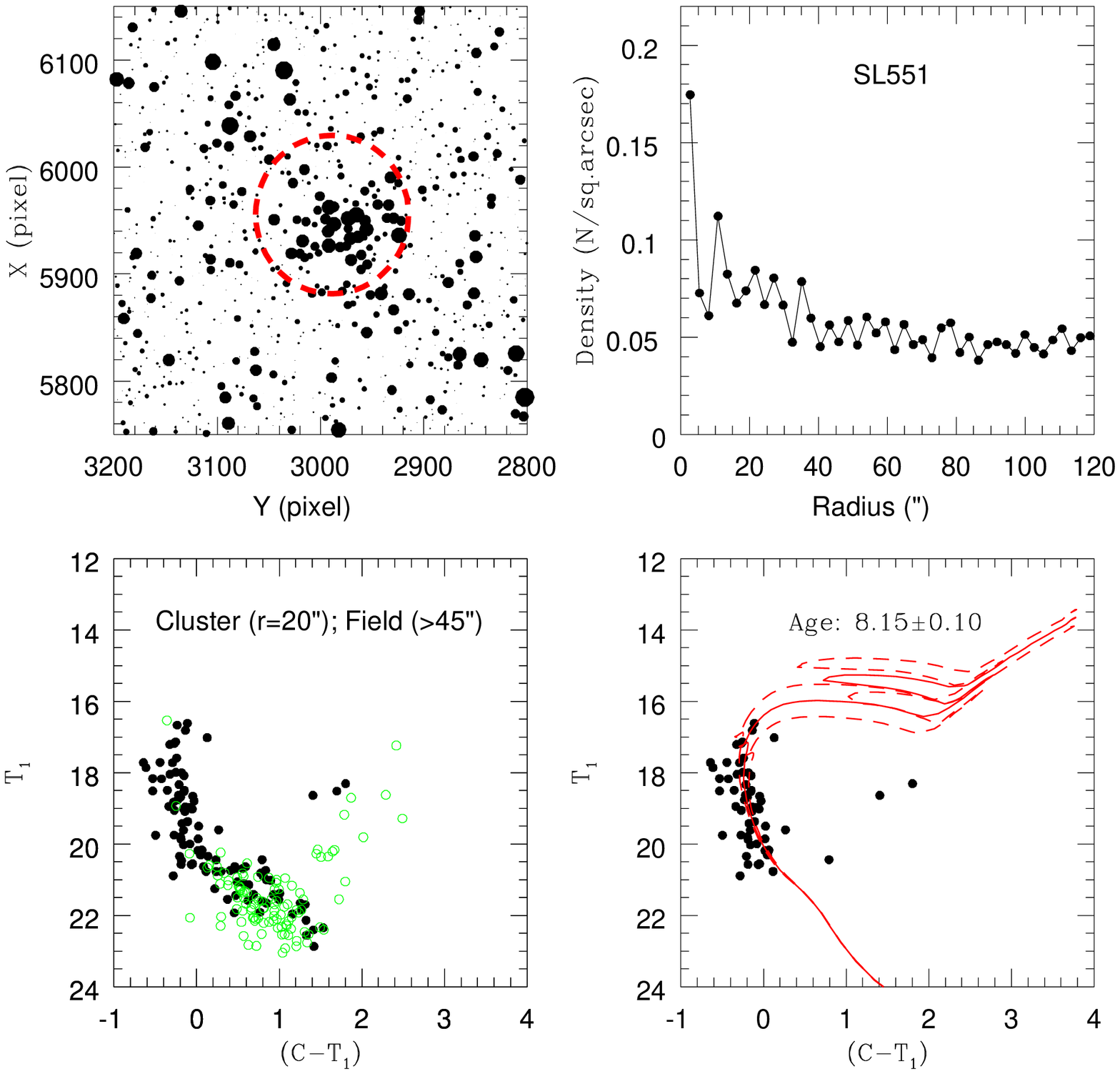}\hspace{1.0cm} 
 \includegraphics[height = .35\textheight, keepaspectratio]{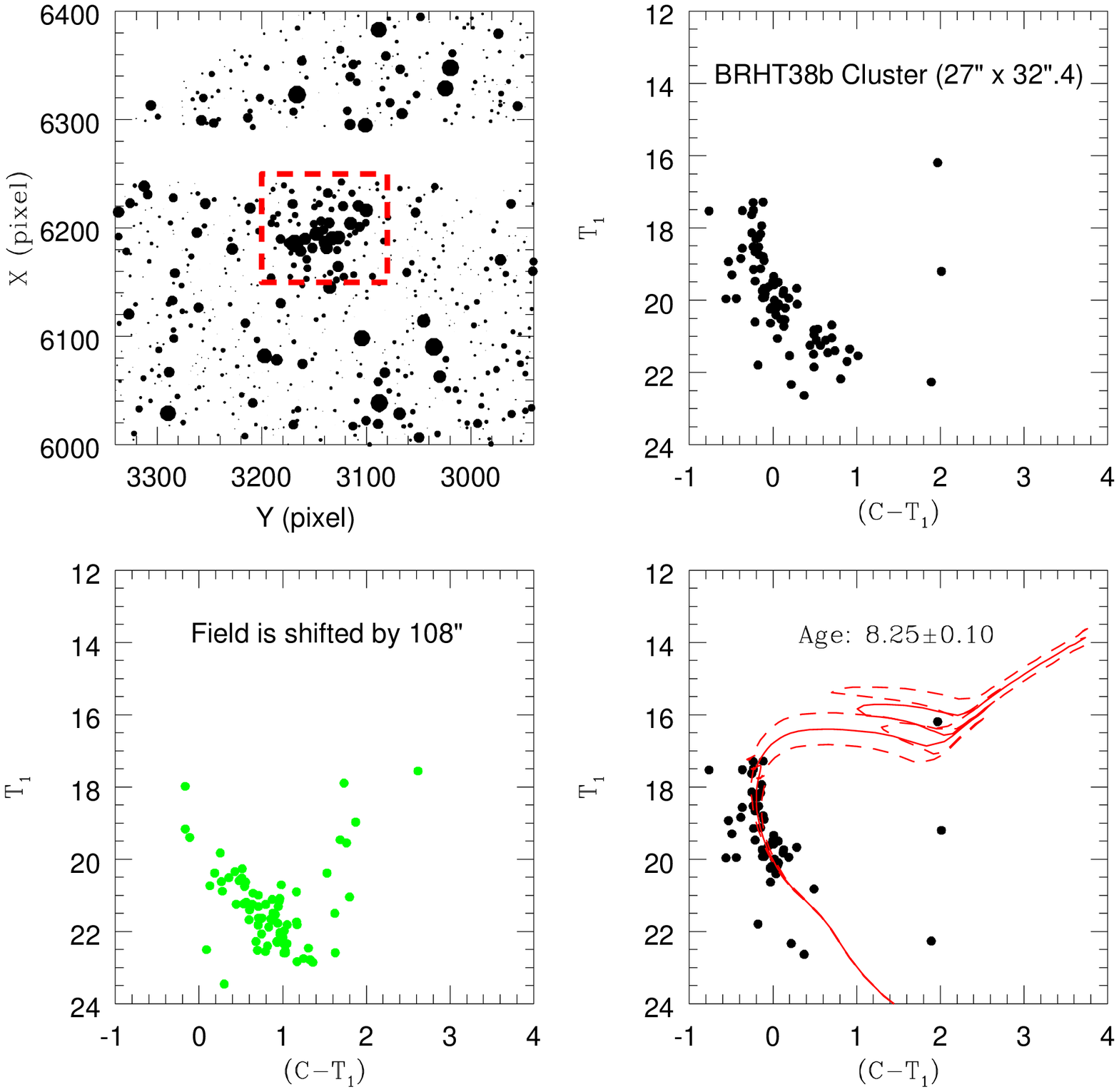} \\
\caption{\small {Double clusters: SL230 \& SL229; SL551 \& BRHT38b. For SL229 the field CMD is of a non-annular, similar sized region located away from the cluster. For SL551 no King profile over plot to RDP is shown. For SL230 and BRHT38b the top-right panel shows the CMD of stars within the estimated cluster size (black filled circles). Whereas, their bottom left-panel shows the CMD of an annular/non-annular, similar sized field region (green filled circles) located away from the cluster. The top-left and bottom-right panel description for these clusters are same as Figure 1.}}
\label{double2}
\end{center} 
\end{figure*}

\begin{figure} 
\begin{center} 
\includegraphics[height=3.5in,width=3.5in]{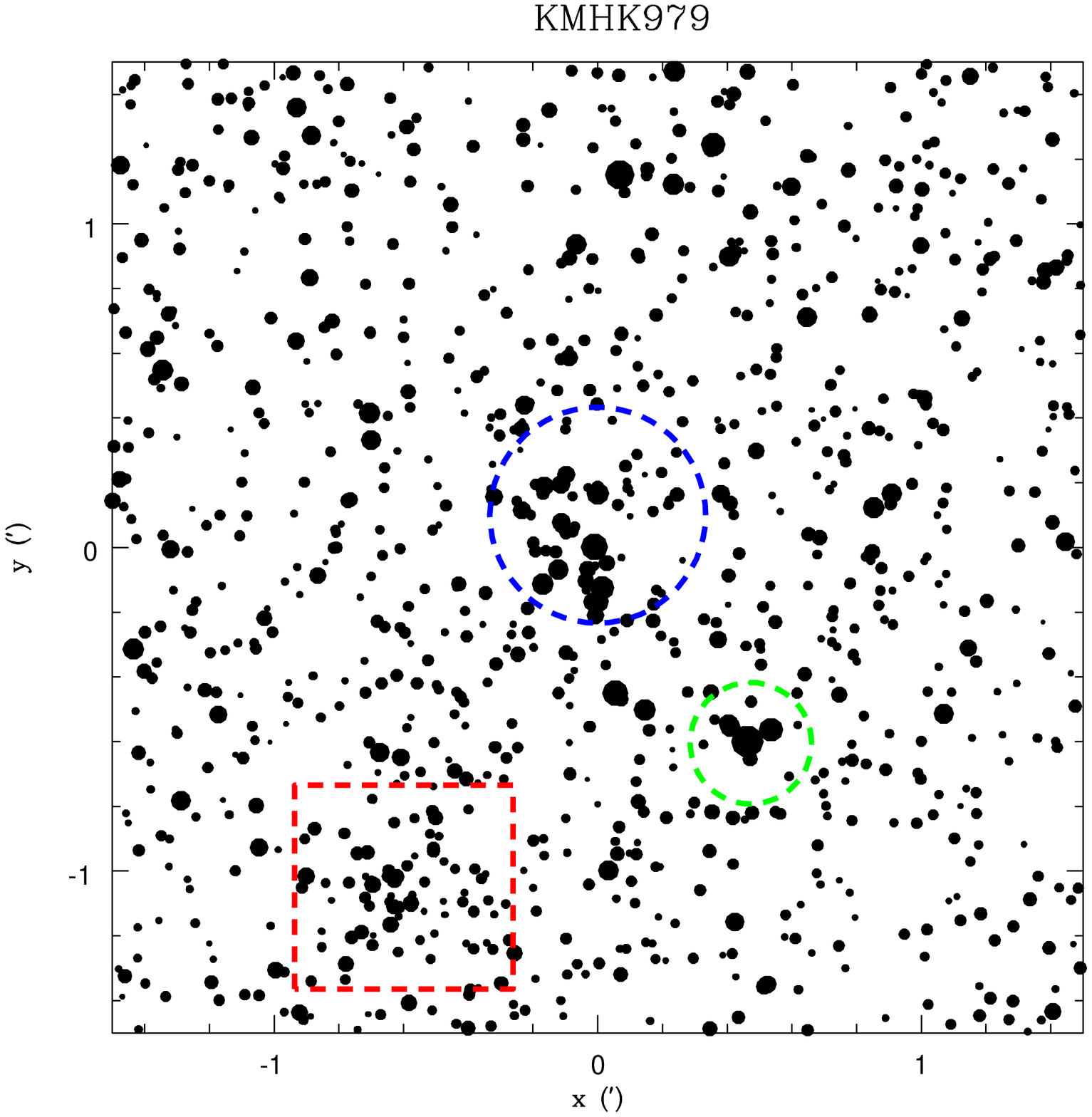}
\caption{\small {OGLE III field of KMHK979. The blue dashed circle and red rectangle denotes the extent of KMHK979 and HS329 respectively, as estimated using our Washington data. The green dashed circle denotes the average size of BSDL1980, estimated from Dieball et al. (2002).}}  
\label{ogle3}
\end{center} 
\end{figure}
\section[]{DOUBLE CLUSTERS}
%{\bf

In this section, we discuss the cases of double clusters. Their corresponding multi panel plots are shown in Figure \ref{double1} (BSDL341 \& H88-52; HS154 \& HS156; KMHK979 \& HS329) and Figure \ref{double2} (SL230 \& SL229; SL551 \& BRHT38b). The OGLE III spatial plot corresponding to KMHK979 is presented in Figure \ref{ogle3}.
%-----------------------
\begin{itemize}
\item BSDL341 and H88-52: These are a pair of a young and an intermediate age clusters, with their centers separated by $\sim$ 60$''$. BSDL341 is the younger ($\sim$280 Myr) of the pair showing a bright upper MS. H88-52 is an intermediate age ($\sim$ 1.1 Gyr) and compact cluster showing prominent MSTO and RC stars, located in the bottom south west direction of BSDL341. 

%-----------------------------------------------

\item HS154 and HS156: These are two clusters with their centers separated by $\sim$ 81$''$. HS154 is the young ($\sim$450 Myr, i.e. log(t)=8.65$\pm$0.10) cluster with a prominent upper MS. \cite{p12} has studied this cluster and estimated an age of log(t)=8.70$\pm$0.10. HS156 is an intermediate age ($\sim$ 1.1 Gyr) cluster showing a prominent MSTO, located in the eastern direction of HS154. \cite{paletal13} and \cite{p14}, both found HS156 to be an intermediate age cluster, aged $\sim$ 1 Gyr. In this study we looked upon HS154 and HS156 from the point of view of a double cluster. Also, we find excellent agreement of our derived ages for both these clusters when compared with their respective previous studies.

%----------------------------------------------
\item KMHK979 and HS329: KMHK979 is a young cluster aged $\sim$80 Myr.  We are unable to estimate a RDP for this cluster. We choose the cluster radius (20$''$) at which the cluster profile becomes well populated. Due to the proximity of clusters in the field, we choose fields of similar dimension in different parts of the observed field, away from the cluster, to clean the cluster profile. The cluster may be younger than our estimated age due to missing bright stars at the center. We have extracted the corresponding OGLE III field for this cluster (Figure \ref{ogle2}). The OGLE field shows clear density enhancement due to bright stars at the center of KMHK979, thus validating the existence of a true young cluster. 

The second member of the pair, HS329 is identified as a dense clump of stars distributed asymmetrically at $\sim$ 68$''$ away towards the south east direction of KMHK979. The CMD of this clumpy region shows the existence of evolved population like RC and RGB stars along with a spread in MSTO. An eye estimated center in the clumpy region is chosen as the cluster center and we estimate that the cluster is spread across an area of (37.$''$8$\times$40.$''$5) around this center, based on the clumpy distribution of the evolved stars. The CMD of the field region also shows the presence of evolved stars. In addition, the whole field suffers from high differential reddening owing to its location near the central region of the LMC, thus making it inconvenient to identify the cluster MSTO efficiently after cleaning. Due to insufficient coverage of field region on the southern side of the cluster, we could not perform annular field subtraction. Field star subtraction using fields of similar dimension in different parts of the observed field (away from the cluster) were hence tried out. A unique determination of age for this cluster was found to be difficult and we suggest that the cluster might be in the age range of 800 Myr to 1 Gyr.

According to \cite{d02} there exists a third cluster, BSDL1980 in this field with coordinates (5$^h$ 29$^m$ 33$^s$, -70$^\circ$ 59$'$ 38$''$). The average radius suggested by them is $\sim$11$''$, smaller when compared to the sizes of KMHK979 and HS329. According to G10, this cluster is a young one ($\sim$20 Myr) and is clearly seen as a small and poor density enhanced spot in the OGLE III spatial plot (Figure \ref{ogle3}) toward the south west direction of KMHK979. However due to the incompleteness of bright stars in our data, we have not been able to estimate the parameters of this cluster.

%-----------------------------------------------
\item SL230 and SL229: These are two young bright clusters with prominent upper MS and MSTOs. The younger cluster SL230 is aged $\sim$80 Myr. Some of the bright stars near the cluster center were saturated and hence the photometry could not be done. It was difficult to construct a RDP for SL230. We thus selected the cluster radius (r=25$''$) as the one at which the cluster features were found to be well populated. It is quite possible that we missed out some bright members of the cluster due to saturation and hence SL230 may be younger than our estimate. SL229 is located in the south west direction of SL230 with its center separated by $\sim$ 66$''$ and is aged around $\sim$320 Myr. Due to insufficient data coverage in the southern direction for SL229, we performed field star subtraction using circular fields in different directions away from the cluster. This cluster been studied by \cite{p12}, and the author gives a same age estimation as us. We studied these clusters as a double cluster, with a first time age estimation of SL230 using deep Washington photometric data along with a reconfirmation of the age of SL229.

%-----------------------------------------------
 
\item SL551 and BRHT38b: These are two young ($\sim$160 Myr) and bright clusters whose upper MS and MSTO are prominently visible in their respective CMDs. BRHT38b is located towards north east direction of SL551 within a separation of $\sim$ 77$''$. For BRHT38b we consider rectangular areas of different dimensions around an eye estimated center in the clumpy region and for which the cluster features get well populated gives us the extent of the cluster. Due to insufficient data coverage in the northern direction of BRHT38b, the cluster CMD is cleaned using rectangular fields of same area located in different parts of the observed field away from the cluster. 
 
%-----------------------------------------------

\end{itemize}

%}
%-----------------------------------------------

%-----------------------------------------------
\section[]{POSSIBLE CLUSTERS/ASTERISMS}
\begin{figure*}
\begin{center}
 \includegraphics[height = .35\textheight, keepaspectratio]{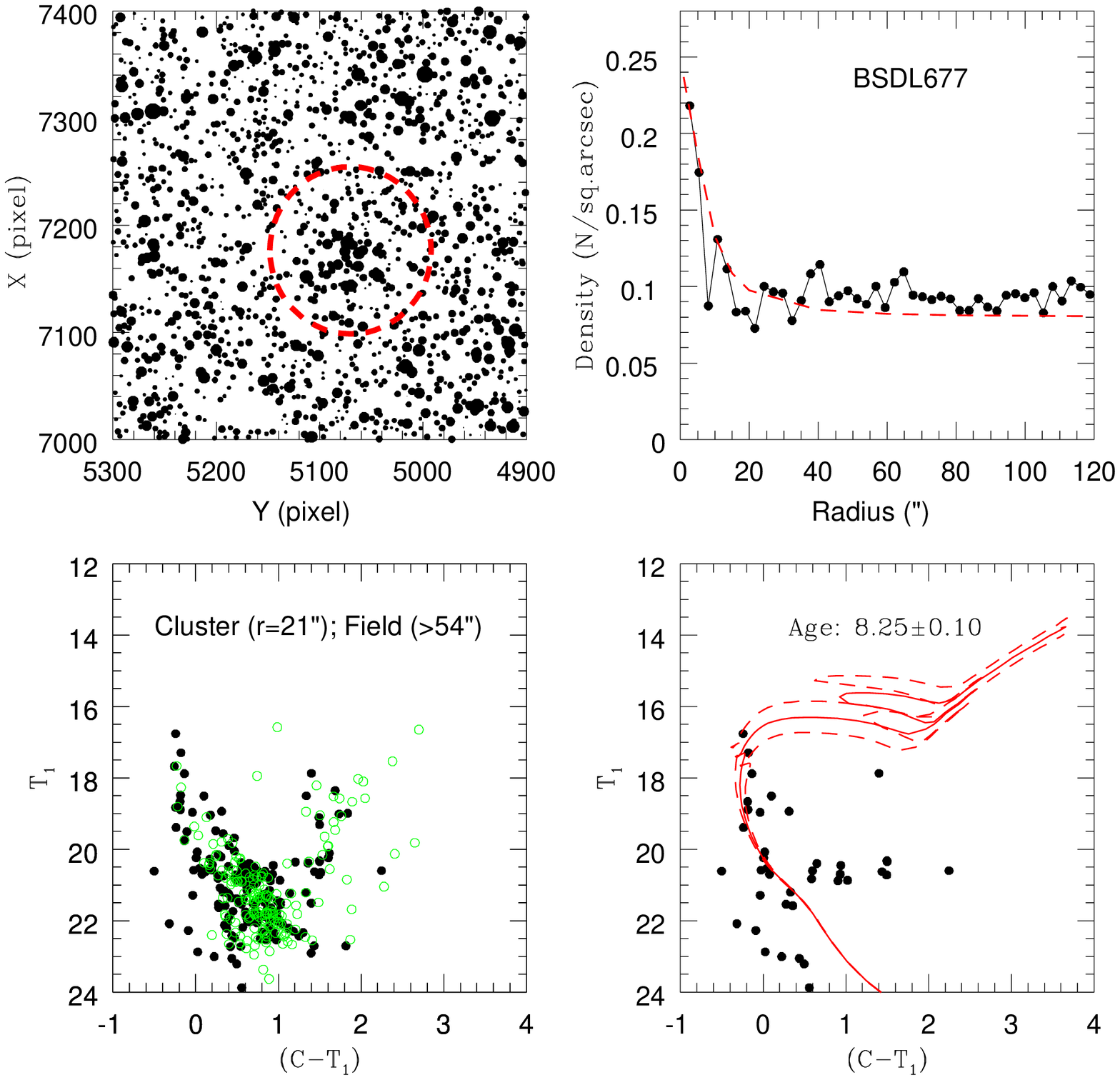}\hspace{1.0cm}
 \includegraphics[height = .35\textheight, keepaspectratio]{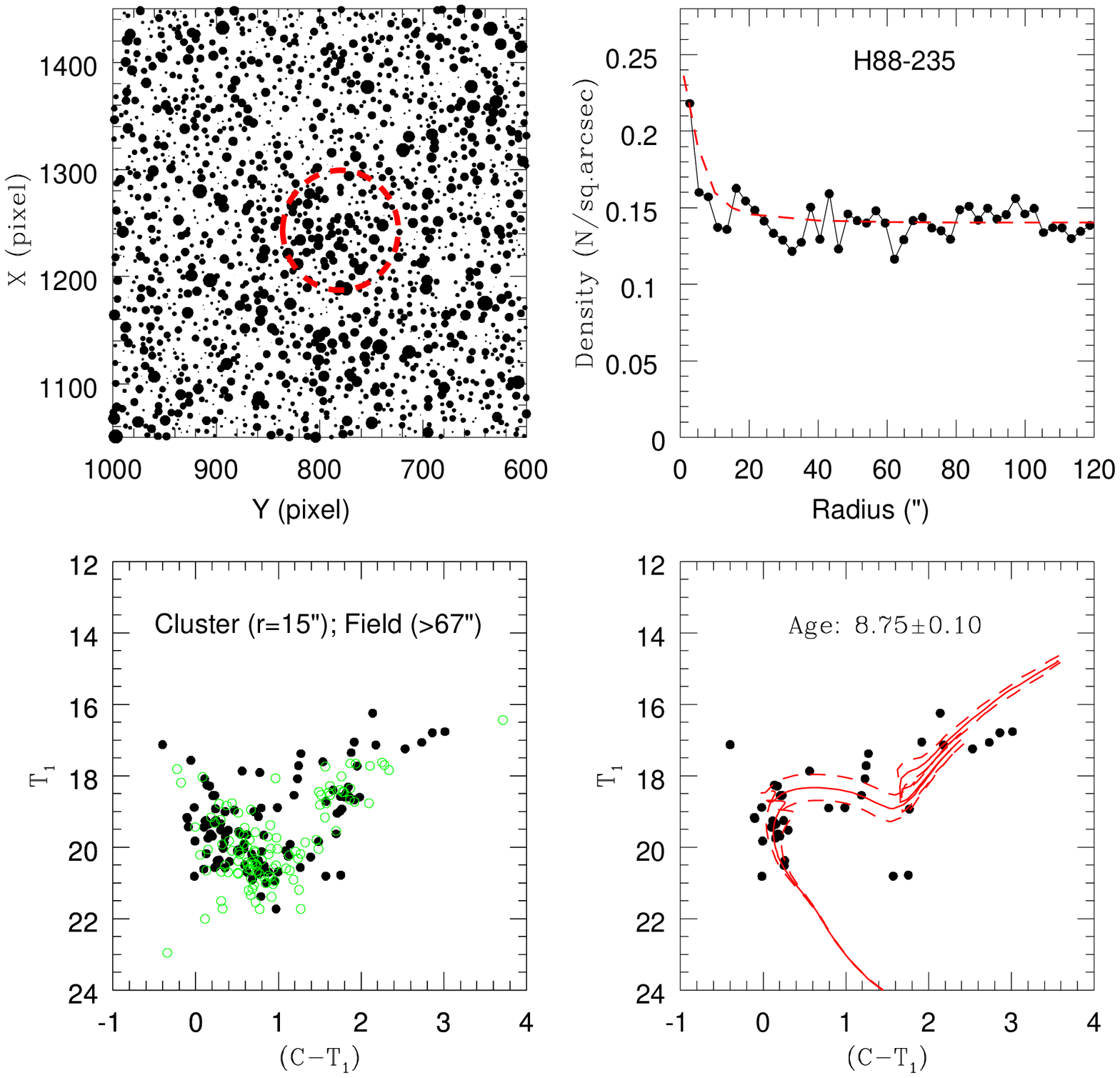} \\
 \includegraphics[height = .35\textheight, keepaspectratio]{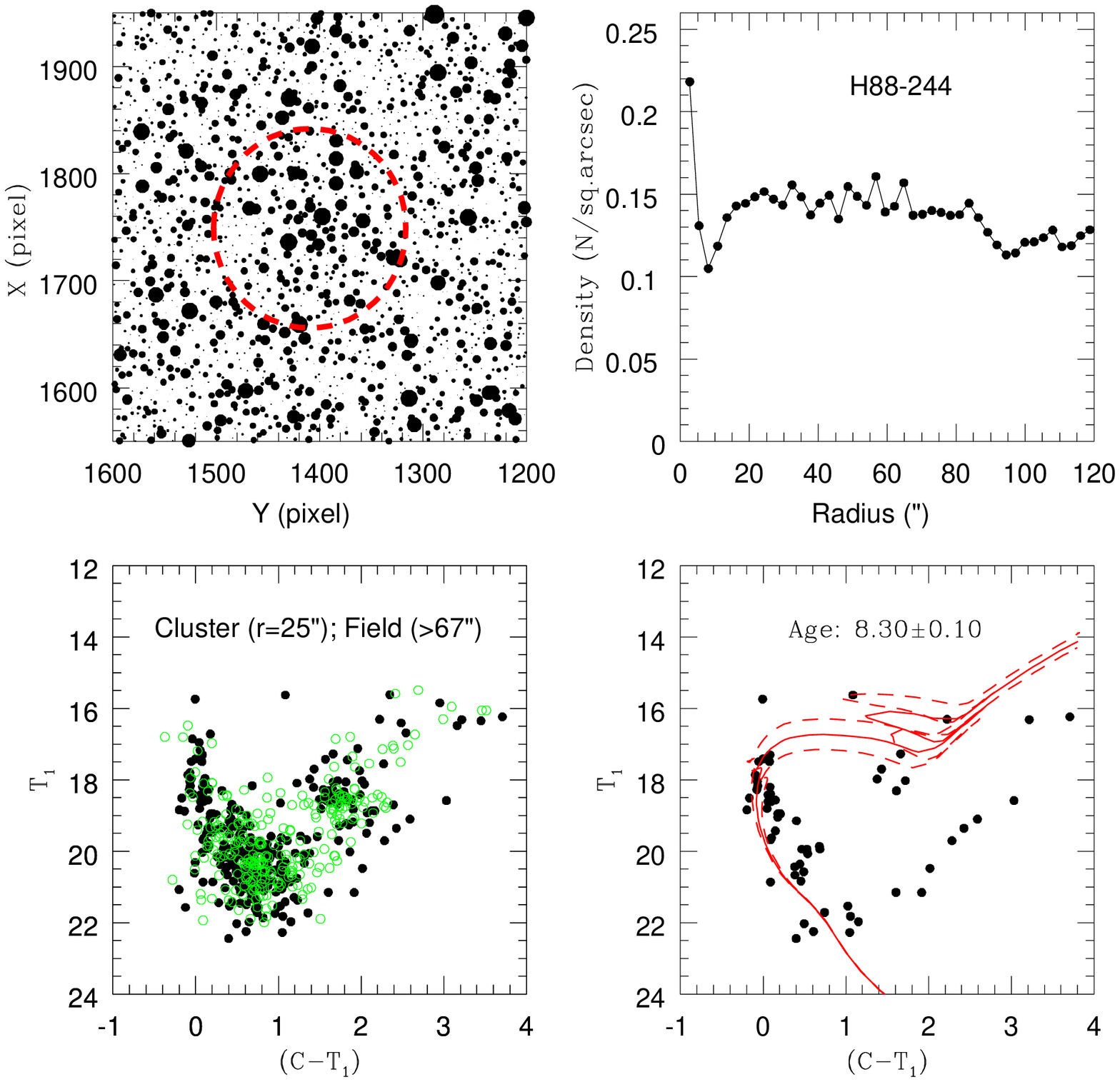} \\
 \includegraphics[height = .35\textheight, keepaspectratio]{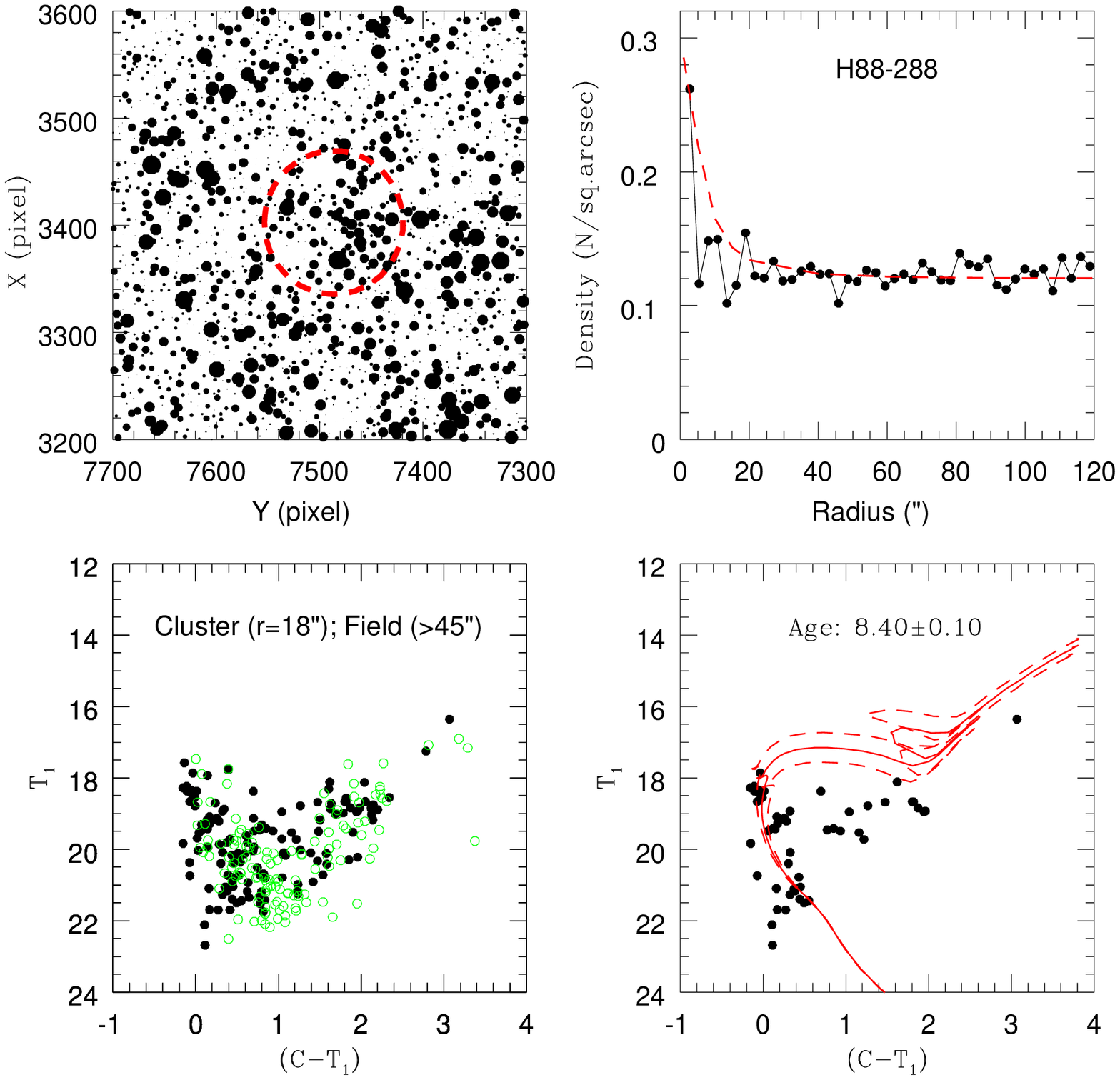}\hspace{1.0cm} 
 \includegraphics[height = .35\textheight, keepaspectratio]{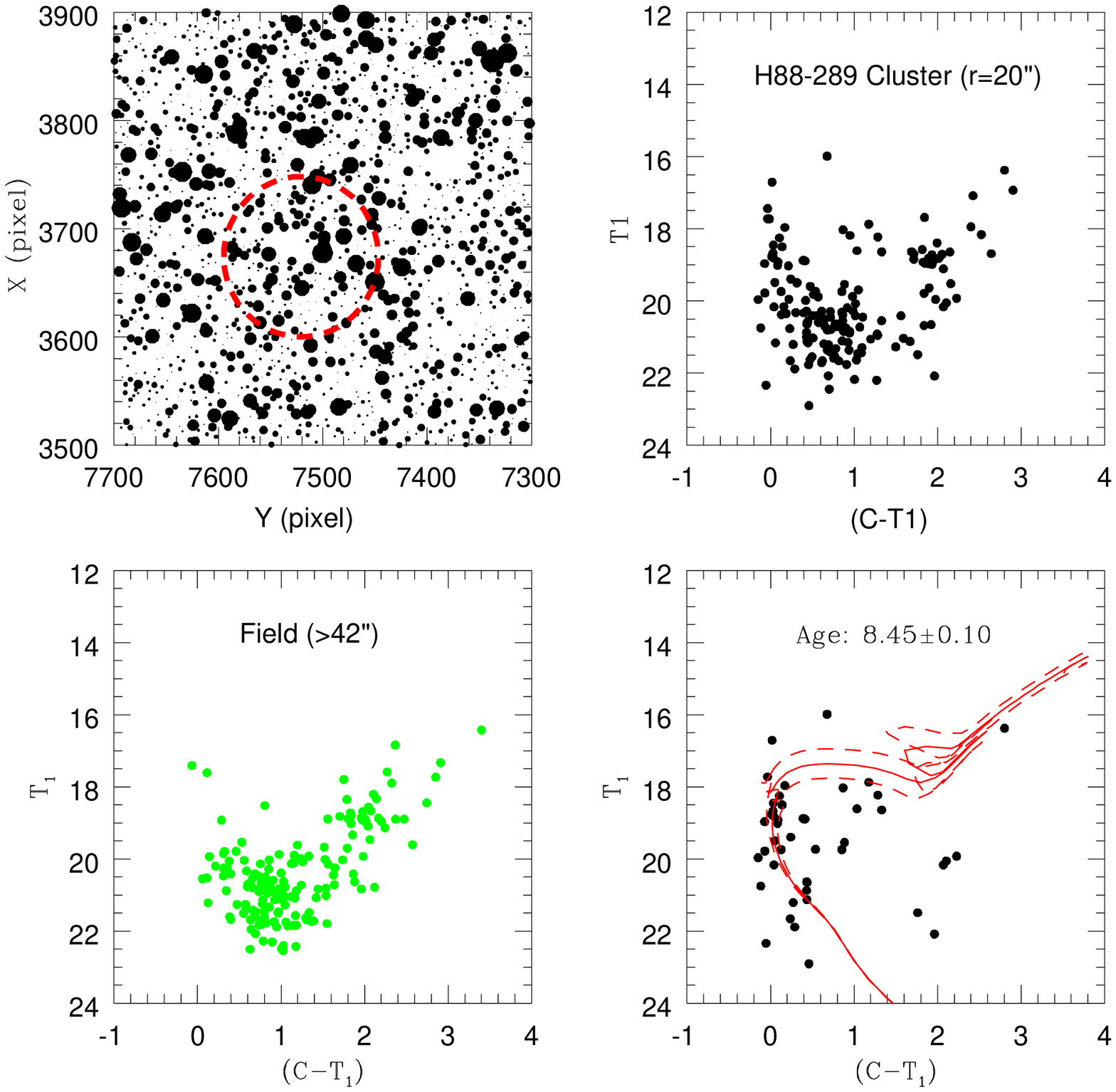}\\
 \caption{\small {Possible clusters/Asterisms: For BSDL677, H88-235, H88-244 and H88-288 the panel description is same as Figure 1, except, that in the case of H88-244 no King profile fit to RDP is shown. For H88-289 the top-right panel shows the CMD of stars within the estimated cluster radius (black filled circles), whereas the bottom-left panel shows the CMD of the annular field (green filled circles). The top-left and bottom-right panel description for H88-289 is same as Figure 1.}}
\label{possi1}
\end{center}
\end{figure*}

\begin{figure*}
\begin{center}
 \includegraphics[height = .35\textheight, keepaspectratio]{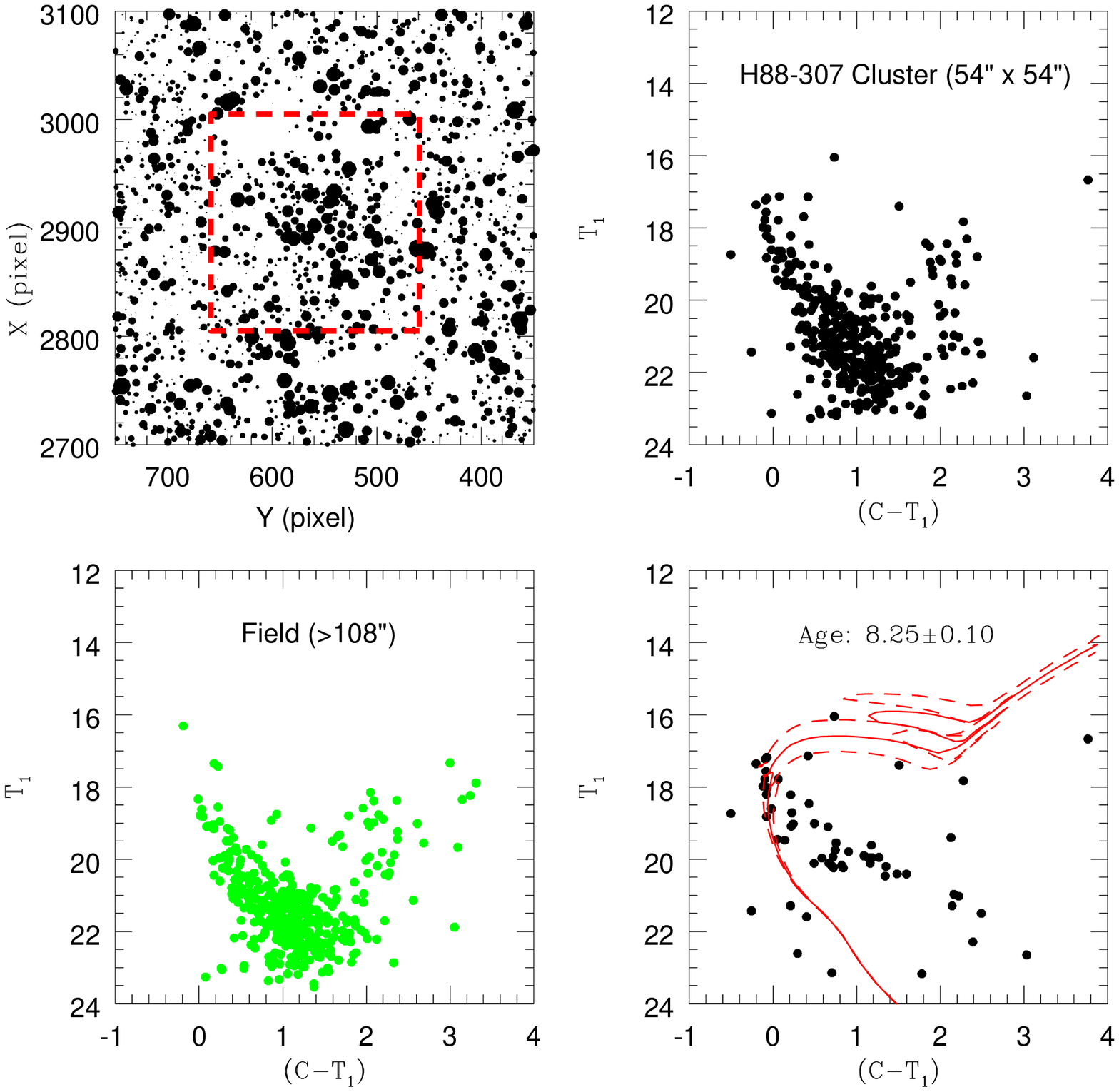}\hspace{1.0cm}
 \includegraphics[height = .35\textheight, keepaspectratio]{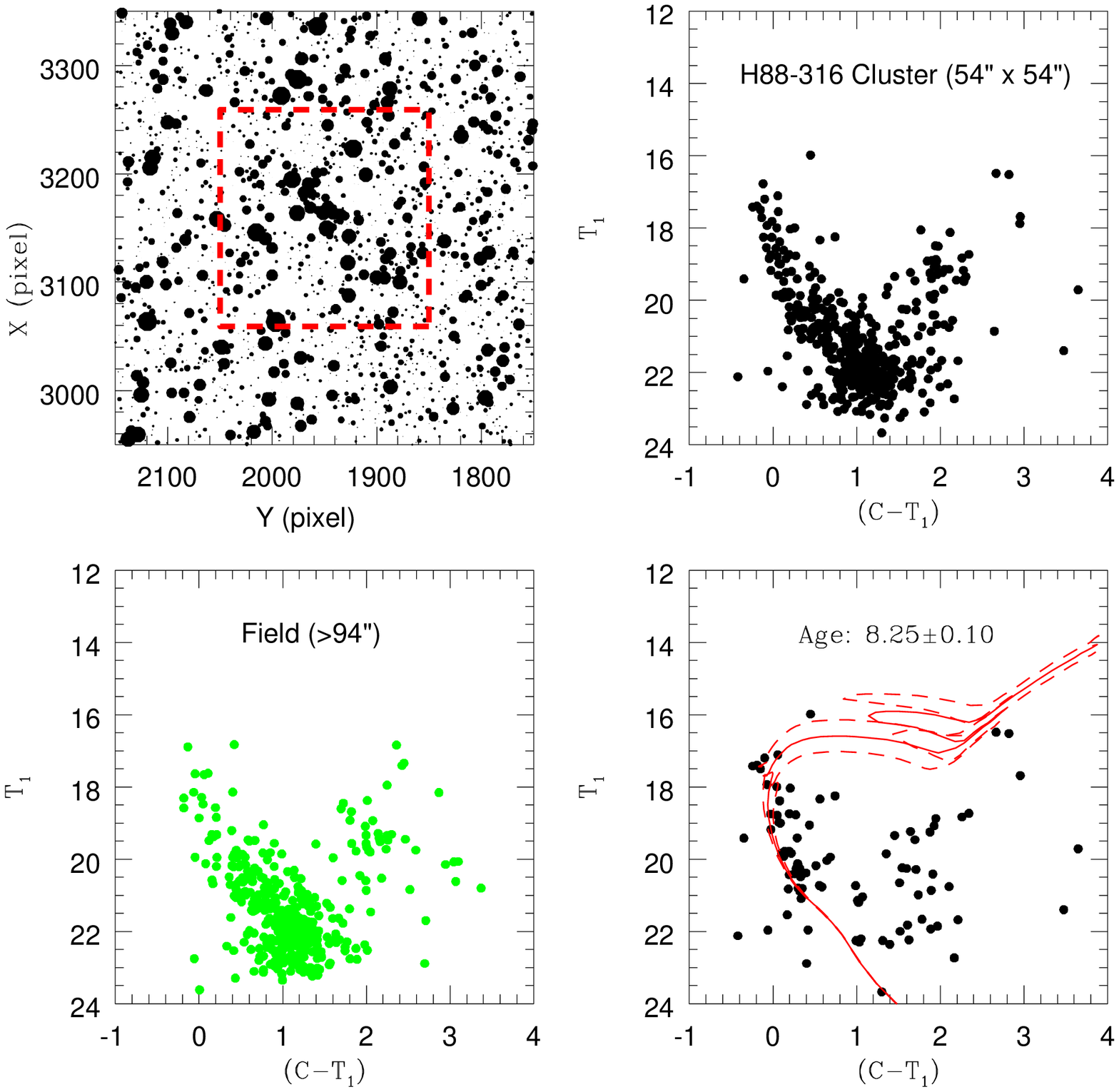} \\
 \includegraphics[height = .35\textheight, keepaspectratio]{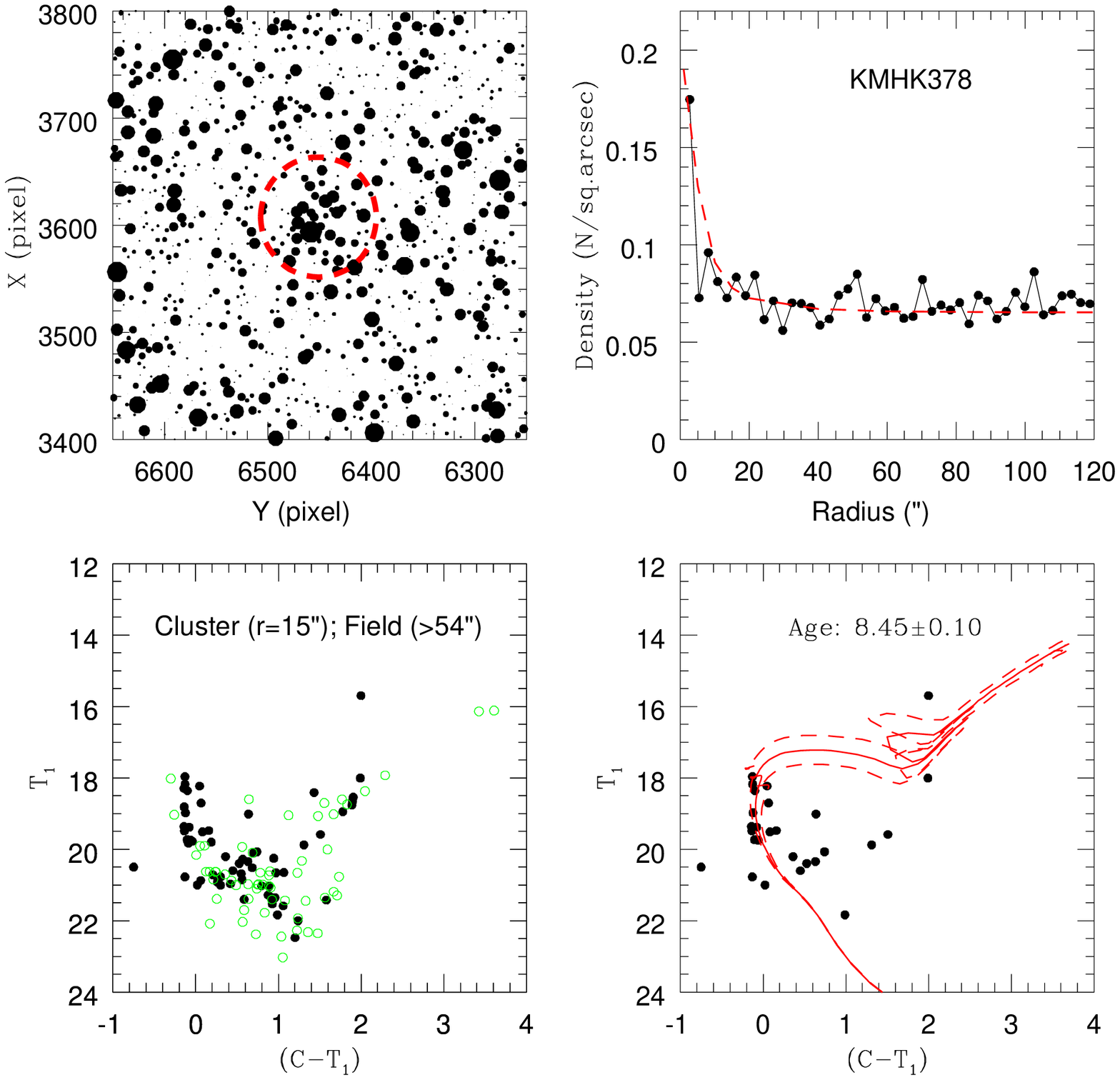} \\
 \includegraphics[height = .35\textheight, keepaspectratio]{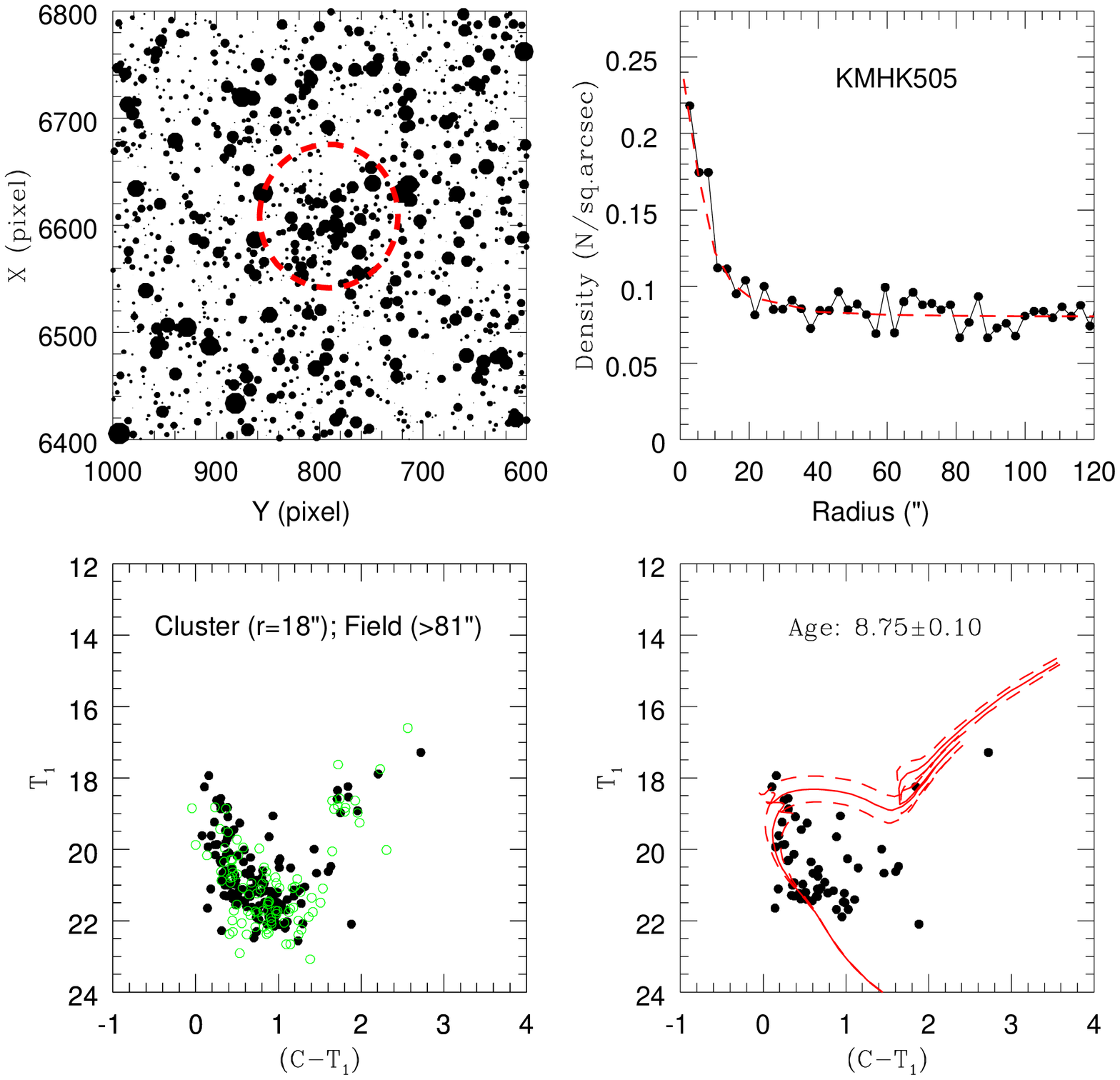}\hspace{1.0cm} 
 \includegraphics[height = .35\textheight, keepaspectratio]{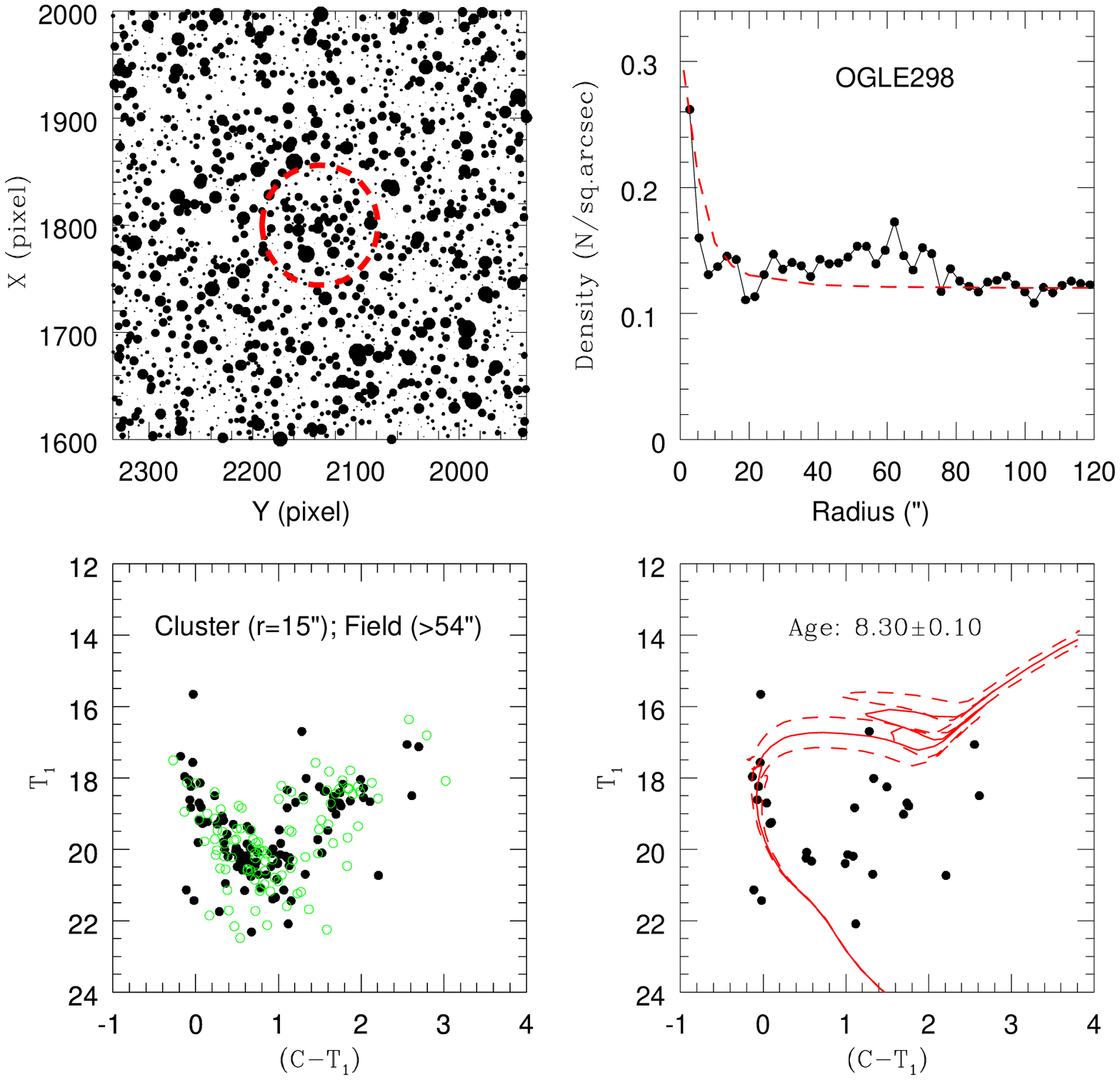}\\
\caption{\small {Possible clusters/Asterisms: For KMHK378, KMHK505 and OGLE298 the panel description is same as Figure 1. For H88-307 and H88-316, the top-right panel shows the CMD of stars within the estimated cluster size (black filled circles), whereas the bottom-left panel shows the CMD of the annular field (green filled circles). The top-left and bottom right-panel description for them are same as Figure 1. }}
\label{possi2}
\end{center}
\end{figure*}

\begin{figure*}
\begin{center}
 \includegraphics[height = .35\textheight, keepaspectratio]{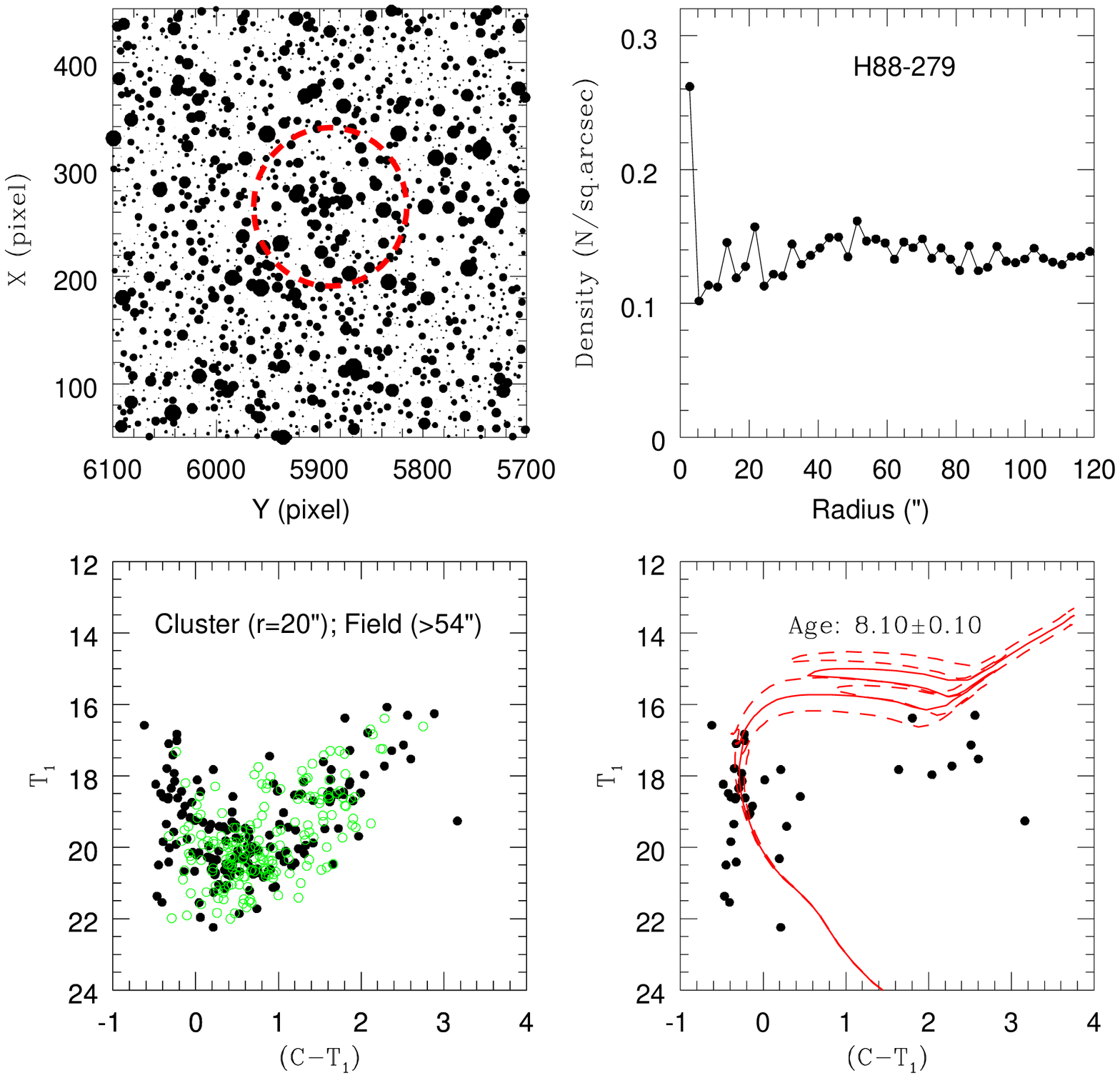}\hspace{1.0cm}
 \includegraphics[height = .35\textheight, keepaspectratio]{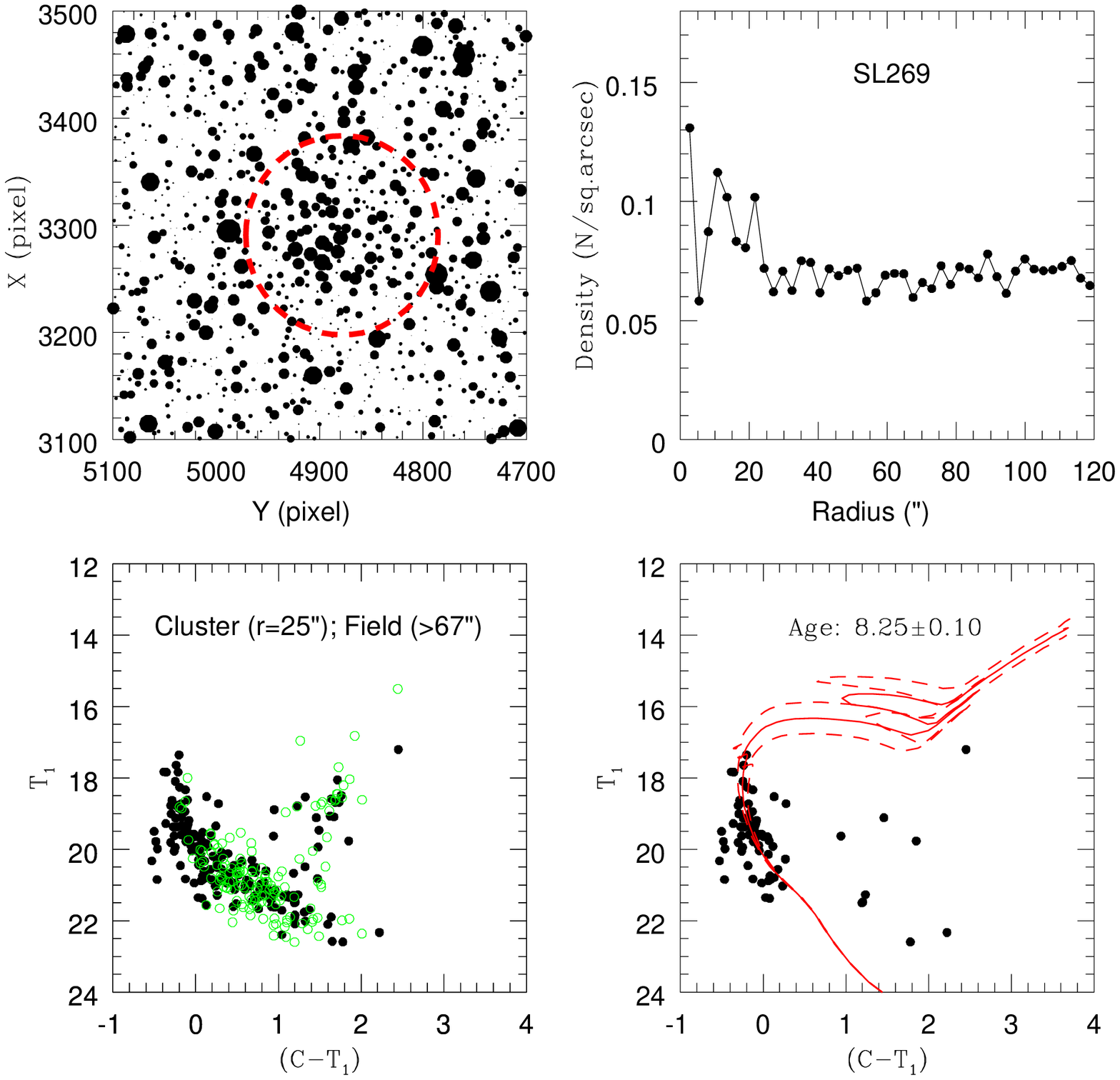}\\
\caption{\small {Possible clusters/Asterisms: For H88-279 and SL269, the panel description is same as Figure 1, except, that for both cases no King profile fit is shown.}}
\label{possi3}
\end{center}
\end{figure*}

\begin{figure*}
\begin{center}
 \includegraphics[height = .35\textheight, keepaspectratio]{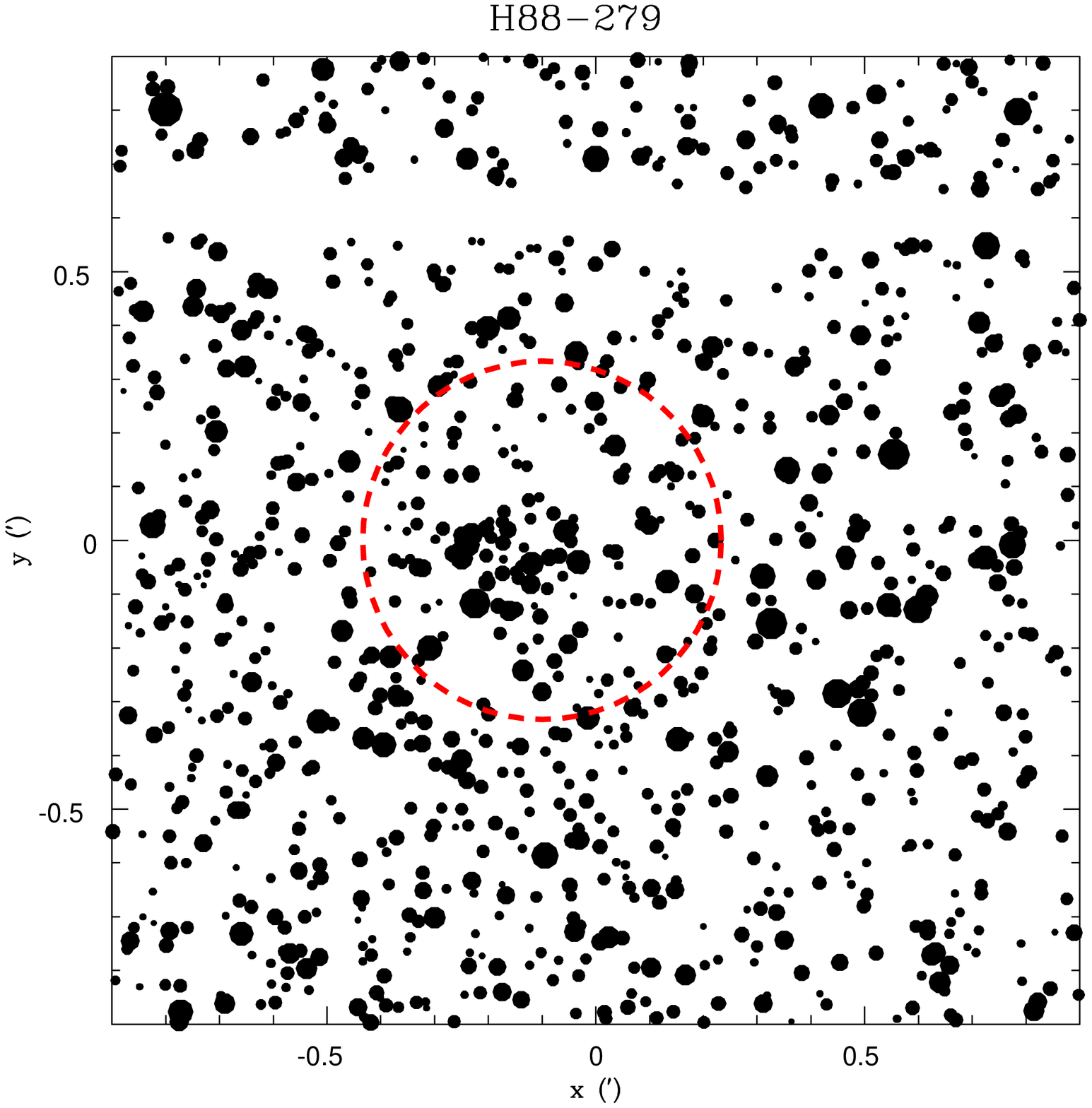}
 \includegraphics[height = .35\textheight, keepaspectratio]{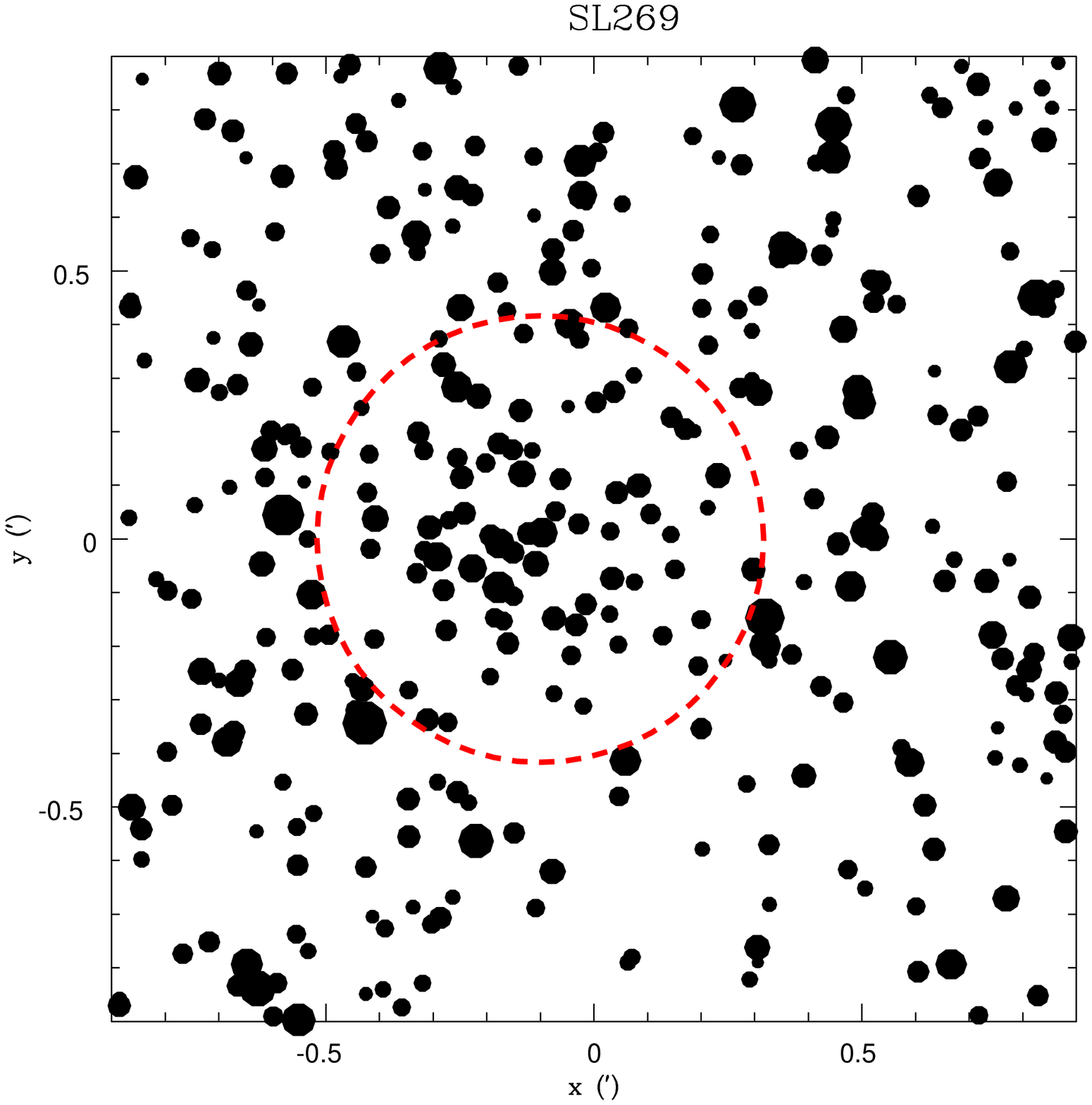}\\
 \caption{\small {The OGLE III schematic charts for H88-279 and SL269. The red dashed circles correspond to their derived size using our Washington data.}}
\label{ogle4}
\end{center}
\end{figure*}

%{\bf
We discuss the individual cases categorized as possible clusters/asterisms below. Their corresponding multi panel plots are shown in Figure 
\ref{possi1} (BSDL677, H88-235, H88-244, H88-288 and H88-289), Figure \ref{possi2} (H88-307, H88-316, KMHK378, KMHK505 and OGLE298) and Figure\ref{possi3} (H88-279 and SL269). The OGLE III spatial plot corresponding to H88-279 and SL269 are presented in Figure \ref{ogle4}.

\begin{itemize}

%-----------------------------------------------

\item BSDL677: The center of the cluster field (as mentioned in B08 catalog) is chosen as the cluster center. The estimated RDP shows a feeble density enhancement within a radius of 21$''$. The CMD of stars within this radius seems to exhibit a poor upper MS (brighter than 19.0 mag and bluer than 0.2 mag) which may belong to the cluster. The upper MS feature becomes unclear after field star subtraction, and visual fitting of isochrone to these few bright stars suggest an age of $\sim$ 180 Myr. Based on such feeble density distribution and unclear upper MS, we conclude that BSDL677 is either a very poor cluster or an asterism.
 
%-----------------------------------------------

\item H88-235: In the spatial plot no clear density enhancement is observed near the expected center of the cluster. We estimated the RDP, which supports the same, and shows only a very poor density enhancement within 15$''$. A CMD within this radius shows a few upper MS stars brighter than 20.0 mag and bluer than 0.6 mag which may belong to the cluster. We also find a good number of such stars distributed in the field, suggesting that the field as well as the cluster have similar MS and the field star subtraction renders a very poor cluster MS in the CMD. Visual fitting of isochrone to the poor upper MS suggest an age of $\sim$ 560 Myr (log(t)=8.75$\pm$0.20). Previous study by P12 mentions an age of $\sim$350 Myr (log(t)=8.55$^{+0.02}_{-0.08}$) for this cluster, which moderately agrees with our estimation within the errors. With almost negligible density enhancement, poor cluster feature and similar SFH as the field, H88-235 is likely to be an asterism.

%------------------------------------------------

\item H88-244: The spatial plot of H88-244 looks almost homogeneous with no significant density enhancement near the expected cluster center. The center of the cluster field (as mentioned in B08 catalog) is chosen as the cluster center and the CMD extracted within the estimated cluster radius (25$''$) seems to show a very poor upper MS (brighter than 19.0 mag and bluer than 0.4 mag) which may belong to the cluster. However the spatial distribution of the bright upper MS stars does not show any significant density enhancement at the location of the
cluster and looks almost homogeneous. Feeble cluster feature, similarity between field and cluster CMDs, as well as large amount of differential reddening within the field region makes it inconvenient to efficiently identify the presence of the cluster. The estimated reddening for this cluster is about 0.25 mag, much higher than the reddening of the corresponding field which is 0.10 mag. If at all a cluster exists, it is a small, poor and young ($\sim$ 200 Myr, i.e. log(t)=8.30$\pm$0.20) one. Previous study by PU00 mentioned an age of $\sim$125 Myr (log(t)=8.10$\pm$0.10) which agrees with our estimate within the errors.
%------------------------------------------------

\item H88-288 and H88-289: H88-289 is located towards the north of H88-288 at a distance of about 74$''$. They seem to show poor density enhancement around their respective cluster centers and exhibit a poor upper MS (brighter than 19.0 mag and bluer than 0.4 mag) within their estimated radii. Their size and age ($\sim$ 250 Myr i.e. log(t)=8.40$\pm$0.20) are almost similar. However H88-289 could be younger than our estimated age, as some bright MSTO/upper MS stars seem to be missing from the central region due to saturation effect. Using integrated photometry, P12 estimated the ages for H88-288 and H88-289 to be log(t)=8.04$^{+0.20}_{-0.04}$ and log(t)=7.80$^{+0.43}_{-0.03}$ respectively, which is younger relative to our estimate.  We realized that the field region for this pair suffers from variable density and reddening. Hence decontaminating the cluster CMD from field stars was found to be difficult for this pair of clusters. We conclude that with such poor density enhancement and cluster features along with issues related to variations in density and reddening within the field region, it is difficult to categorize H88-288 and H88-289 as true clusters.

\item H88-307: It is difficult to observe any prominent dense clump of stars within the central region. The cluster and the field regions show similar features in the CMD, thus posing difficulty to efficiently identify the presence of a cluster and estimate its corresponding parameters. The center of the cluster field (as mentioned in B08 catalog) is chosen as the cluster center. A comparison of the cluster and field CMD suggests a very poor upper MS (brighter than 19.0 mag and bluer than 0.4 mag). The cleaned CMD has very few MS stars. If at all a cluster is present, it could be poor, young ($\sim$ 180 Myr) located within an area of (54$''$$\times$54$''$) around the cluster center. \cite{d02} mention the presence of two more clusters within the same field. These are BSDL2768 (5$^h$ 40$^m$ 39$^s$, -69$^\circ$ 15$'$ 29$''$) and H88-306 (5$^h$ 40$^m$ 24$^s$, -69$^\circ$ 15$'$ 10$''$). B08 lists BSDL2768 as an association. G10 gives an estimation of its age as $\sim$ 60 Myr (log(t)=7.80, with 0.30$\leq$ error $<$0.50). For H88-306, \cite{p14} estimated an age of $\sim$125 Myr (log(t)=8.10$\pm$0.10). The coordinates of both BSDL2768 and H88-306 suggest that they could lie within the estimated area for the central cluster H88-307. However, given the broad and marginally dense stellar distribution coupled with similar SFH with respect to the field, we are not able to detect, identify and estimate the parameters for each of these clusters individually. 

%------------------------------------------------

\item H88-316: The spatial plot indicates an asymmetric distribution of bright stars around the expected cluster center. The cluster and the field regions show similar features in the CMD, thus posing difficulty to efficiently identify the presence of a cluster and estimate its corresponding parameters. The center of the cluster field (as mentioned in B08 catalog) is chosen as the cluster center. By comparing the CMDs of the central region with field regions at different annular radii, we conclude that there could be a cluster located within an area
of (54$''$$\times$54$''$) around the cluster center. The poor MS feature (brighter than 19.0 mag and bluer than 0.4 mag) identified in the central region is found to be $\sim$ 180 Myr (log(t)=8.25$\pm$0.20), suggesting that if at all a cluster is present, it is a poor and young located within the mentioned area. The age for this cluster has been previously estimated by G10 as $\sim$ 100 Myr (i.e log(t)=8.00, with 0.30$\leq$ error $<$0.50) whereas P12 estimated the cluster to be much younger, $\sim$ 18 Myr (log(t)=7.27$^{+0.30}_{-0.17}$). Our age estimation agrees well with that of G10 within the errors.

%---------------------------------------------------

\item KMHK378: The spatial distribution shows a small and feebly enhanced stellar distribution spot near the central region. The CMD extracted within the estimated cluster radius (15$''$) shows a poor upper MS (brighter than 20.0 mag and bluer than 0.4 mag) which could belong to the cluster. The MS feature does not get prominent for larger radii and is retained even after field star subtraction. The bright upper MS stars are found to be compactly distributed around cluster center. Thus there is a possibility that KMHK378 is a small, poor and young ($\sim$ 280 Myr i.e. log(t)=8.45$\pm$0.20) cluster candidate. According to \cite{d02}, another cluster, KMHK372 is present in the same field, with coordinates (4$^h$ 58$^m$ 07$^s$, -69$^\circ$ 48$'$ 16$''$) whereas B08 lists it as an association. The age estimated by G10 for KMHK378 is $\sim$ 25 Myr (log(t)=7.40, error $\leq$0.30) and is very similar to that estimated by P12, log(t)=7.37$^{+0.12}_{-0.14}$. For KMHK372, G10 estimated an age of $\sim$ 250 Myr (log(t)=8.40, with 0.30$\leq$ error $<$0.50). Given its coordinate, KMHK372 could lie within the estimated radius of KMHK378. However looking at the spatial plot, its difficult to identify them individually considering the density distribution near center is very poor. It is probable that we are estimating the parameters for KMHK378 and KMHK372 put together.

%---------------------------------------------------

\item KMHK505: Marginal density enhancement around the cluster center is observed in the spatial plot, which corresponds to a peak in the RDP within a radius of 18$''$. The cluster feature extracted within this estimated radius shows a poor and broad upper MS (brighter than 20.0 mag and bluer than 1.0 mag). The feature stays even after field star subtraction. A spatial distribution of these upper MS stars shows a density enhancement at the location of the cluster. The corresponding OGLE III field does not show any significant density enhancement within the cluster region. The reason could be the age of this cluster and the absence of bright giants. Based on our analysis we infer that KMHK505 could be a poor, small and young ($\sim$ 560 Myr) possible cluster candidate. 

%-----------------------------------------------

\item OGLE298: The spatial plot of OGLE298 appears very homogeneous without any significant density enhancement near the expected cluster center. The center of the cluster field (as mentioned in B08 catalog) is chosen as the cluster center and the CMD extracted within the estimated cluster radius (15$''$) seems to show a very poor upper MS (brighter than 19.0 mag and bluer than 0.4 mag) which may belong to the cluster. However the spatial distribution of the bright upper MS stars does not show any significant density enhancement at the location of the
cluster and looks almost homogeneous. Feeble cluster feature, similarity between field and cluster CMDs, as well as large amount of differential reddening within the OGLE298 cluster field makes it inconvenient to efficiently identify the presence of the cluster. The estimated reddening for this cluster is about 0.25 mag, much higher than the reddening of the corresponding field which is 0.10 mag. We conclude that if at all a cluster exists, it is a very small, poor and young ($\sim$ 200 Myr i.e. log(t)=8.30$\pm$0.20) one. The cluster has been previously studied by PU00, who claim the cluster to be much younger ($\sim$ 20 Myr i.e. log(t)=7.30$\pm$0.20). We detect only one bright MS star and could possibly derive a younger age if we consider it as a cluster member.

%------------------------------------------------

\item H88-279: The spatial plot of H88-279 appears quite homogeneous with no significant density enhancement near the expected center of the cluster. We considered an area of (54$''$$\times$54$''$) around the central region of the observed field. Comparison of CMDs of suspected cluster region and different field regions helped in identifying a poor cluster MS, brighter than 19.0 mag and bluer than 0.2 mag. For cluster areas greater than the above value, there is no change in the observed cluster MS. This suggests that there may be a cluster in the suggested location. In order to locate the cluster center, we made a spatial distribution of the bright upper MS stars and found a small compact distribution of them near the suspected cluster location. The cluster MS in the CMD was found to be well populated including stars within 20$''$ radius. At larger radii there is not much change observed in the extracted cluster feature. In order to cross check the presence of a young cluster in this field we extracted OGLE III data corresponding to this cluster, with similar area as our data. This schematic chart shows a feeble density enhancement due to bright stars around the cluster center (Figure \ref{ogle4}). We thus suggest that H88-279 is possibly a small, poor and young ($\sim$ 125 Myr, i.e. log(t)=8.10$\pm$0.20) cluster candidate. Earlier study by PU00, using OGLE II data mentioned an age of $\sim$ 100 Myr (log(t)=8.00$\pm$0.10), whereas P12 estimated relatively younger age of $\sim$ 74 Myr (log(t)=7.87$^{+0.06}_{-0.04}$). Our estimation is in good agreement with G10 and in moderate agreement with P12 within the errors.
 
%------------------------------------------------

\item SL269: The central region of the cluster does not look compact but dispersed resulting in an uneven RDP. An upper MS (brighter than 20.0 mag and bluer than 0.2 mag) is observed when the CMDs of the cluster and the field region are compared. The cluster radius is selected to be the one (25$''$) at which the upper MS becomes well populated. It is observed that the cluster features stays and appears prominent even after field star subtraction. A spatial distribution of these upper MS stars shows a density enhancement at the location of the cluster. The cluster is a young one, and the corresponding OGLE III field (Figure \ref{ogle4}) shows a marginal density enhancement within the cluster region. We conclude that SL269 may be a poor, small and young ($\sim$ 180 Myr) possible cluster candidate. 

%--------------------------------------------------               

\end{itemize}
%}

%---------------------------------------------------

%----------------------------------------------

%-------------------------------------------------


\begin{thebibliography}{}

\bibitem[\protect\citeauthoryear{Baumgardt et al.}{2013}]{baetal13}
Baumgardt H., Parmentier G., Anders P., Grebel E.K. 2013, MNRAS, 430, 676

\bibitem[\protect\citeauthoryear{Bessell \& Brett}{1988}]{bb88}
Bessell M. S., Brett J. M. 1988, PASP, 100, 1134

\bibitem[\protect\citeauthoryear{Bica et al.}{2008}]{betal08}
Bica E., Bonatto C., Dutra C.M., Santos J.F.C., Jr. 2008, MNRAS, 389, 678  

\bibitem[\protect\citeauthoryear{Bonatto et al.}{2010}]{boetal10}
Bonatto C., Ortolani S., Barbuy B., Bica E. 2010, MNRAS, 402, 1685

\bibitem[\protect\citeauthoryear{Canterna}{1976}]{c76}
Canterna R., 1976, AJ, 81, 228

\bibitem[\protect\citeauthoryear{Dieball et al.}{2002}]{d02}
Dieball A., M\"{u}ller H., Grebel E.K. 2002, A\&A, 391, 547

\bibitem[\protect\citeauthoryear{Geisler}{1996}]{g96}
Geisler D., 1996, AJ, 111, 480

\bibitem[\protect\citeauthoryear{Geisler et al.}{1997}]{getal97}
Geisler D., Bica E., Dottori H., Claria J.J., Piatti A.E., Santos J.F.C., Jr.
1997, AJ, 114, 1920

\bibitem[\protect\citeauthoryear{Geisler \& Sarajedini}{1999}]{gs99}
Geisler D., Sarajedini A. 1999, AJ, 117, 308

\bibitem[\protect\citeauthoryear{Glatt et al.}{2010}]{gletal10}
Glatt K., Grebel E.K., Koch A. 2010, A\&A, 517, A50

\bibitem[\protect\citeauthoryear{Grocholski et al.}{2006}]{groetal06}
Grocholski A. J., Cole A. A., Sarajedini A., Geisler D., Smith V. V.
2006, AJ, 132, 1630

\bibitem[\protect\citeauthoryear{Hunter et al.}{2003}]{huntetal03}
Hunter A.D., Elmegreen B.G., Dupuy T.J., Mortonson M., 2003, AJ, 126, 1836

\bibitem[\protect\citeauthoryear{Jannuzi, Claver \& Valdes}{2003}]{jetal03}
Jannuzi B.T., Claver J., Valdes F. 2003, The NOAO
Deep Wide-Field Survey MOSAIC Data Reductions,
{\small http://www.noao.edu/noao/noaodeep/ReductionOpt/frames.html}

\bibitem[\protect\citeauthoryear{Marigo et al.}{2008}]{metal08}  
Marigo P., Girardi L., Bressan A., Groenewegen M.A.T., Silva L., Granato G.L.
2008, A\&A, 482, 883

\bibitem[\protect\citeauthoryear{Olszewski et al.}{1991}]{olsetal91}
Olszewski E. W., Schommer R.A., Suntzeff N.B., Harris H.C. 1991, AJ, 101, 515

\bibitem[\protect\citeauthoryear{Palma et al.}{2013}]{paletal13}
Palma T., Clari\'{a} J.J., Geisler D. , Piatti A. E., Ahumada A.V. 2013, A\&A, 555, A131

\bibitem[\protect\citeauthoryear{Piatti et al.}{2003}]{petal03}
Piatti A.E., Bica E., Geisler D., Clari\'{a} J.J. 2003b, MNRAS, 344, 965

\bibitem[\protect\citeauthoryear{Piatti et al.}{2009}]{petal09}
Piatti A.E., Geisler D., Sarajedini A., Gallart C. 2009, A\&A, 501, 585

\bibitem[\protect\citeauthoryear{Piatti}{2011}]{p11}
Piatti A.E. 2011, MNRAS, 418, L40

\bibitem[\protect\citeauthoryear{Piatti et al.}{2011}]{petal11}
Piatti A.E., Clari\'a J.J., Parisi M.C., Ahumada A.V. 2011, PASP, 123, 519

\bibitem[\protect\citeauthoryear{Piatti}{2012}]{p12}
Piatti A.E. 2012, A\&A, 540, A58

\bibitem[\protect\citeauthoryear{Piatti \& Geisler}{2013}]{pg13}
Piatti A.E., Geisler D. 2013, AJ, 145, 17

\bibitem[\protect\citeauthoryear{Piatti}{2014}]{p14}
Piatti A.E. 2014, MNRAS, 440, 3091

%\bibitem[\protect\citeauthoryear{Piatti et al.}{2012}]{petal12}
%Piatti A.E., Geisler D., Matulena R., 2012, AJ, 144,100

\bibitem[\protect\citeauthoryear{Pietrzy\'{n}ski \& Udalski}{2000}]{pu00}
Pietrzy\'{n}ski G., Udalski A. 2000, Acta Astron., 50, 337

\bibitem[\protect\citeauthoryear{Popescu et al.}{2012}]{pop12}
Popescu B., Hanson M.M., Elmegreen B.G., 2012, ApJ, 751, 122

\bibitem[\protect\citeauthoryear{Saha et al.}{2010}]{saetal10}
Saha A., Olszewski E.W., Brondel B., et al. 2010, AJ, 140, 1719

\bibitem[\protect\citeauthoryear{Stetson, Davis \& Crabtree}{1990}]{setal90}
Stetson P.B., Davis L.E., Crabtree D.R. 1990, in ASP Conf. Ser. 8,
CCDs in Astronomy (San Francisco: ASP), 289

\bibitem[\protect\citeauthoryear{Subramaniam \& Subramanian}{2010}]{smsn10}
Subramaniam A., Subramanian S. 2010, ASInC, 1, 107

\bibitem[\protect\citeauthoryear{Subramanian \& Subramaniam}{2009}]{snsm09}
Subramanian S., Subramaniam A. 2009, A\&A, 496, 399

\bibitem[\protect\citeauthoryear{Udalski et al.}{2008}]{uetal08}
Udalski A., Soszy\'{n}ski I., Szyma\'{n}ski M.K., et al. 2008, Acta Astron., 58, 89
%Kubiak M., Pietrzy\'{n}ski G., Wyrzykowski L., Szewczyk O.,2,1, Ulaczyk K., R. Poleski R.

\bibitem[\protect\citeauthoryear{Zaritsky et al.}{2004}]{zetal04}
Zaritsky D., Harris J., Thompson I. B., Grebel E. K., 2004, AJ, 128, 1606



\end{thebibliography}
\end{document}